\documentclass[apj]{emulateapj}

\usepackage{ifthen}
\newboolean{preprint}
\setboolean{preprint}{false}

\newcommand{\CII}{\hbox{{\rm C}\kern 0.1em{\sc ii}}}
\newcommand{\CIV}{\hbox{{\rm C}\kern 0.1em{\sc iv}}}

\newcommand{\FeI}{\hbox{{\rm Fe}\kern 0.1em{\sc i}}}
\newcommand{\FeII}{\hbox{{\rm Fe}\kern 0.1em{\sc ii}}}
\newcommand{\SiII}{\hbox{{\rm Si}\kern 0.1em{\sc ii}}}
\newcommand{\AlII}{\hbox{{\rm Al}\kern 0.1em{\sc ii}}}
\newcommand{\NiII}{\hbox{{\rm Ni}\kern 0.1em{\sc ii}}}
\newcommand{\CrII}{\hbox{{\rm Cr}\kern 0.1em{\sc ii}}}

\newcommand{\ZnII}{\hbox{{\rm Zn}\kern 0.1em{\sc ii}}}
\newcommand{\NII}{\hbox{{\rm N}\kern 0.1em{\sc ii}}}
\newcommand{\OI}{\hbox{{\rm O}\kern 0.1em{\sc i}}}
\newcommand{\MgI}{\hbox{{\rm Mg}\kern 0.1em{\sc i}}}
\newcommand{\MgII}{\hbox{{\rm Mg}\kern 0.1em{\sc ii}}}

\newcommand{\HI}{\hbox{{\rm H}\kern 0.1em{\sc i}}}
\newcommand{\HII}{\hbox{{\rm H}\kern 0.1em{\sc ii}}}
\newcommand{\lya}{\hbox{{\rm Ly}\kern 0.1em$\alpha$}}
\newcommand{\Ly}{\hbox{{\rm Ly}\kern 0.1em$\alpha$}}
\newcommand{\Ha}{\hbox{{\rm H}\kern 0.1em$\alpha$}}
\newcommand{\Hb}{\hbox{{\rm H}\kern 0.1em$\beta$}}

\newcommand{\mpy}{\hbox{$M_{\odot}$~yr$^{-1}$}}

\newcommand{\msun}{\hbox{$M_{\odot}$}}

\newcommand{\kms}{\hbox{km~s$^{-1}$}}
\newcommand{\nn}{\nonumber}

\newcommand{\mmin}{$10^{11}$}
\newcommand{\mmax}{$1.5\times 10^{12}$}
\newcommand{\esfr}{\hbox{$\epsilon_{\rm sfr}$}}
\newcommand{\epsin}{\hbox{$\epsilon_{\rm in}$}}
\newcommand{\ein}{\hbox{$0.7$}}
\newcommand{\torb}{\hbox{$2\times 10^7$}} 
\newcommand{\rout}{0.6}

\newcommand{\be}{\begin{equation}}
\newcommand{\ee}{\end{equation}}
\newcommand{\bea}{\begin{eqnarray}}
\newcommand{\eea}{\end{eqnarray}}

\newcommand{\ifm}[1]{\relax\ifmmode#1\else$\mathsurround=0pt #1$\fi}

\newcommand{\ltsima}{$\; \buildrel < \over \sim \;$}
\newcommand{\lsim}{\lower.5ex\hbox{\ltsima}}
\newcommand{\gtsima}{$\; \buildrel > \over \sim \;$}
\newcommand{\gsim}{\lower.5ex\hbox{\gtsima}}
\newcommand{\prop}{\propto}

\newcommand{\Mv}{\hbox{$M_{\rm h}$}}
\newcommand{\Rv}{R_{\rm vir}}
\newcommand{\Vv}{V_{\rm h}}

\newcommand{\Vm}{V_{\rm max}}
\newcommand{\Vrot}{\Vm}

\newcommand{\Mmi}{M_{\rm min}}
\newcommand{\Mma}{M_{\rm max}}
\newcommand{\Msh}{\Mma}
\newcommand{\Mg}{M_{\rm g}}
\newcommand{\Ms}{M_{\star}}
\newcommand{\fg}{f_{\rm g}}
\newcommand{\sfr}{\dot{M}_\star}

\def\ssa{p}
\def\ssb{q}
\def\ssc{m}
\def\ssd{n}
\def\sse{s}
\def\ssf{t}

\slugcomment{Received Dec. 9 2009; Accepted June 1 2010}

\shorttitle{The role of an accretion floor}
\shortauthors{Bouch\'e et al.}

\begin{document}

\title{The Impact of cold gas accretion above a mass floor on galaxy scaling relations}

\author{N. Bouch\'e\altaffilmark{1,2},  
A. Dekel\altaffilmark{3}, 
R. Genzel\altaffilmark{1},
S. Genel\altaffilmark{1},  
G. Cresci\altaffilmark{1,4}, 
N. M. F\"orster Schreiber\altaffilmark{1}
, K. L. Shapiro\altaffilmark{5}
, R. I. Davies\altaffilmark{1}
, L. Tacconi\altaffilmark{1}
}

\altaffiltext{1}{Max Planck Institut f\" ur  extraterrestrische Physik, Giessenbachstrasse, D-85748 Garching, Germany}
\altaffiltext{2}{Department of Physics, University of California, Santa Barbara, CA 93106, USA}
\altaffiltext{3}{Racah Institute of Physics, Hebrew University, Jerusalem, 91904, Israel}
\altaffiltext{4}{Osservatorio Astrofisico di Arcetri, 50125 Florence, Italy}
\altaffiltext{5}{Department of Astronomy, Campbell Hall, University of California, Berkeley, 94720 CA, USA}

\begin{abstract}
Using the cosmological baryonic accretion rate and normal star formation (SF) efficiencies,
we present a very simple  model for star-forming galaxies  (SFGs)
that accounts for the   mass and redshift dependences of the SFR-Mass  and Tully-Fisher (TF)
 relations from $z\sim 2$ to the present.
The time evolution follows from the fact that
each modeled galaxy approaches  a steady state where the SFR 
follows the (net) cold gas accretion rate.
The key feature of the model is a halo mass floor $\Mmi\simeq 10^{11}$~\msun\
below which accretion is quenched in order to simultaneously account for 
the observed slopes of the SFR-Mass and TF relations.
The same successes cannot be achieved via a star-formation threshold (or delay)
nor by varying the SF efficiency or the feedback efficiency.
Combined with the mass ceiling for cold accretion  due to virial shock heating,
the mass floor $\Mmi$ explains  galaxy ``downsizing", where  more massive galaxies
formed earlier and over a shorter period of time.
It turns out that the model also accounts for the observed galactic baryon and gas fractions as a function of 
 mass and time, and the  cosmic SFR density,
which are all resulting from the mass floor $\Mmi$.
The model helps to understand that it is
the cosmological decline of accretion rate that drives the decrease of cosmic SFR density
between $z \sim 2$ and $z=0$ and 
the rise of the cosmic SFR density from $z\sim6$ to $z\sim2 $
 allows us to put a constraint on our main parameter $\Mmi\simeq 10^{11}$~\msun.
Among the physical mechanisms that could be responsible for the mass floor, we view that
 photo-ionization feedback (from first in-situ hot stars) lowering the cooling efficiency is likely to play a large role.
 \end{abstract}


\section{Introduction}
\label{section:introduction}

 To a good first order approximation,
 galaxies are either blue and active or red and passive, as indicated by the color bi-modality.
In the past few years,
it has been realized that blue star-forming galaxies (SFGs) lie on 
 a tight relationship between their stellar mass $\Ms$ and
 star-formation rate (SFR) \citep{BellE_05a,ElbazD_07a,NoeskeK_07a,DaddiE_07a,DroryN_08a,ChenY_09a,SantiniP_09a,
PannellaM_09a,DamenM_09a,OliverS_10a}.  This SFR-$M_\star$ relationship is  analogous 
 to the red sequence for passively evolving galaxies
and is sometimes referred to as  the {\it SFR sequence}. 
Every multi-wavelength survey has shown that the specific SFR
(sSFR$\equiv$SFR/$\Ms$) is higher for lower mass SFGs \citep[e.g.][]{BrinchmannJ_04a,BellE_05a,NoeskeK_07a,DroryN_08a,OliverS_10a}, i.e.,
 the mass index of the SFR-$\Ms$ relationship is less than unity. 
The SFR sequence has evolved by a factor  20 at a given stellar mass
from $z\sim2$ to the present time.
This strong evolution of the SFR sequence implies that
distant  $z \sim 2$ SFGs with 
$M_\star > 10^{10.6}\msun$ had SFRs  in excess of 100~\mpy 
\citep{ShapleyA_03a,ErbD_06c,GrazianA_07a,DaddiE_07a}.
Locally, such elevated SFRs are a  natural outcome of  merger-driven starbursts.
However, the tightness of this relation, with rms scatter of less than $0.3$~dex,
indicates that  SFR is not driven  
by merger-induced starbursts but rather by a continuous mass-dependent 
process that is gradually declining with time.

The mean mass dependence and time evolution of the relation between SFR and 
stellar mass can be summarized by the expression
\begin{eqnarray}
\dot{M}_\star &=& 150 \,\Ms{}_{,11}^{\ssa}\, (1+z)_{3.2}^{\ssb}\, \mpy\,,
\label{eq:sfrmass}
\end{eqnarray}
where 
$M_{\star,11} \equiv M_\star/10^{11}$\msun, 
$(1+z)_{3.2} \equiv (1+z)/3.2$,   $\ssa \simeq 0.8$, and $\ssb \simeq 2.7$
 in the redshift range $z=0-2$.~\footnote{
There are marginal indications for a variation of $\ssa$ from near 0.7
at $z=0-1$ \citep{BrinchmannJ_04a,NoeskeK_07a} to about
0.9 at $z \sim 2$ \citep{DaddiE_07a,SantiniP_09a,PannellaM_09a}.}
Reproducing the characteristic mass and time dependencies of the SFR sequence 
is a challenge for models of galaxy formation
\citep[e.g.][]{DaveR_08a,DamenM_09b}.

Equation~\ref{eq:sfrmass} is important as it is very
reminiscent of the halo mean growth rate, which has been shown to be
$\dot{\Mv} \propto \Mv{}^{\sse}(1+z)^{\ssf}$
with a mass index $\sse$ greater than unity $\sse\simeq1.1$ \citep{NeisteinE_08a,GenelS_08a,McBrideM_09a}
that is set
 by the shape of the initial dark matter (DM) power spectrum \citep{NeisteinE_06a,BirnboimY_07a}.
The similarity motivates a closer investigation, and
 we will show that, indeed, there is a strong intimate connection
between the growth of halos and the SFR sequence.

Another relevant scaling relation for (disky) SFGs is the well-known Tully-Fisher \citep[e.g.][]{TullyB_77a} relation,
which correlates stellar mass $\Ms$ and maximum circular velocity $V_{\rm max}$.
The TF relation appears to be already in place at high-redshifts $z\geq1$ \citep{KassinS_07a,PuechM_08a,EpinatB_09a,CresciG_09a} and
has evolved  by a factor of 2.5 at a given mass  \citep{CresciG_09a} from
$z\simeq2.2$ to the present. It is of the form: 
\be
M_\star \propto {\Vrot}^{\ssc}(1+z)^{\ssd} ,
\label{eq:TFR}
\ee
where $\ssc \simeq 4$ to a 10\% accuracy locally \citep[e.g.][]{TullyB_00a,McGaughS_05a,MeyerM_08a}.
As for the SFR sequence, Equation~\ref{eq:TFR} is reminiscent of the virial relation for DM halos, namely,
$\Mv \propto \Vv^{3}$.
Thus, both the SFR sequence and the TF relation behave similarly to their DM counterparts, but have
 slightly different mass indices, i.e., they
are tilted with respect to their DM counterparts.

Aside from our knowledge on these global properties,
our detailed understanding of individual SFGs has also improved greatly
thanks to spatially resolved kinematic studies of the ionized gas in $z \sim 2$ SFGs 
\citep[e.g. SINS survey,][]{ForsterSchreiberN_06a,ForsterSchreiberN_09a,GenzelR_06a,GenzelR_08a,ShapiroK_08a,ShapiroK_09a,CresciG_09a}.
The SINS survey, consisting of 80 $z=2$ SFGs, has revealed
 that 30\%--50\% of SFGs with $\Ms\sim10^{10.5\hbox{--}11.5}$ are gas-rich thick rotating disks
 \citep[see also][]{VanstarkenburgL_08a,WrightS_07a},
with the rest either dominated by dispersion velocities or showing 
the signature of recent mergers 
\citep[e.g.][]{LawD_07a,LawD_09a,ForsterSchreiberN_09a}.
The high-redshift SFG disks are different from local spirals in many ways.
In particular, they are thick and correspondingly of high velocity dispersion 
\citep{ForsterSchreiberN_06a,GenzelR_08a,CresciG_09a}, 
and they are often made of giant SF `clumps' \citep[][F\"orster Schreiber, Shapley et al. in
prep.]{CowieL_95b,GenzelR_06a,ElmegreenD_07a}.
This feature indicates wild gravitational instability,
which is a likely outcome of a high gas density 
\citep{NoguchiM_98a,ImmeliA_04a,BournaudF_07a,ElmegreenB_08a,DekelA_09b}.
Indeed, direct evidence for high gas fractions at high redshifts is now 
mounting \citep{ErbD_06b,DaddiE_08a,TacconiL_10a,DaddiE_10a} thanks to 
rapid progress in the sensitivity of mm interferometers.

The global scaling relations and the recent surveys of galaxy kinematics
 raise several outstanding questions:
(1) What causes the high SFRs at $z\simeq2$?
(2) What drives the evolution of the SFR sequence and TF relation from $z\sim2$
to the present?
(3) What drives the cosmological evolution of the average SFR 
density?
(4) Why did more massive galaxies form their stars before less massive SFGs?
(5) Why are $z=2$ SFGs so gas rich?

In this paper, we address these questions  in the context of the cosmological growth of 
DM halos. 
In particular,  we construct a very simple model that ties
the SFR sequence and TF relation to their DM counterparts.
The important feature of the model is the suppression of the accretion below a
mass floor at $\Mmi \sim 10^{11}$~\msun, which, as we will show,
 has several other important consequences.
We focus our attention at $z\simeq2$, i.e., when the universe was just 3 Gyr old.

This paper is organized as follows.  
In Section~\ref{section:model}, we present the model.
In Section~\ref{section:results}, we show that the SFGs reach a quasi-steady state and
demonstrate the impact of the mass floor $\Mmi$ on the SFR sequence  and the TF relation.
In Section~\ref{section:addresults}, we find that
the mass floor naturally 
delays the star-formation activity and  leads to ‘downsizing’.
The model also simultaneously accounts 
for the baryonic and gas fractions as a function of mass and time.
In Section~\ref{section:madau}, we put direct constraints on the numerical
value of $\Mmi$ from the cosmological history of star-formation density.
In Section~\ref{section:conclusions}, 
we discuss the model limitations and  present our conclusions.
Finally, in Section~\ref{section:accretionfloor}, we discuss the possible 
origin of the mass floor.
We use throughout the standard $\Lambda$CDM cosmology with the parameters
$\Omega_{\rm m}=0.3$, $\Omega_{\Lambda}=0.7$, $h=0.7$ and $\sigma_8=0.8$.

\section{The reservoir model}
\label{section:model}

In this Section we present our `reservoir' model 
and its two major ingredients (Section~\ref{section:bathtub}), namely,
the accretion efficiency, and the SF efficiency.
We discuss the mass and redshift dependences of these two parameters
in Section~\ref{section:massdependence}.
The model ingredients are then summarized in Section~\ref{section:overview}.
We will use this model as {\it a learning tool} to gain insights on
the role played by several key physical parameters.

\subsection{The basic equation}
\label{section:bathtub}

We consider a galaxy with its dark halo as a reservoir that is fed by a source
and is emptied into a drain. 
The source represents the amount of newly accreted cold gas, and the drain is the 
gas consumption into stars as well as outflows.
The basic equation of our model is the differential equation expressing
the conservation of gas mass,
\begin{eqnarray}
\dot M_{\rm gas}&=& \dot M_{\rm gas, in} - (1-R) \dot{M}_\star 
- \dot M_{\rm gas, out} \ , 
  \label{eq:bathtub}
\end{eqnarray}
where $\dot M_{\rm gas,in}$ is the gas accretion rate,  
$(1-R)\dot{M}_\star$ is the net SFR corrected for the 
recycled fraction, and $\dot M_{\rm gas, out}$ is the mass outflow rate.
For our purposes, $R$ is a kept time-independent given
its slow dependence on stellar population age $T$ for $T>10^9$~Gyr \citep{BruzualG_03a},
i.e., at $z<4$. We discuss the limitations of this assumption in Appendix~A.

To be as general as possible, 
we included an additional drain in our model (Equation~\ref{eq:bathtub})
representing any  outflowing gas.
Observationally, the outflow rate  $\dot M_{\rm gas,out}$ 
is observed to be roughly proportional to the SFR,  
$\dot M_{\rm gas,out}=a\;\times\;$SFR,
with $a\simeq1.0$ \citep{HeckmanT_00a,MartinC_05a,RupkeD_05a,ErbD_08a}.
We note, however, that some fraction of the supernova (SN) driven winds is likely to be recycled,
especially in the most massive halos, where much of the outflowing gas falls
back into the galaxy and can boost the stellar mass at $z=0$ by 33\% 
\citep[simulations by][]{OppenheimerB_08a}.
Hence, the net outflow rate $\dot M_{\rm gas,out}$ may be less than the SFR, i.e.,
$\dot M_{\rm gas,out}\lsim$SFR.

Regardless of the numerical value of the proportionality constant $a$
between outflow rate and SFR, 
Equation~\ref{eq:bathtub} can be re-written as
\be
\dot M_{\rm gas}= \dot M_{\rm gas,in} - \alpha \sfr \, ,  \nn
\ee
where $\alpha$ includes the corrections for recycling and outflows. 
This equation  expresses the trivial fact that the
gas reservoir $M_{\rm gas}$ will be filled up or get emptied depending on the
relative power of the source and the drain terms.
As discussed in section~\ref{section:steadystate},
the system will self-regulate itself to a steady state 
($\dot M_{\rm gas}\simeq 0$)
where  SFR is proportional to the cosmological accretion rate.

We stress that the feedback term will  have little impact on the scaling relations,
such as the SFR sequence, or the sSFR.
Indeed, if half of the accreting gas is entrained and expelled
 by the SN-driven winds, the SFR will be lowered by a factor of 2.
As a result, if the SFR is lower by a factor of 2, the stellar mass
 $\Ms=\int \rm d t$~SFR$(t)$ will be lower by the same factor,
and the SFR sequence will remain unchanged.

\subsubsection{Halo growth and gas accretion}
 \label{section:gasaccretion}

The gas replenishment term $M_{\rm gas,in}$
 in Equation~\ref{eq:bathtub} is required observationally.
For example, the G-dwarf problem \citep{VandenberghS_62a,SchmidtM_63a} 
calls for a significant amount of newly accreted gas 
\citep[e.g.][]{LarsonR_74a}.  Furthermore, the gas depletion
timescale in local massive galaxies is  $\sim$~few Gyr \citep[e.g.][]{WongT_02a,JamesP_08a}, i.e.,
much shorter than the time required to build their stellar masses.
Similarly, for $z \sim 2$ SFGs, the gas consumption timescale of less than
0.5~Gyr is shorter than their typical stellar ages of 1--2~Gyr  
\citep[e.g.][]{ErbD_08a,TacconiL_10a}, thus
requiring intense gas accretion.

Cold gas accretion is also required theoretically  
as it is a natural consequence of the `cold-accretion' regime \citep{BirnboimY_03a,KeresD_05a}
when the cooling time is shorter than the dynamical time \citep{WhiteS_91a}.
Furthermore, since the cold gas is not shock-heated to the virial temperature,
high gas accretion efficiency is a natural outcome of
 efficient penetration of cosmological cold streams into the inner galaxies at high
redshift, as seen in hydrodynamical cosmological simulations 
\citep{OcvrikP_08a,DekelA_09a}.

In this context, it is the halo growth rate of DM halos 
that is regulating the baryonic accretion.
The halo growth rate is by now well understood based on $N$-body
simulations and the extended Press-Schechter (EPS) analytic formalism
\citep{EfstathiouG_85a,WechslerR_02a,vandenBoschF_02a,
SpringelV_06a}. 
An approximation for the average mass growth rate of DM halos
of virial mass $\Mv$ at redshift $z$ in cosmological $N$-body 
simulations \citep{vandenBoschF_02a,GenelS_08a,McBrideM_09a}, 
which is also understood using the extended Press-Schechter (EPS)
analytic formalism \citep{NeisteinE_08a}, 
is obtained by the fitting function
\begin{equation}
\dot {\Mv} \simeq 510 \,\Mv{}_{,12}^{\sse} (1+z)_{3.2}^{\ssf} \, \mpy\,, 
\label{eq:growth}
\end{equation}
where  $\Mv{}_{,12}\equiv \Mv/10^{12}$~\msun, $(1+z)_{3.2}\equiv (1+z)/3.2$,
 $\ssf\simeq 2.2$, and  $\sse\simeq 1.1$, with the estimates for $\sse$ ranging from 1.08 to 1.14
\citep{NeisteinE_08a,GenelS_08a,McBrideM_09a}

Given the average halo growth rate, the corresponding average gas accretion 
rate is 
\begin{eqnarray}
\dot M_{\rm gas,in} &=& \epsin\, f_{\rm b}\, \dot{\Mv} \nn\\
  &\simeq& 90  \, \epsin\, f_{{\rm b},0.18}\, \Mv{}_{,12}^{1.1}\,               (1+z)_{3.2}^{2.2} \,\mpy\, ,
\label{eq:gasaccretion}
\end{eqnarray}
where $f_{\rm b,18}\equiv f_{\rm b}/0.18$ is the cosmic baryonic fraction, 
and $\epsin \lsim 1$ is the accretion efficiency expressing  
the effective fraction of the baryonic matter that is actually accreted as cold 
gas into the galaxy.

The accretion efficiency  \epsin\ is a very important parameter
and must be within the range 0.5--1.0 for  several reasons.
Theoretically, $\epsin$ is high under the cold-accretion regime \citep{DekelA_09a}.
Observationally, this efficiency is supported by the fact that the observed 
SFR in massive galaxies at $z \sim 2$ matches within a factor of two
the maximum predicted gas accretion rate \citep{GenelS_08a,DekelA_09a}.
In addition, \citet{GenelS_08a} showed that the halo  major merger fraction 
in the mass range of interest $\log \Mv=12$ is 
low and consistent with the low merger fractions in the SINS survey \citep{ShapiroK_08a,ForsterSchreiberN_09a},
 provided that the accretion efficiency \epsin\ is greater than 50\%\
\citep[see Fig.~1 of][]{GenelS_08a}.
Recently, \citet{BauermeisterA_10a} compared the inferred `external' gas accretion to the DM accretion rate
and found this efficiency needs to be $>70$\%.
In light of all these results, we will adopt a fiducial value of $\epsin=0.7$.

As mentioned in the Introduction,
this baryonic  growth rate (Equation~\ref{eq:gasaccretion})
 is    very reminiscent of the SFR-Mass relationship (Equation~\ref{eq:sfrmass}),
given the similar  near-linear increase of the accretion rate with  mass.
However, the mass index $\sse$ in the accretion rate
 is slightly larger than unity, while the mass index $\ssa$ in the 
SFR sequence   is somewhat smaller then unity.
The clear implication from this difference is that the specific accretion
rate increases with mass for the DM component, while the
sSFR decreases with mass. 
Thus, if there is a connection between Equation~\ref{eq:gasaccretion}
and Equation~\ref{eq:sfrmass}, the difference   has to be explained (see   Section~\ref{section:results}).

\subsubsection{Star formation rate}

Star formation, a very complex, local and inefficient process
(giant molecular
clouds (GMCs) typically turning 1-2\%\ of the gas mass into stars in a free-fall 
time) is the primary drain of the reservoir (in Equation~\ref{eq:bathtub}).
On galaxy scales, the amount of gas consumed  is well described by 
the empirical relation:
\begin{eqnarray}
\Sigma_{\rm SFR}&=& \esfr \Sigma_{\rm gas}/t_{\rm dyn}, \; \; \hbox{i.e.,} \nn\\
\hbox{SFR} &=& \esfr M_{\rm gas}/(t_{\rm dyn})	  \label{eq:KS},
\end{eqnarray}
where $t_{\rm dyn} = R_{1/2}/V_{\rm c}$ is the galaxy dynamical time,
and $\esfr$ is the SFR efficiency parameter.
For a marginally unstable disk with Toomre~$Q$ parameter $Q\sim1$, it can be shown
\citep{MartinC_01a,KrumholzM_07a}
that Equation~\ref{eq:KS} is equivalent to the traditional Kennicutt-Schmidt (KS) relation
\citep{SchmidtM_59a,KennicuttR_98a},
which relates the surface densities of gas and SFR via
$\Sigma_{\rm SFR} \prop \Sigma_{\rm gas}^{1.5}$.
 
The locally inferred value by \citet{KennicuttR_98a} for the SFR efficiency is $\esfr \simeq 0.02$ and 
the most recent advancements in constraining the KS relation at $z\sim2$ show no evidence for any evolution
\citep{GenzelR_10a,DaddiE_10b}. In addition, at $z \sim 2$, we know that massive disky SFGs, with $\Ms> 10^{10.5}\msun$,
 extend to half-light radii of $R_{1/2} \sim 4$~kpc \citep{BoucheN_07b} 
and they rotate with circular velocities $V_{\rm c} \sim $200~\kms\ 
\citep{ForsterSchreiberN_06a}.
Expressing the quantities in these units, the orbital time can be written as  
\begin{equation}
t_{\rm dyn}= \torb {\rm yr}\, \left (\frac{R_{1/2}}{4\,\hbox{kpc}} \right )\, \left (\frac{V_{\rm c}}{200\,\kms}\right )^{-1} \, .
\label{eq:torb}
\end{equation}
At $z>2$, we assume that the orbital time scales with the halo dynamical time,
$\Rv/\Vv$, which is a fixed fraction, about 18\%, 
of the Hubble time at the given redshift, thus proportional to $(1+z)^{-3/2}$.

We note that our results on the scaling relations presented in \ref{section:orig:sfrmass} are completely independent
of the SF efficiency (i.e., \esfr\ and $t_{\rm dyn}$).
 This is because the SFR is going to be driven
by the cosmological accretion rate, as explained 
in Section~\ref{section:steadystate}.

\subsection{Mass dependence}
\label{section:massdependence}

\begin{figure*}
\centering
\plottwo{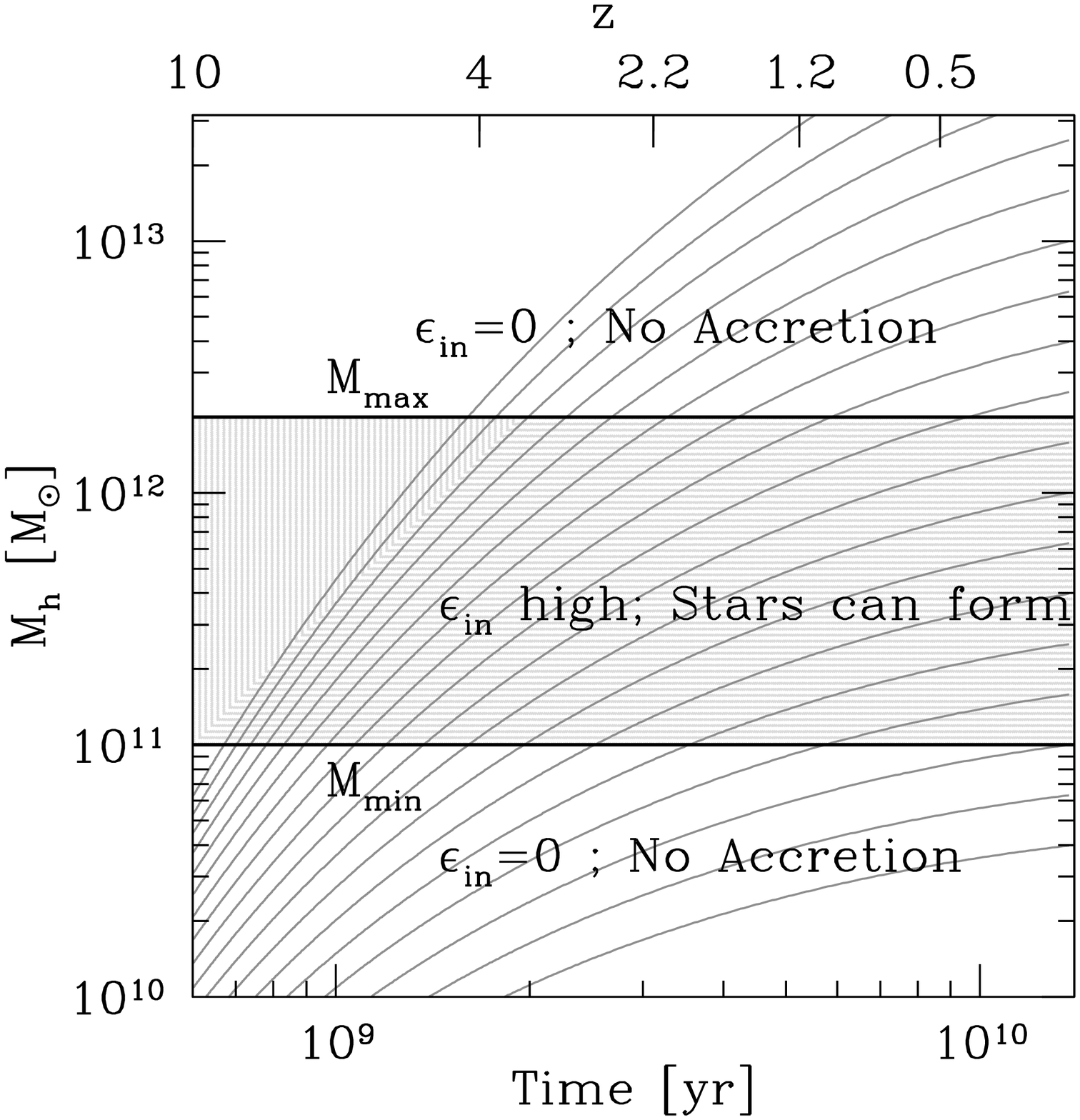}{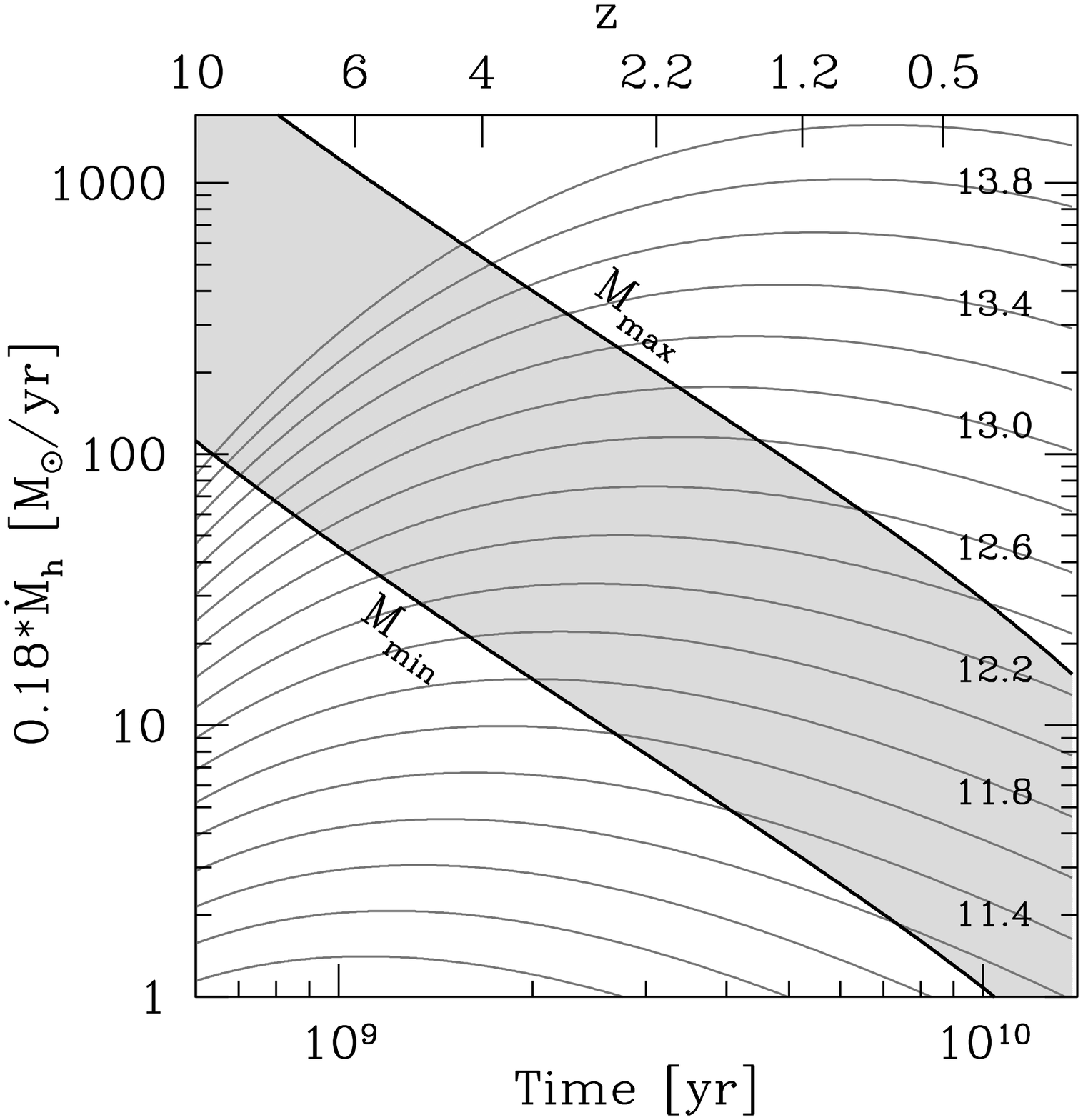}
\caption{Left: The growth history
$\Mv(z)$ for each modeled halo as a function of redshift calculated according to Equation~\ref{eq:growth}.
Above the quenching mass $\Msh$, accretion is increasingly hot.
Below the quenching mass $\Msh$, cold accretion brings in gas
efficiently with \epsin.
Importantly, below a minimum mass $M_{\rm min}$, accretion is inefficient.
Right: The maximum baryonic accretion rate inferred from the growth rate 
for each modeled halo. The final halo mass $\Mv(z=0)$ is labeled.
More massive halos reach the $M_{\rm min}$ threshold first and cross $\Msh$ 
 over a shorter time scale (downsizing; see Section~\ref{section:downsizing}).
  \label{fig:massgrowth}}
\end{figure*}

The two efficiency parameters $\esfr$ and $\epsin$ appearing in 
Equations~\ref{eq:gasaccretion}--\ref{eq:KS} may in principle vary 
with mass and redshift. 
The SF efficiency $\esfr$ does not appear to evolve with redshift.
Indeed, the KS relation seems to be in place
as early as $z=2$ and shows no significant evolution between $z=0$ and $z=2$
 \citep{BoucheN_07a}.

On the other hand,  the accretion efficiency \epsin\ must be a strong function 
of halo mass \citep[e.g.][]{WhiteS_91a}.
At the massive end $\Mv\gsim \Msh \simeq 10^{12}\msun$, 
and at low redshifts, 
the cold flows are not expected to be in a form of narrow, dense streams 
and they therefore fail to penetrate through the shock-heated halo gas.
This may play a significant role in quenching star formation and causing
the transition of blue galaxies onto the red sequence
\citep{DekelA_06a,CattaneoA_06a}.
To model this effect, we set a ceiling for cold-gas accretion at all redshifts
\footnote{Cold accretion can occur above this mass threshold at redshifts
beyond $z=2$ when the filament cross-section is much smaller than the virial radius.
However, the number density of such halos with mass $10^{13}$ (and above) is low.},
\be 
\epsin(\Mv)=0 \;\; \hbox{if $\Mv > \Msh $.}
\ee

At the low-mass end, 
there is strong evidence for a drop in the efficiency of galaxy formation
\citep[e.g.][]{ShankarF_06a,vandenBoschF_07a,BaldryI_08a,KravtsovA_09a,
MosterB_10a,GuoQ_10a}.
In the spirit of our simple toy model approach,
 we make the ansatz,
\be
\epsin(\Mv)=0 \;\; \hbox{if $\Mv < \Mmi$,}
\ee
i.e.,  we set a sharp mass floor for $\epsin(\Mv)$ below $M_{\rm min}$.
For any reasonable physical mechanism that may be responsible for the
suppression of accretion and SFR in low-mass halos
(see Section~\ref{section:accretionfloor}),
modeling the effect as a sharp cutoff is clearly a very crude
approximation, but it is useful as a first attempt in 
capturing the key features with a minimum 
number of parameters.

This accretion floor means that the effective accretion of cold gas is 
suppressed in halos less massive than $\Mmi$.  
Halos more massive than $\Mmi$ accrete baryons that are embedded
in other merging halos, which could themselves be more massive or less 
massive than $\Mmi$, as well as smooth gas unbound to any halo \citep{StewartK_08a,GenelS_10a}.
The relative contribution of these different contributions  
leaves sufficient margins to meet an accretion efficiency of \ein.

The accretion efficiency $\epsin$ at a given mass is likely to decline in time, 
as the overall fractions of stars and hot gas ($\gg 10^4$K) in the inter-galactic medium
(IGM) grow. 
We model this by incorporating a redshift dependence $\epsin(z)=\epsin\;f(z)$, 
to allow for a decrease in cold accretion efficiency from $z=2$ to $z=0$.
For simplicity reasons, we use a function linear in time with the boundary conditions $f(z=2.2)=1$ and $f(z=0)=0.5$.
Note that our main results on the slope of the scaling relations are independent of this assumption 
since a change in $\epsin$ changes both SFR and $\Ms$  by the same amount.
However, as we discuss in Section 3.4, this affects the evolution of the SFR sequence.

Figure~\ref{fig:massgrowth}(left) shows the mass growth histories for our modeled halos
  and the model accretion efficiency $\epsin(\Mv)$   with the
upper and lower cutoffs at $\Msh$ and $\Mmi$. 
Figure~\ref{fig:massgrowth}(right) shows the maximum baryonic accretion rate
$f_B\;\dot{\Mv}$ available for each halo as a function of time.
More massive halos reach the $M_{\rm min}$ threshold first and cross $\Msh$ 
 over a shorter time scale, leading to  downsizing (see section~\ref{section:downsizing}).

\subsection{Overview of model ingredients}
\label{section:overview}

\begin{table*}
\centering
\begin{tabular}{lcccc}
Model 	&   $M_{\rm max}$   & $M_{\rm min}$  	&   Floor	& 	Feedback \\
	&	\msun		& \msun		&		&	$a$ \\
\hline \\
accFloor	&   \mmax		& \mmin		&   accr.	&    0 \\
noMmin	&   \mmax		& n.a.	&   n.a.		&    0 \\
sfrFloor	&    \mmax		& \mmin		&   sfr.	&    0  \\	
accFloor$+$	&   \mmax		& \mmin		&   accr.	&    0.6 \\
\hline
\end{tabular}
\caption{Definition of our fiducial models. \label{table:models}}
\end{table*}

At this stage, it is worth summarizing our model and its parameters.
The model parameters are as follows:
\begin{enumerate}
\item The   key assumption in our model is the halo mass floor for cold gas accretion $\Mmi$, 
on the order of $10^{10}-10^{11}~\msun$.
As we will show in Sections~\ref{section:orig:sfrmass} and \ref{section:madau},
  $M_{\rm min}\simeq$\mmin~\msun\ will be
required to match observations.
\item 
The halo mass ceiling for cold gas accretion $\Mma$, set at
a value of \mmax\,$\msun$ to match the characteristic threshold 
for virial shock heating \citep{DekelA_06a} and for best reproduction of the 
observed features of the galaxy bimodality \citep{CattaneoA_06a}.
\item
The cold gas accretion efficiency $\epsin(\Mv,z)$.
As illustrated in Figure~\ref{fig:massgrowth}, it vanishes below $\Mmi$ and above 
$\Mma$, and  is set to \ein\ at $z>2$.
Below $z<2$,  we use $\epsin(z)=f(z)\times\ein$, 
where $f(z)$ is a function that is  decreasing linearly with time with the boundary conditions
$f(2.2)=1$ and $f(0)=0.5$.
\item
The SFR efficiency $\esfr = 0.02$ set from  the observed KS relation, 
and the orbital timescale 
$t_{\rm dyn} = \torb~{\rm yr}\, (1+z)_{3.2}^{-1.5}$.
\end{enumerate}

  Our fiducial model described above is the ``accretion floor" model,
dubbed hereafter `accFloor', where the accretion of cold gas, and therefore
star formation, are totally suppressed below $M_{\rm min}$.
We emphasize that our terminology `accretion floor' means that the {\it effective} accretion of cold  baryons is suppressed
in halos with $\Mv(z)<M_{\rm min}$.  Either these baryons are prevented from cooling or they are prevented from entering and reaching the central object.

In Section~\ref{section:results}, 
we will analyze two radically different models for comparison
(listed in Table~\ref{table:models}):
(1) in the other extreme `noMmin' model, the mass floor is not applied at all,
i.e., we allow both gas accretion and star formation to proceed unperturbed
below $M_{\rm min}$ and
(2) in the intermediate `sfrFloor' model, we do allow gas accretion below
$M_{\rm min}$, but forbid star formation there. In this case, the cold gas
accumulates until the halo becomes more massive than $M_{\rm min}$, and it then
turns into stars.
We also investigate the impact of the outflow term
$\dot M_{\rm out}$ in Equation~\ref{eq:bathtub} with a model we term `accFloor+'.
However, as described in Section~\ref{section:bathtub}, this term
will have no impact on the scaling relations.

\begin{figure*}
\centering
\includegraphics[width=5.5cm]{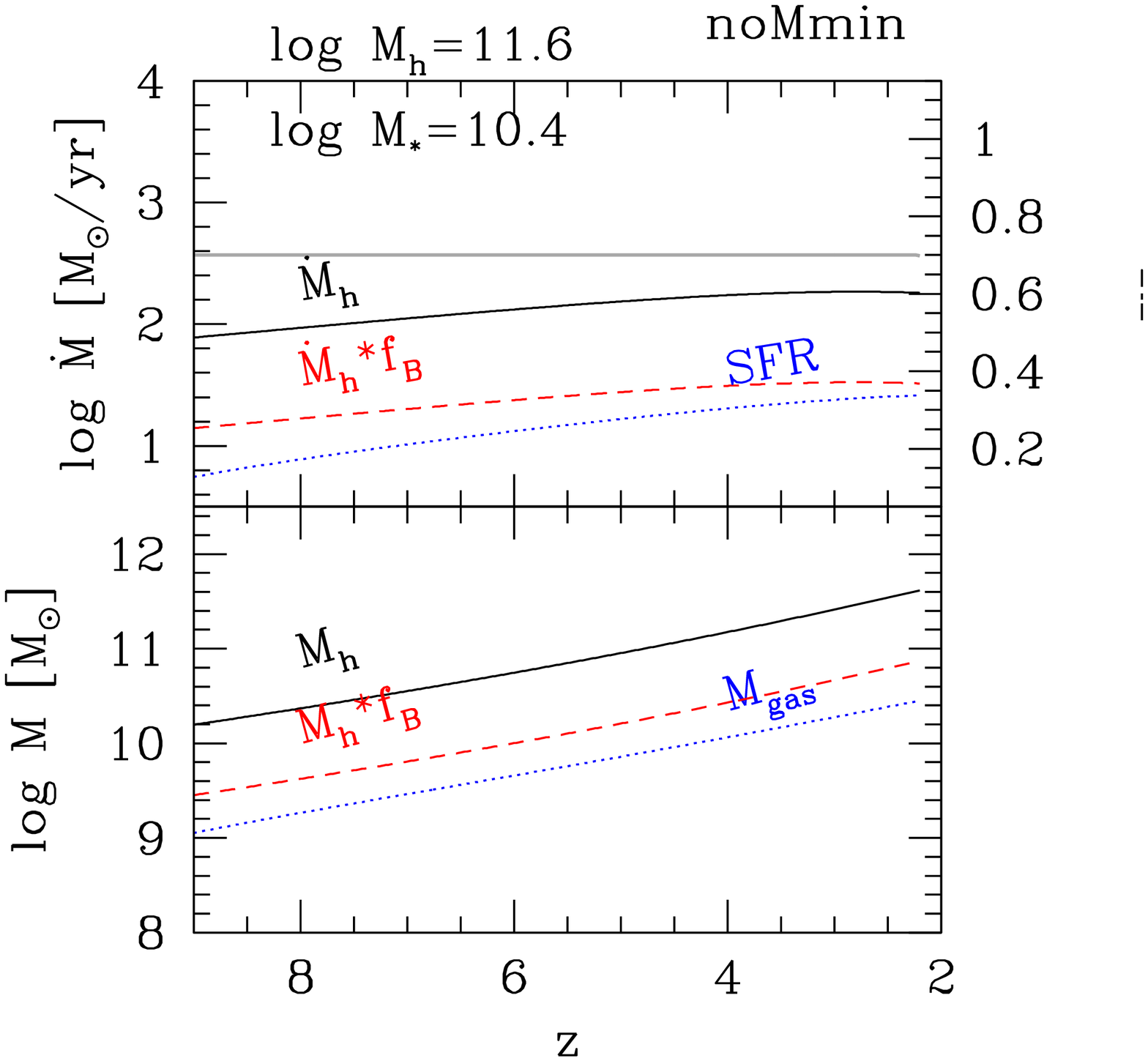}
\includegraphics[width=5.5cm]{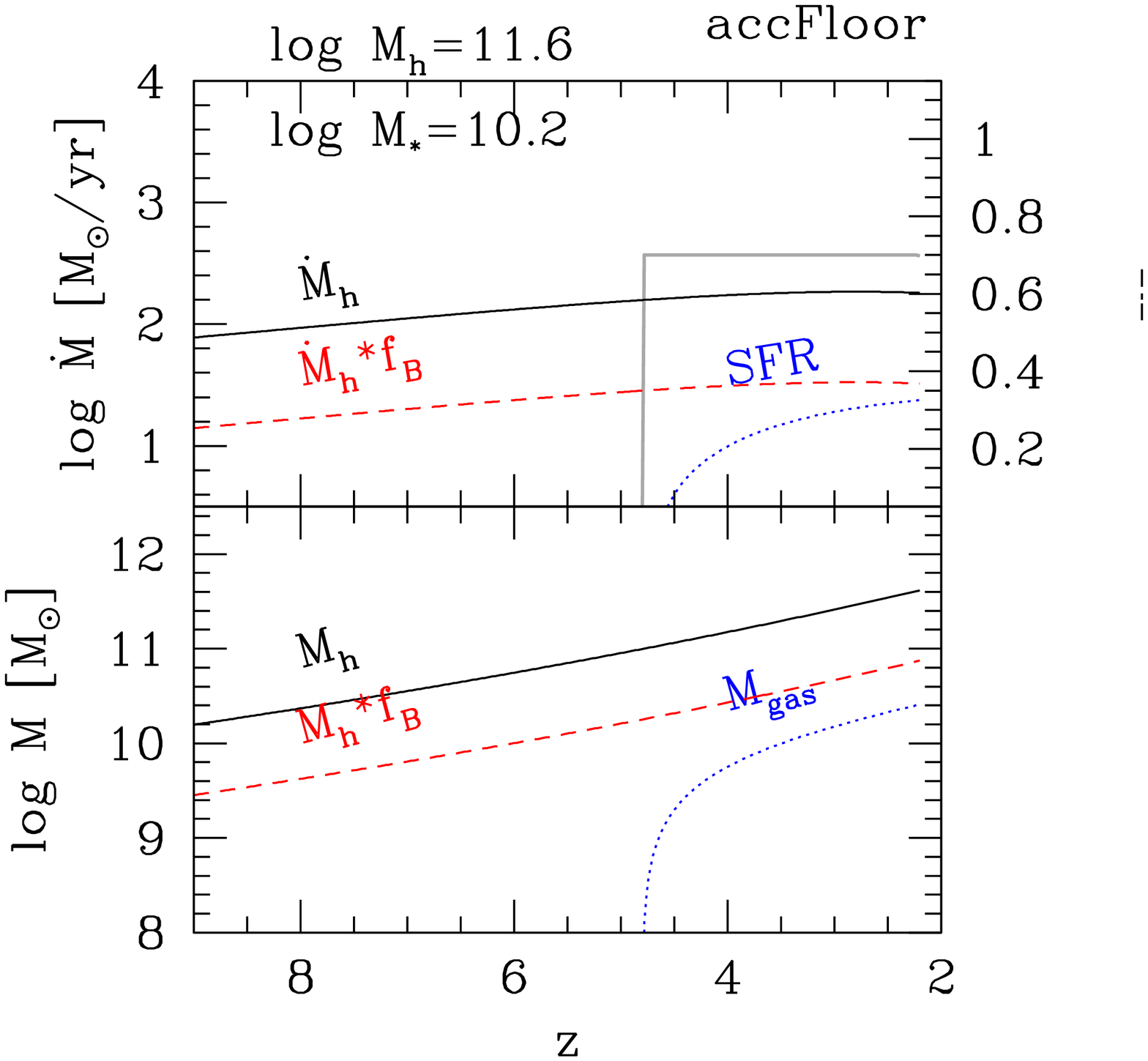}
\includegraphics[width=5.5cm]{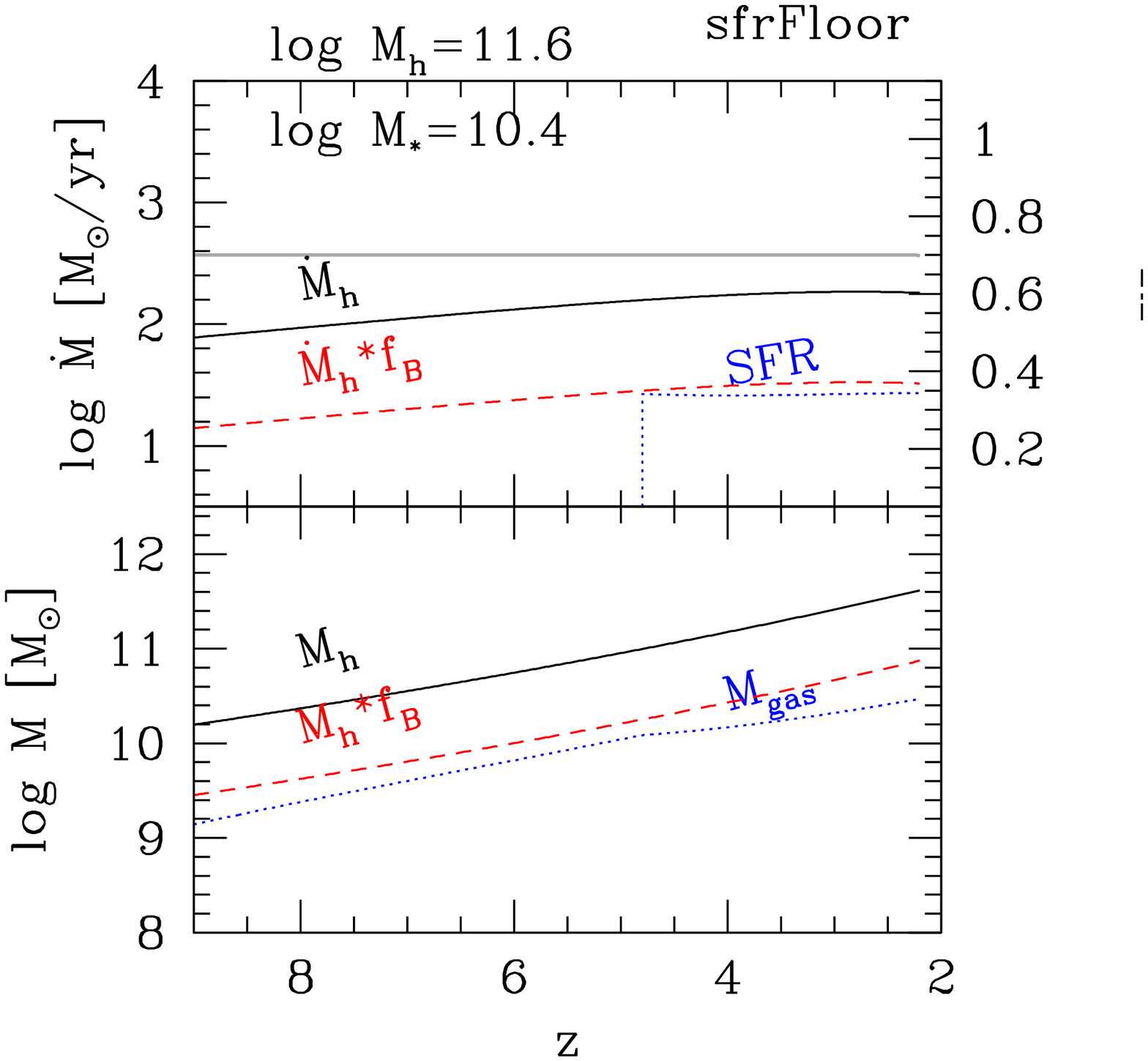}
\caption{For a $\log \Mv=11.6$ halo at $z=2.2$, we show the behavior of each of our three models, 
namely when  $M_{\rm min}=0$ (`noMmin', left panel), 
when $\epsin=0$ below the minimum mass floor (`accFloor', middle panel),
 and when $\esfr=0$ below the SF floor $M_{\rm min}$ (`sfrFloor', right panel).
The top panels show the redshift evolution of
the DM accretion rate $\dot{\Mv}$ (solid line),
the  baryonic accretion rates ($\dot{\Mv}*f_B$) (dashed line)
and the SFR (dotted line).
The bottom panels show the halo mass ($\Mv$) (solid line),
the maximum baryonic mass ($\Mv\times f_B$) (dashed line) and the gas mass ($\Mg$) (dotted line).
The effect of $M_{\rm min}$ is clearly apparent in each case. 
The `sfrFloor' model will lead to the same stellar mass $\Ms$ as the `noMmin' model,
 given that the SF is simply delayed, and the amount of gas accreted remains
the same.
\label{fig:efficiency}}
\end{figure*}

{Figure~\ref{fig:efficiency} highlights the main differences between
the three alternative models, `noMmin' (left), `accFloor' (middle), and `sfrFloor' (right).
In the top panels, we show the assumed variation of $\epsin$ with redshift (thick solid lines, right axis)
and the resultant evolution of the accretion rate (solid line), baryonic accretion rate (dashed line), and
the SFR (dotted line) as a function of redshift.
In the bottom panels, we show the resultant evolution of the DM halo mass (solid line),
the maximum baryonic mass (dashed line), and the cold gas mass (dotted line).
These three models differ significantly in their initial behavior.
Compared to the `noMmin' model, the `accFloor' and `sfrFloor' models add  a (mass-dependent) delay in the evolution of SFR.
In contrast to the `accFloor' model, the `sfrFloor' model accumulates cold gas till the halo reaches $\Mmi$.
Both of these models lead to no SF below $M_{\rm min}$, 
but have rather distinct physical interpretations and consequences.
}

\subsection{Numerical approach}

{Using this simple reservoir set up (Equations \ref{eq:bathtub} and \ref{eq:KS}), 
we solve for its gas and stellar content at
each redshift as follows.
We first construct average mass growth histories $\Mv (M_0,z)$
for the main progenitors of halos of masses in the range 
$M_0 = 10^{10.5-14}~\msun$ today, as in Figure~\ref{fig:massgrowth},
following \citet{NeisteinE_08a} (see their Equation8).
We then numerically integrate Equations~\ref{eq:bathtub}--\ref{eq:KS} 
for $M_{\rm gas}$ (and $\dot M_\star$) from $z=10$ to any $z$ until $z=0$,
with the initial conditions $M_{\rm gas}=M_\star=0$ at $z=10$. 
The stellar mass $M_\star$ is updated at each time step according to
$M_\star=(1-R)\int \hbox{SFR}\;\rm d t$, where $R=0.52$
is the recycled gas fraction for a Chabrier initial mass function (IMF) \citep{BruzualG_03a}
\footnote{For our purposes, $R$ is a constant given
its slow dependence on stellar population age $T$ for age $T>10^9$~Gyr.}. 
}

\section{Results}
\label{section:results}

\subsection{Steady state}
\label{section:steadystate}

\begin{figure}
\plotone{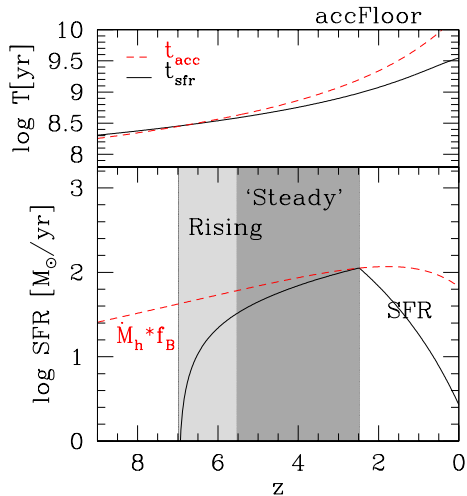}
\centering
\caption{For our fiducial model ('accFloor'), the bottom panel shows the 
SFR (solid line) and the maximum accretion rate ($\epsin\;f_B\times \dot{\Mv}$) (dashed line).
The top panel shows that below $z\simeq 5.5$, the accretion time scale ($t_{\rm acc}\equiv\Mv/\dot{\Mv}$)
 is longer than the SF time scale ($t_{\rm sfr}\equiv M_{\rm gas}/$SFR).
The reservoir model reaches rapidly a {quasi}-steady state in which the SFR 
scales with the accretion rate.
\label{fig:steadystate}}
\end{figure}

Before showing the numerical results, it is worth 
understanding the key behavior of the reservoir model
from a qualitative inspection of Equations \ref{eq:bathtub} and \ref{eq:KS}.
When $M_{\rm gas}$ in the galaxy is still low, the SFR is low by the KS law
(Equation~\ref{eq:KS}), and it is smaller than the accretion rate dictated by the
cosmological environment.
The gas reservoir then gradually fills up until the SFR becomes
comparable to the accretion rate (see Figure~\ref{fig:steadystate}). The galaxy 
enters a {\it quasi}-steady state where the SFR is essentially set by the accretion 
rate, SFR$ \simeq \dot M_{\rm gas,in}$.
Would the gas reservoir be temporarily overfilled (for its current SFR), the SFR would then be larger than
the accretion rate, and the galaxy will return to the quasi-steady state.

Figure~\ref{fig:steadystate} demonstrates the quasi-steady state behavior.
For our fiducial model ('accFloor'), the bottom panel shows the 
SFR (solid line) and the maximum accretion rate ($f_B\;\dot{\Mv}$) (dashed line).
The reservoir model reaches rapidly a {\it quasi}-steady state in which the SFR 
scales with the accretion rate.

{Because the accretion rate varies with redshift,
the steady state can only be achieved  when the timescale associated with SFR
is comparable to or shorter than the accretion timescale.
The SFR timescale, using Equation~\ref{eq:torb}, is
\begin{eqnarray}
t_{\rm sfr}&\equiv&\frac{M_{\rm gas}}{\dot M_{\star}}\nn\\
&\simeq&  \left ({t_{\rm dyn,7}}/{2}\right )\, \epsilon_{{\rm sfr},0.02}^{-1}\,(1+z)_{3.2}^{-1.5}\, \hbox{Gyr} \, ,
\end{eqnarray}
where $t_{{\rm orb},7} \equiv t_{\rm dyn}/10^7$~yr,
and $\epsilon_{{\rm sfr},0.02} \equiv \esfr/0.02$.
From Equation~\ref{eq:growth}, the accretion timescale is
\be 
t_{\rm acc}\equiv \Mv/\dot{\Mv}
\simeq 2~ \Mv{}_{,12}^{-0.1}\, (1+z)_{3.2}^{-2.2} \hbox{Gyr} \, .
\ee
Therefore, the condition that the SFR timescale be shorter than the accretion timescale
\begin{eqnarray}
t_{\rm sfr} \leq  t_{\rm acc} \label{eq:steadystate}
\end{eqnarray}
is met at redshifts $z \lsim 7 $.
}

{The steady-state solution SFR$\propto \dot M_{\rm gas,in}$
has several important implications.
First,  it implies that the SFR$(z)$ is driven by the (net) accretion rate 
as illustrated on Figure~\ref{fig:steadystate}.
As a result, at around $z\sim 2$, 
SFR maintains a high and slowly varying value
for a few Gyrs (Figure~\ref{fig:massgrowth}(b)), which means that the average SFR is expected
to be comparable to the instantaneous SFR.
This is indeed supported by the SINS sample at $z \sim 2$,
where the  birthrate parameter, defined as the ratio of instantaneous 
to past SFR, is estimated to be $b \simeq 1.2$
\citep{ForsterSchreiberN_09a}.
}

{Second, the resulting SFR, gas and stellar masses
are thus independent of the initial conditions given the rapid growth
of the gas mass (see Figure~\ref{fig:efficiency}).
}

{Lastly, when the steady state condition (Equation\ref{eq:steadystate})
is not met,  the gas reservoir is being filled up at a rate faster than its consumption rate.
In this case, SFR is time dependent,  of the form $t^{a}$, and sSFR is proportional to $1/t$. }

\subsection{Effect of $\Mmi$ on scaling relations} 
 \label{section:orig:sfrmass}

\begin{figure*}
\centering
\ifthenelse{\boolean{preprint}}{
\includegraphics[width=7.5cm]{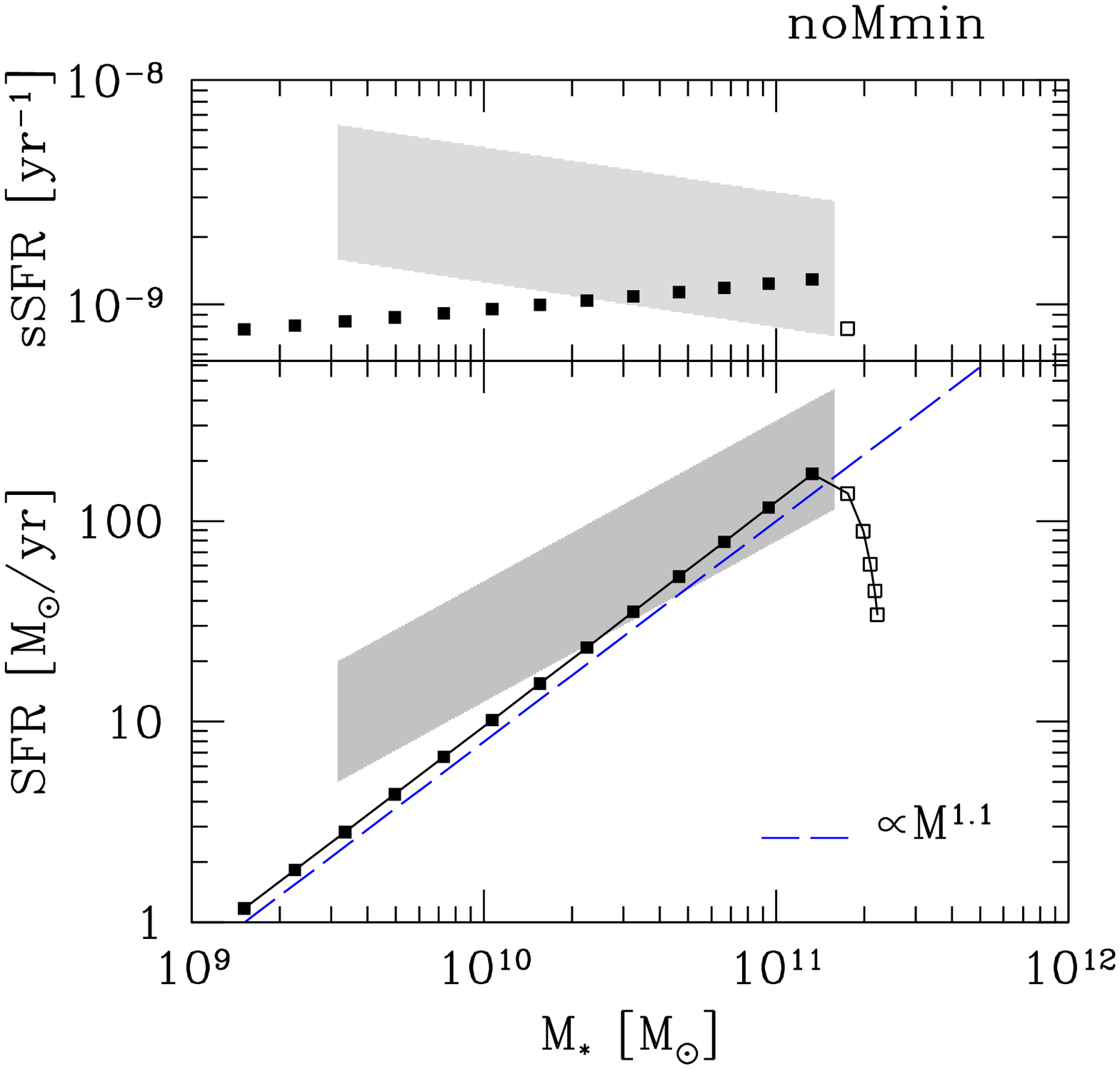}
\includegraphics[width=7.5cm]{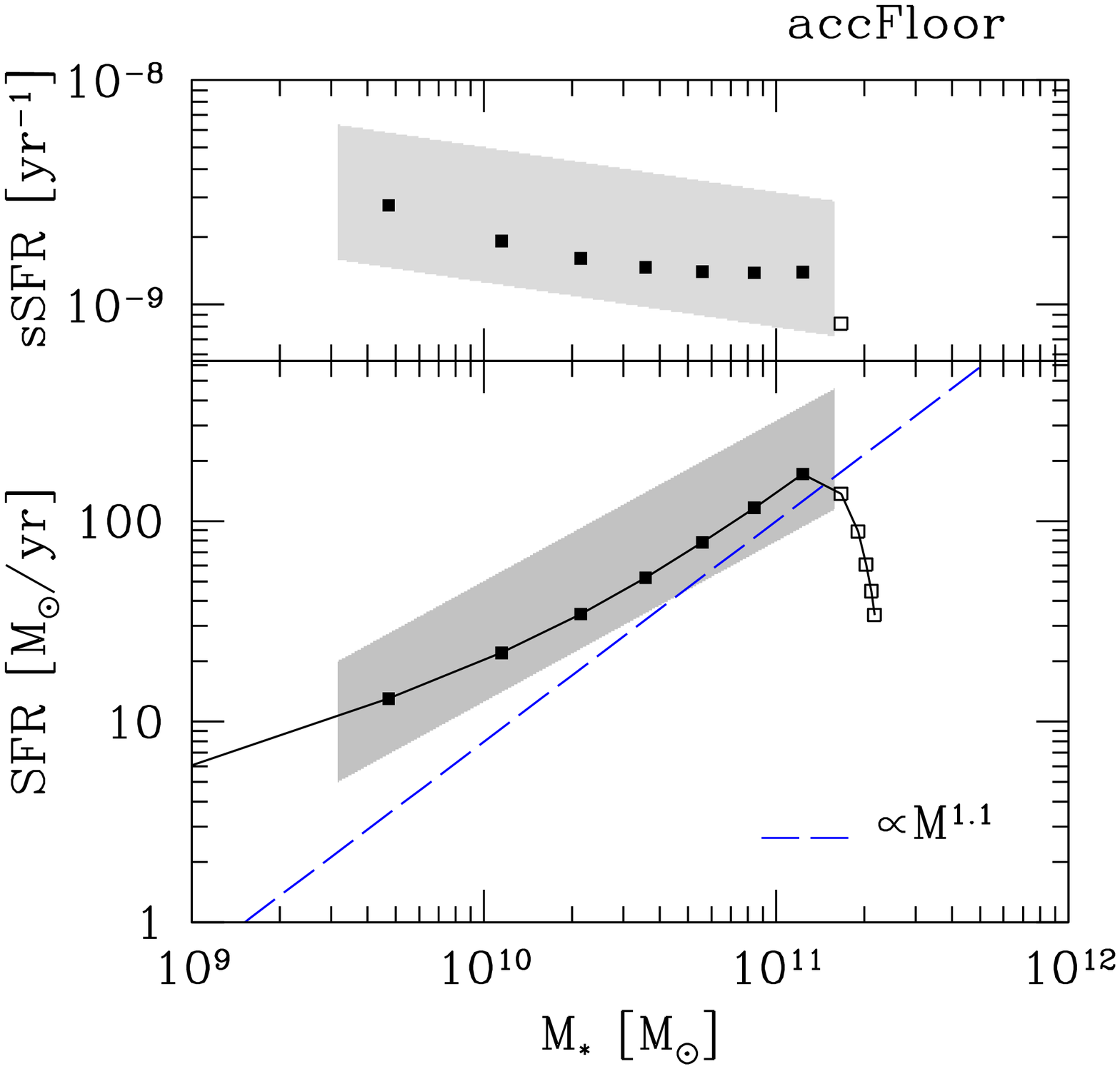}
\includegraphics[width=7.5cm]{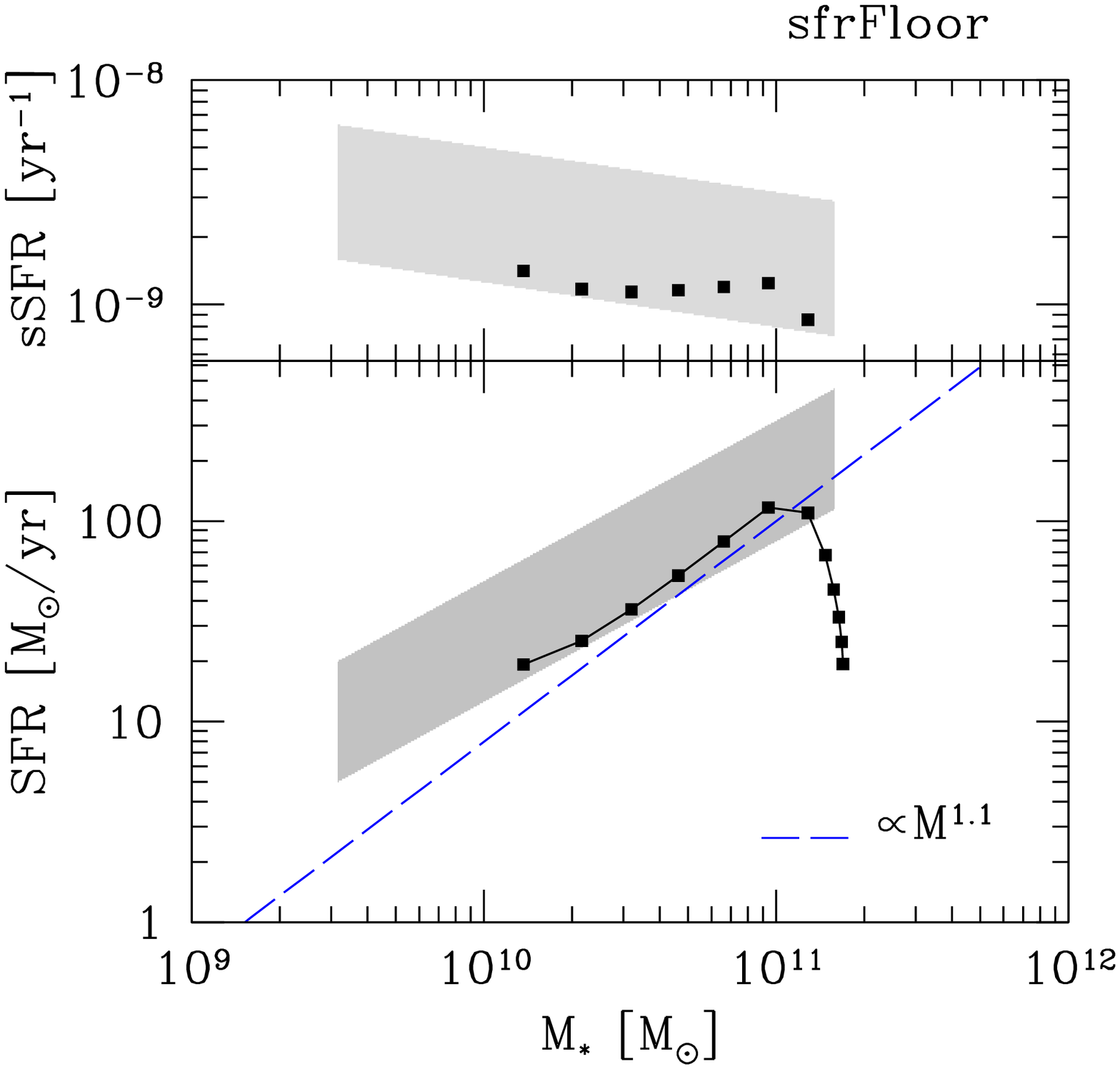}
}{
\includegraphics[width=5.5cm]{f7.eps}
\includegraphics[width=5.5cm]{f8.eps}
\includegraphics[width=5.5cm]{f9.eps}
}
\caption{The $z=2$ SFR sequence  predicted from our three fiducial models
with no mass floor $M_{\rm min}$ (left),    an accretion floor (middle), and
   an SF floor (right).
The modeled points are shown as filled (open) squares, for halos below (above) the
virial shock mass $\Msh$.
The observed  SFR sequence from \citet{PannellaM_09a} and \citet{DaddiE_07a} is shown as gray shaded areas.
 The dotted line in each panel  shows a line of slope $1.1$ expected
from the global accretion rate (Equation~\ref{eq:growth}).
The `accFloor' model provides the best match
to the observed SFR sequence with SFR$\propto \Ms^{0.8}$.}
\label{fig:bluesequence:z2}
\end{figure*} 

\begin{figure*}
\centering
\ifthenelse{\boolean{preprint}}{
\includegraphics[width=7.5cm]{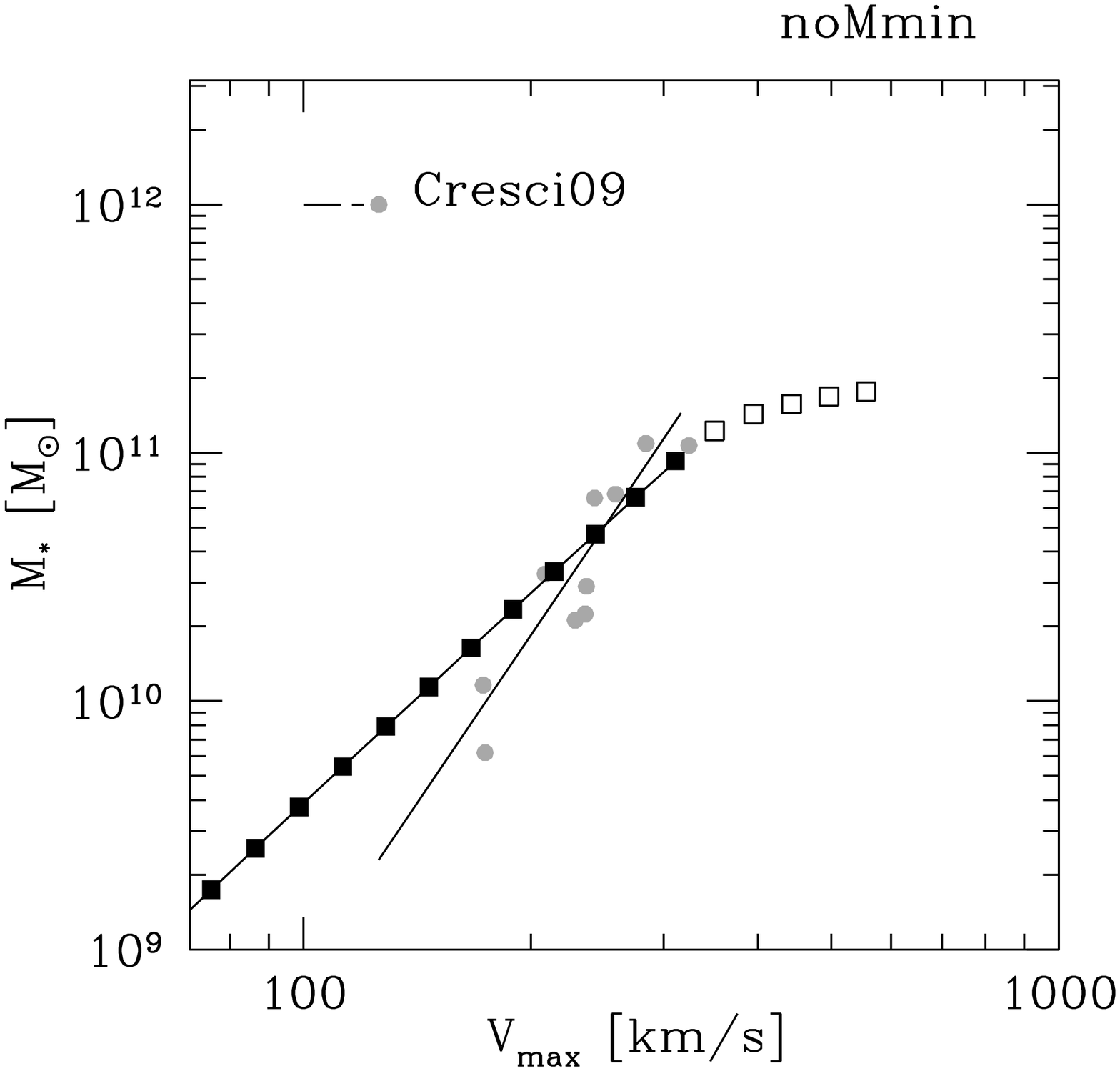}
\includegraphics[width=7.5cm]{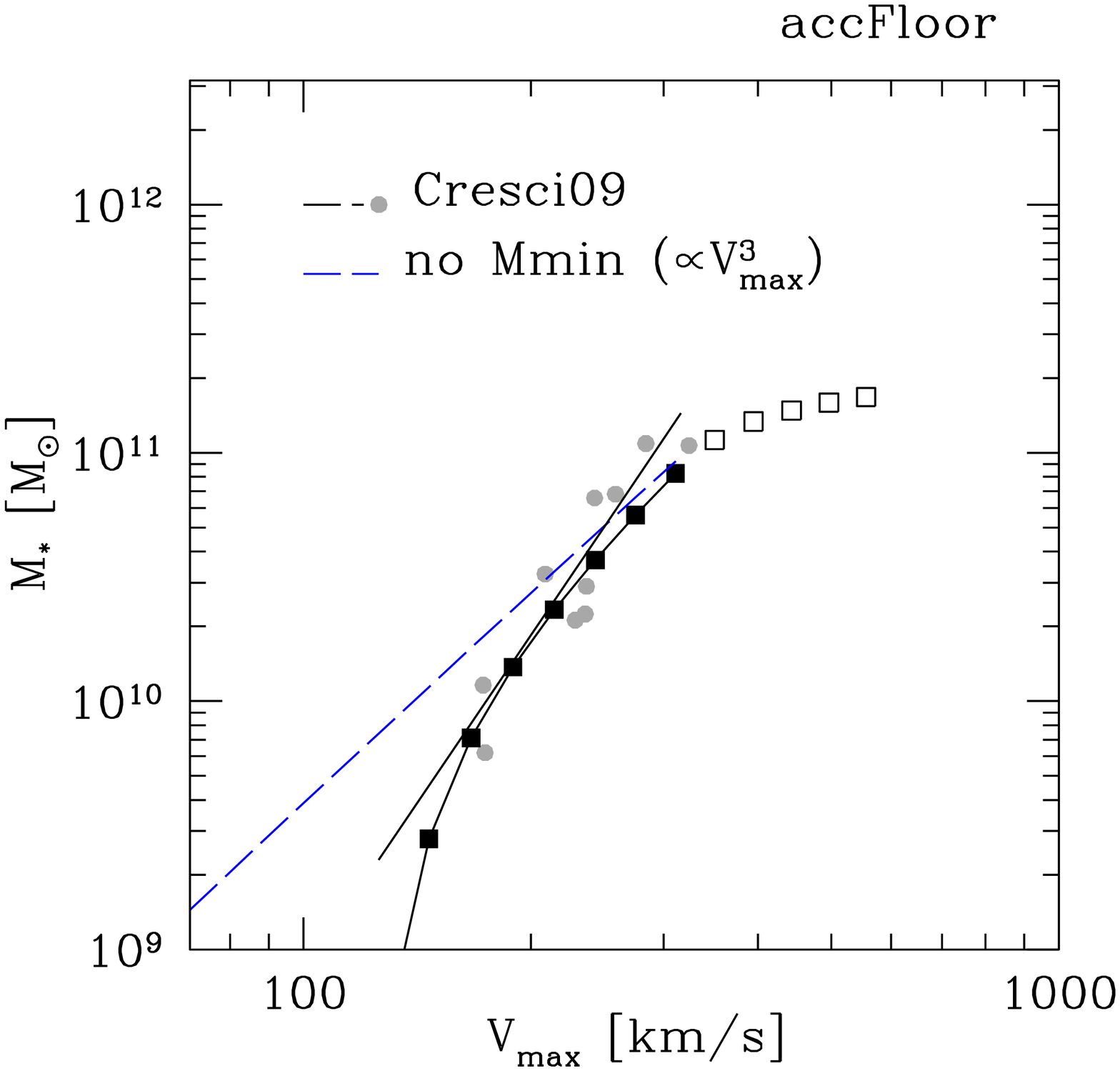}
\includegraphics[width=7.5cm]{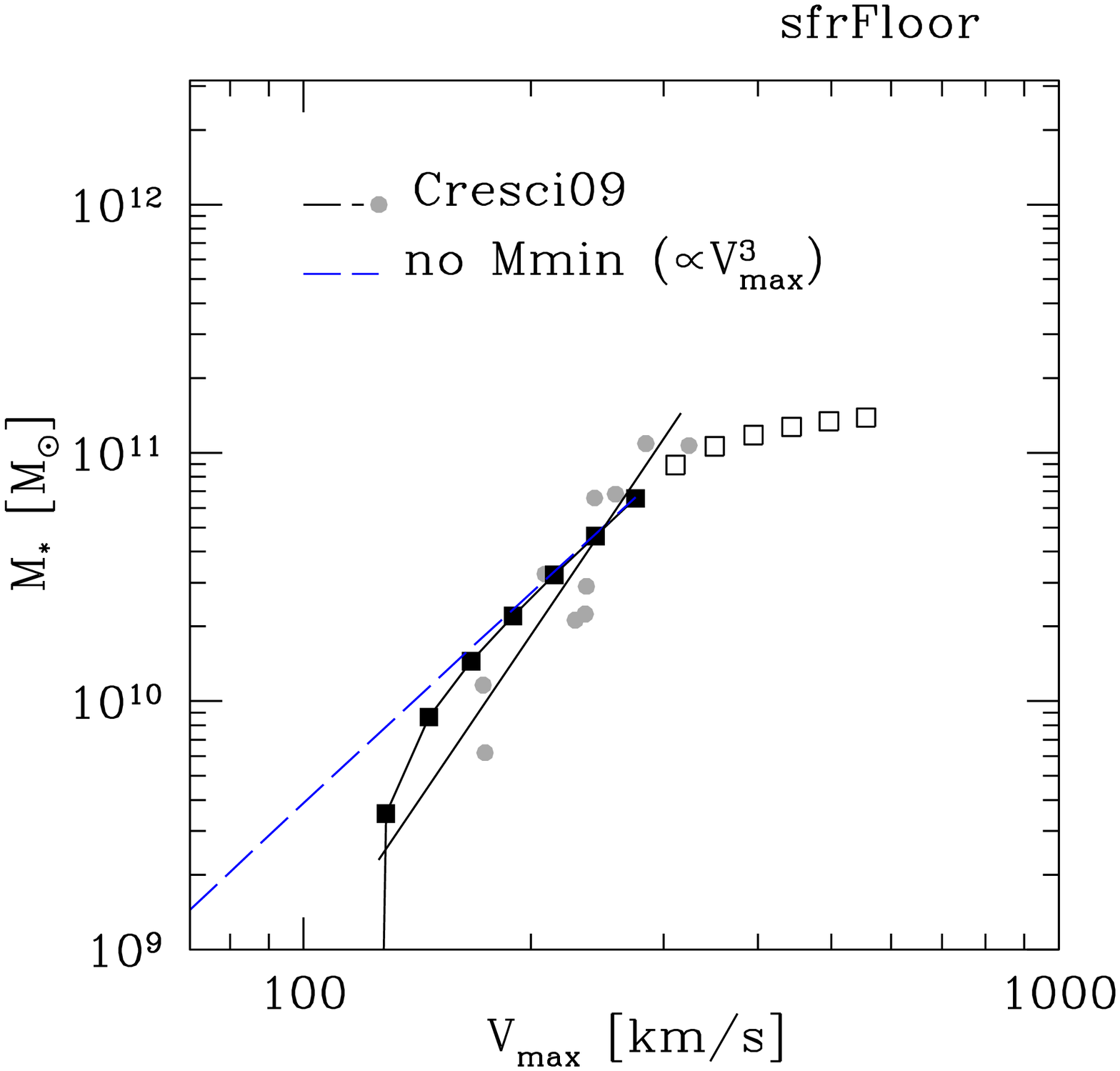}
}{
\includegraphics[width=5.5cm]{f10.eps}
\includegraphics[width=5.5cm]{f11.eps}
\includegraphics[width=5.5cm]{f12.eps}
}
\caption{The $z=2$ Tully-Fisher relation
predicted from our simple model
with no mass floor $\Mmi$  (left), with an accretion floor  (middle), and
with an SF floor (right).
The calculation for a model with no $\Mmi$ is shown
as   dashed lines. 
The data from \citet{CresciG_09a} are shown as gray circles.
The `accFloor' model provides the best match
to the observed  TF relation with $\Ms \propto \Vrot^{\sim 4}$ (solid line).
}
\label{fig:TFR:z2}
\end{figure*}

{Our goal is to better understand the global scaling relations
of  $z\simeq2.2$ SFGs, and 
the arguments presented in this Section are, however, applicable at $z=0$, $z=1$ as well.
Thus, we take the naive point of view that the slopes of the scaling relations do not evolve
with redshift and we stress  that our aim 
 is to demonstrate the key role played by the mass floor $\Mmi$
in connecting the baryonic scaling relations to their DM counterparts.
Any evolution in the slopes of the SFR sequence or TF relation will lead
to an evolution of this mass floor. }

Figure~\ref{fig:bluesequence:z2} shows the SFR sequence  predicted by 
 the three fiducial models (``noMmin'', ``accFloor'' and ``sfrFloor''), in comparison with the observed
SFR sequence at that redshift from \citet{PannellaM_09a} and 
\citet{DaddiE_07a}. 
The data show clearly that the sSFR is higher for lower mass galaxies
 (the mass index $\ssa<1$), whereas 
the expectation from the DM relation would predict an inverted trend 
since its mass index $\sse>1$.
The slope $\sse=$1.1 associated with the predicted
accretion rate as a function of halo mass is shown as the dashed line.
Figure~\ref{fig:TFR:z2} compares the corresponding $z=2$ TF relation 
 with the observed relation from \citet{CresciG_09a}. 
Similarly, the data show a tilt from the slope 3 associated with the virial relation
$\Mv \propto \Vv^3$ (dashed line).

These two figures highlight our first significant results:
the `accFloor' model provides the best match
to the two scaling relations.
The mass floor $\Mmi$ affects the slopes of both
 the SFR sequence and the TF relation in the right direction.
It tilts the SFR sequence away from the $1.1$ dark accretion rate,
and steepens the TF relation away from the $3$ virial slope. 
This effect can easily be  explained.

When a mass floor is not applied, 
once the system reaches the steady state solution, 
 SFR follows the halo accretion
$\dot \Ms \propto \dot {\Mv} $
and thus scales with halo mass according to  
$\dot \Ms \propto \Mv{}^{1.1}$.
Given that the halo mass is  $\Mv =\int_0^t\, \dot{\Mv}\, {\rm d}t$,
and that the stellar mass is $\Ms(t) \prop \int_0^t\, \dot \Ms\, {\rm d} t$,
stellar mass and halo mass are proportional to each other: $\Ms \propto \Mv$.
 Hence,  we expect in this case  $\dot \Ms \propto \Ms^{1.1}$,
as seen in the left panel of Figure~\ref{fig:bluesequence:z2}.

When a mass floor for accretion is imposed (model accFloor),
accretion (hence SF) is suppressed until the time $t_{\rm min}$ when the halo reaches 
$\Mmi$, which is a decreasing function of $\Mv$, and thus
of $M_\star$. The stellar mass is now an integral from 
$t_{\rm min}(\Mv)$ to $t$: $\Ms(t) \prop \int_{t_{\rm min}}^t\, \dot \Ms\, dt$.
The less massive systems had less time to grow stars
and $\Ms$ is now smaller 
at a given halo mass, by a larger factor for a smaller halo. 
This effect is responsible for the required 
tilt both in the SFR sequence and the TF relation
as seen in the middle panels of Figs.~\ref{fig:bluesequence:z2} and \ref{fig:TFR:z2}.

In the sfrFloor model, all the gas that has accumulated in the galaxy until
$t_{\rm min}$ turns into stars after $t_{\rm min}$, so the total stellar mass
after that time is expected to be the same as in model `noMmin', as seen
in the right panels of Figs.~\ref{fig:bluesequence:z2} and \ref{fig:TFR:z2}.

{
In summary, a mass-dependent delay in the gas accretion is found to be necessary
to better reproduce the observed slopes of  the scaling relations.
In order for the time delay to have a significant effect, we
find that $\Mmi$ needs to be within 1~dex of $\Mma$, i.e., $\Mmi\sim$~\mmin~\msun.
In Section~\ref{section:madau}, we will put a stronger constraint
on the numerical value of $\Mmi$.
The physical origin of the mass floor $\Mmi$ is addressed in Section~\ref{section:accretionfloor}.
}

{
This mass-dependent delay can only be achieved in our model by the mass floor $\Mmi$.
Indeed, these results are completely independent of the SF efficiency $\esfr$
and of the feedback term in Equation~\ref{eq:bathtub}.
A change in these two parameters will lead to a change in both $\Ms$ and SFR, i.e., leaving the SFR sequence intact.
A change in the accretion efficiency $\epsin$ via its mass dependence is the only
mechanism that we found to affect the slope of the scaling relations.

}

\subsection{Connections between the two relations}
\label{section:connections}

It is very interesting to realize that the TF relation and 
SFR sequence are in fact tied to each other. Indeed,
the steady-state solution $\dot \Ms \propto \Mv^{\sse}$,
together with  the SFR sequence $\dot{\Ms}\propto \Ms^{\ssa}$
implies the TF relation ($\Ms \propto \Vv^{\ssc}$). 
Indeed, using the virial relation $\Mv \propto \Vv^3$, we find
\be
\ssc =3 \sse/\ssa \, .
\ee
With the indices being $\sse \simeq 1.1$ and $\ssa \simeq 0.8$, the TF slope is indeed 
$\ssc \simeq 4$ as observed \citep[e.g.][]{MeyerM_08a,McGaughS_05a}.
{Note that this argument holds for various values of the indices $\sse$ and $\ssa$
as long as $\sse/\ssa\simeq 4/3$. As noted in Section~\ref{section:introduction},
there are marginal indications that $\ssa$ varies from near 0.7
at $z=0-1$ \citep{BrinchmannJ_04a,NoeskeK_07a} to about
0.9 at $z \sim 2$ \citep{DaddiE_07a,SantiniP_09a,PannellaM_09a}.
These subtleties do not change our arguments given that
all observational evidence show that $\ssa<1$ (sSFR higher at lower mass), 
while the DM counterpart has a mass index $\sse>1$ (specific accretion rate higher at higher mass).
  Would this redshift dependence of the mass index $\ssa$ 
 be confirmed, it  would point toward a redshift evolution of $\Mmi$.
}

Given that the slopes of these two scaling relations are coupled together and that, 
as we will show, the accretion rate regulates the evolution of the
SFR sequence, we expect the accretion rate to regulate the evolution
of the TF relation as well. However, a more detailed analysis of the TF zero-point
normalization  requires  knowledge of the explicit
relation between the virial velocity of the halo, which we crudely referred
to so far as $\Vv$, and the observed maximum rotation velocity  
$\Vrot$. 
Early models of galaxy formation assumed halo
adiabatic contraction due to the infall of baryons into the halo center
\citep{BlumenthalG_86a,MoH_98a}, which gives typically 
$\Vrot \simeq 1.6$--$1.8\, \Vv$ for a wide range of halo concentrations $c=[5,30]$ and disk mass fractions $m_d=[0.02,0.2]$.
However, such models often fail to reproduce the observed TF zero point
\citep[e.g.][]{NavarroJ_00a,DuttonA_07a}.
Disk formation models can simultaneously fit the TF relation and the 
galaxy luminosity function if there was actually a slight halo expansion,
where $\Vrot \simeq 1.2\, \Vv$ \citep{DuttonA_07a}.
Evidence against adiabatic contraction is mounting also at $z \sim 2$
\citep{BurkertA_09a}.
We find that a similar factor of about 1.2 between $\Vrot$ and 
$\Vv$ provides a good match to the TF zero point at $z \sim 2$.

\subsection{Redshift dependence}
\label{section:evolution}

\begin{figure*}
\centering
\plottwo{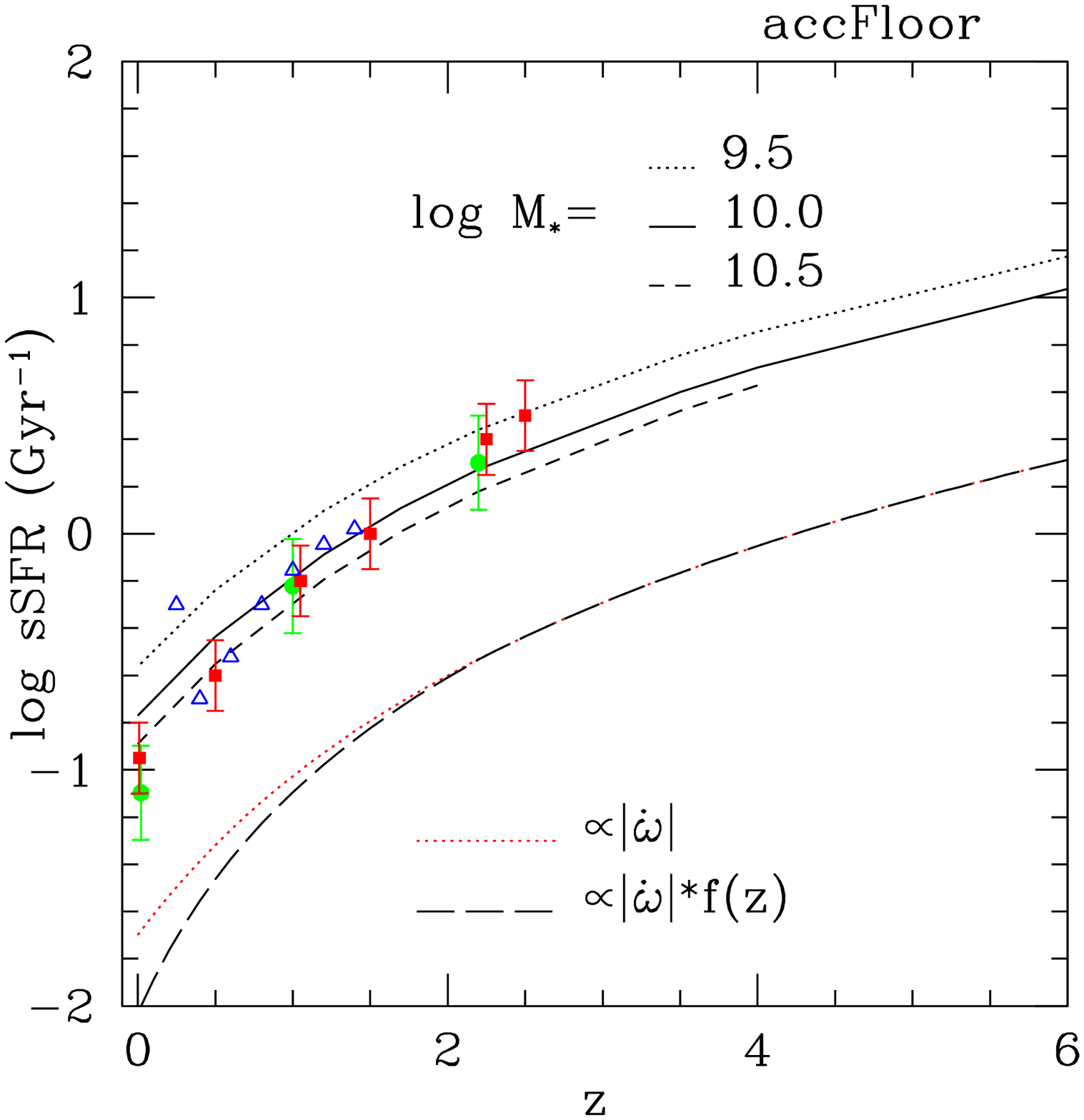}{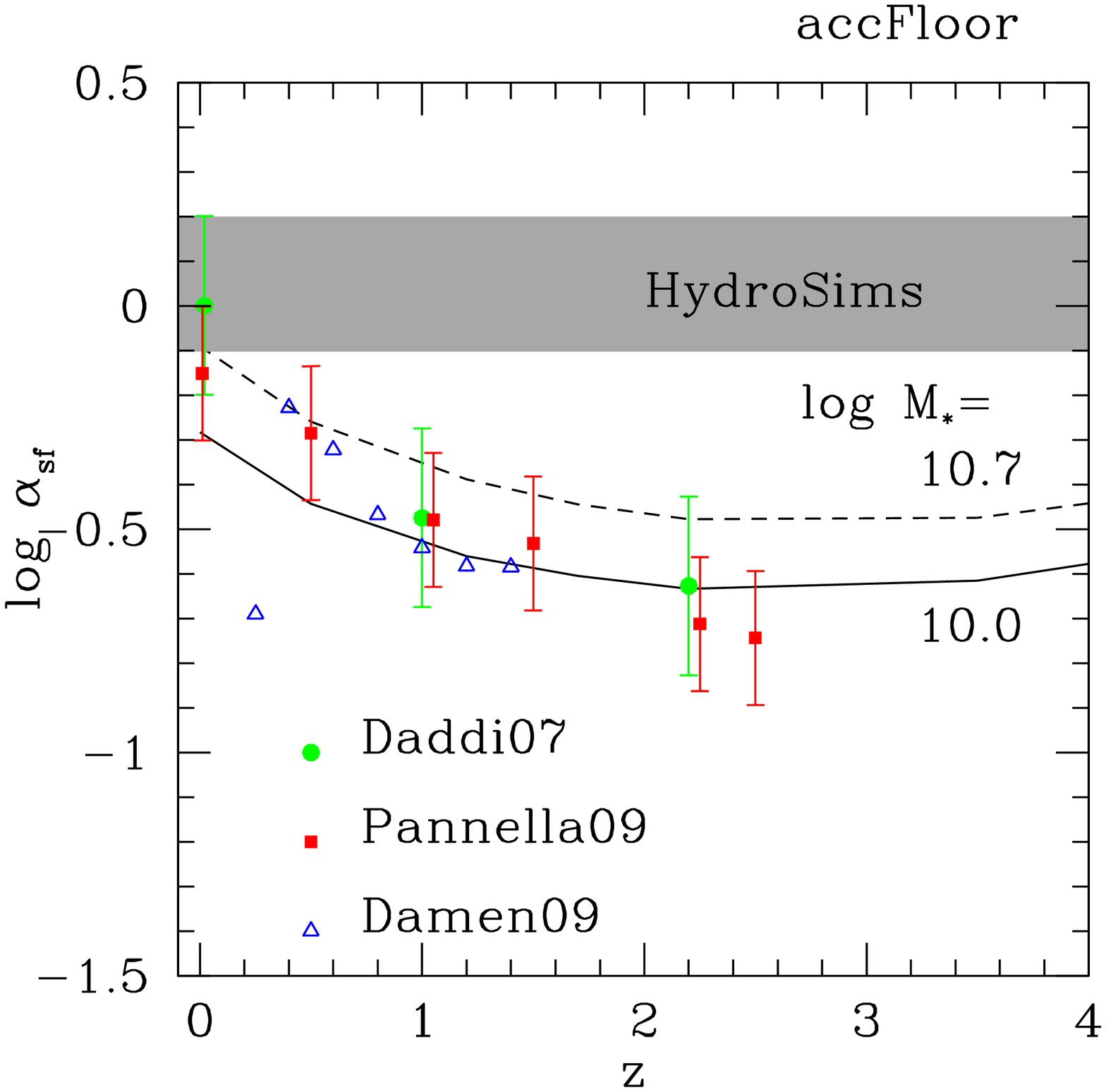}
\caption{Left: Evolution of specific star formation rate (sSFR$\equiv$SFR$/\Ms$) at a fixed $M_\star$ from our fiducial model (`accFloor').
The dotted line  shows the time dependence of the accretion rate 
through the function $\dot \omega \propto (1+z)^{2.2}$ (see text).
The long-dashed line shows the slight modification of the introduced function $f(z)$
(defined in Section~\ref{section:overview}). 
Right: Evolution of the star formation activity parameter $\alpha_{\rm sf}\equiv(\Ms/\dot \Ms)/(t_H-1\hbox{Gyr})$.
 In both panels, we show the  observations by \citet{DaddiE_09a}  (solid circles), \citet{DamenM_09b} (open triangles)
and \citet{PannellaM_09a} (solid squares) for galaxies with $\log M_\star=10$--10.5.
The shaded area shows the hydrosimulation results of \citet{DaveR_08a}. 
The evolution of sSFR and $\alpha_{\rm sf}$ is mostly driven by the time dependence of the
accretion.
\label{fig:ssfr}}
\end{figure*}

\begin{figure*}
\centering
\plottwo{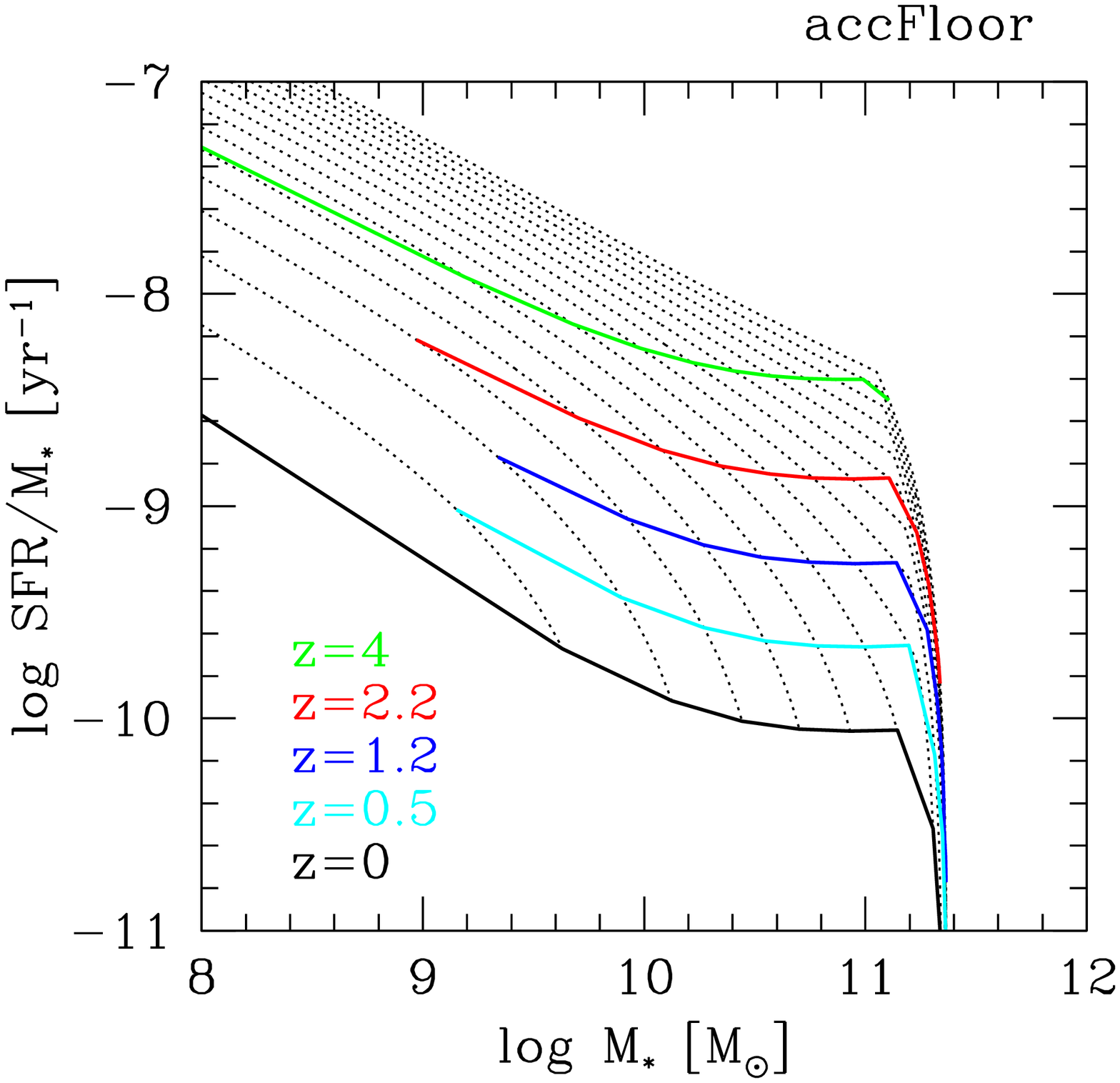}{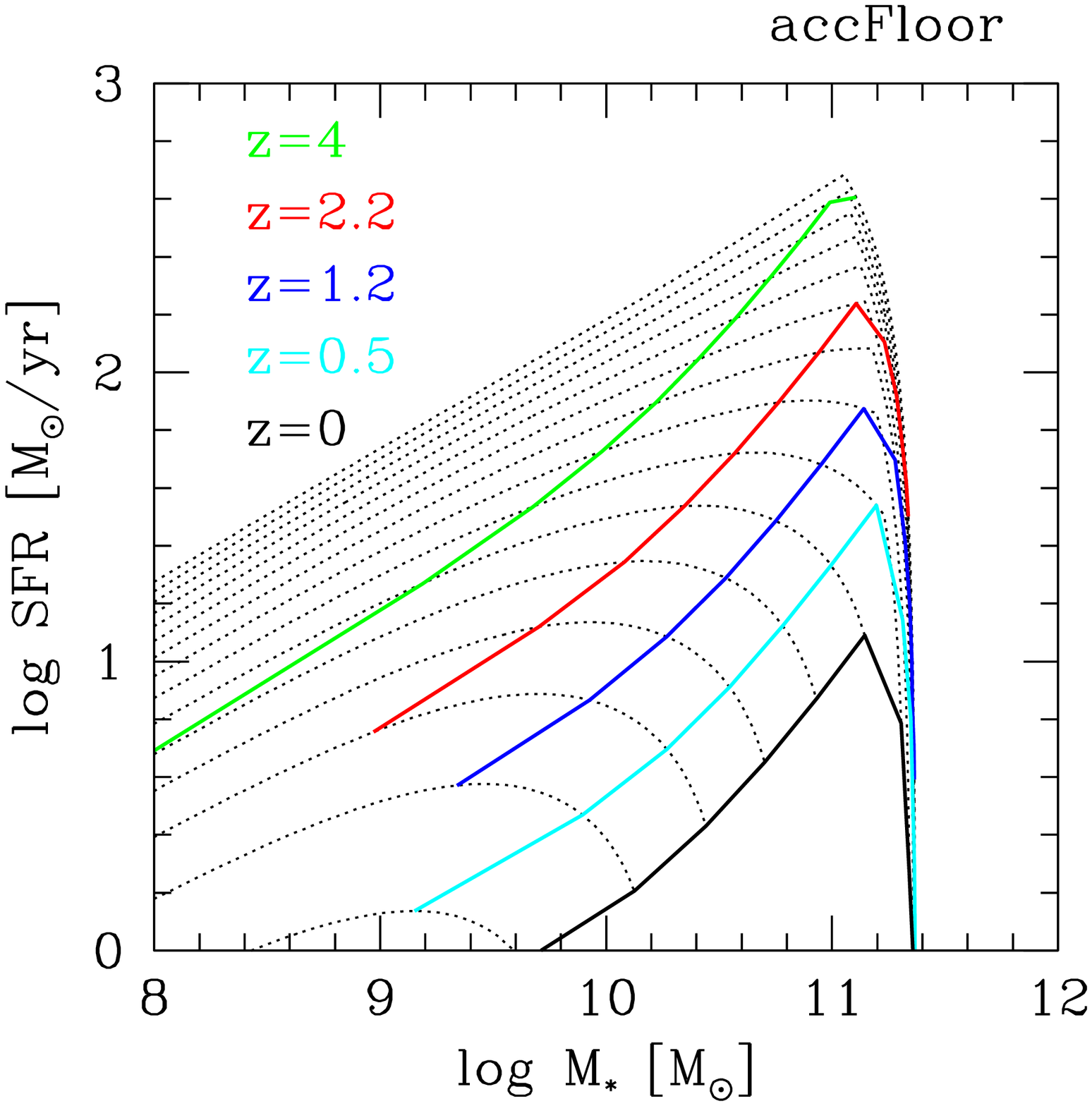}
\caption{Left: Evolution of the sSFR-$\Ms$ relation.
Right: Evolution of the  main sequence.
In both panels, 
the  solid lines show the isochrones of the scaling relations
at $z=4, 2.2, 1.2$ and $z=0$.
The dotted lines show the 
individual tracks for the simulated halos.
}
\label{fig:ssfr:evolution}
\end{figure*}

According to  Equation~\ref{eq:sfrmass},
the observed SFR-$\Ms$ relation 
(equivalently the sSFR) at a given stellar mass)
is declining with time from $z=2$ to the present by a factor $\sim 20$.
Not only does our model reproduce this evolution, but it can help us understand
 its origin.

Figure~\ref{fig:ssfr}(a) shows the model sSFR as a function of redshift for galaxies
of  different stellar  masses, $M_\star=10^{9.5}$, $10^{10}$ and $10^{10.5}~\msun$,
in comparison with observed sSFR for $M_\star$ in the range 
$10^{10}-10^{10.5}~\msun$ by \citet{DaddiE_09a,DamenM_09b} and \citet{PannellaM_09a}.
One sees that our toy model reproduces the variation of sSFR with redshift 
relatively well to $z=2$. We discuss the recent results at $z=4$ to $z=7$ in the Appendix.

The redshift dependence of the sSFR is driven by redshift evolution of
the average accretion rate, Equation~\ref{eq:growth}, which is 
$\propto (1+z)^{2.2}$. Simply put, this factor is driven by the expansion
of the universe.
This can be clearly seen using the EPS formalism, 
where the factor $(1+z)^{2.2}$  originates directly from the derivative of the linear 
growth factor $\dot D$ \citep{NeisteinE_08a,BirnboimY_07a}.
Indeed, under the EPS formalism,  the specific accretion rate is 
$\dot{\Mv}/\Mv=- s(\Mv)\dot \omega$,
where $s(\Mv)$   is the self-invariant mass variable,
and $\omega(z) = 1.69/D(z)$ is the self-invariant time variable.
$s(\Mv)$ is a function of the variance of the initial density fluctuations
on mass scale $\Mv$, $\sigma(\Mv$) \citep[see Equation A5-A6 in][]{BirnboimY_07a}

The dotted line in Figure~\ref{fig:ssfr}(a) demonstrates that the time evolution
of the sSFR (Equation~\ref{eq:gasaccretion}) is indeed roughly proportional to 
$\dot \omega$,
except that the predicted decline with time of the overall accretion rate
is a bit slower than the decline in sSFR at $z<1$ (Equation~\ref{eq:sfrmass}).
An improved fit to the evolution at low redshifts is achieved by a small
modification of the form $\dot \omega\times f(z)$ (long dashed line, defined in
Section~\ref{section:overview}), attempting to correct 
for the gradual decline with time of the cold gas fraction in the overall
baryonic accretion.
This confirms that a gradual decrease in the accretion efficiency parameter, 
by a factor of $\sim 2$ between $z=2$ and $z=0$, helps recovering the
observed evolution of the sSFR.

\citet{DaveR_08a} parameterized the evolution of the SFR sequence at a given 
$M_\star$ in terms of a star-formation activity parameter, 
defined as $\alpha_{\rm sf}=(M_\star/\dot M_\star)/(t_H-1$Gyr). 
He pointed out that most models predict that galaxies have a growth rate comparable to $t_H$, 
the Hubble time, i.e., with a constant $\alpha_{\rm sfr}\simeq 1$, whereas observations 
reveal a growth of $\alpha_{\rm sf}$ by a factor of $\sim 3$ from $z=2$ to 
$z=0$ (see Fig~\ref{fig:ssfr}).  
They used this apparent discrepancy to 
argue that the initial stellar mass function (IMF) must be evolving in time.
Figure~\ref{fig:ssfr}(b) shows the evolution of $\alpha_{\rm sf}$
predicted by our fiducial `accFloor' model for $\log M_\star = {10.0}$ and $10.7\,\msun$. 
An increase of $\alpha_{\rm sf}$ by a factor of $\sim3$ between $z=2$ and $z=0$, not far 
from the observational result, without any evolution in the IMF,
 is a general feature of the time dependence of the accretion efficiency.

The SFR sequence and sSFR($\Ms$) at any given redshift 
are simply isochrones in their respective parameter plane. 
Our model gives us the opportunity to show the actual tracks of individual
galaxies, which are shown in Figure~\ref{fig:ssfr:evolution}
 for the sSFR-$\Ms$ (left) and  the SFR-$\Ms$  (right) relations.


\section{Additional role of $\Mmi$}
\label{section:addresults}

Given that our toy model seems to reproduce  the  
scaling relations and their evolution rather well,
we now compare the predictions of the  model to observations involving 
the `downsizing' effect (Section~\ref{section:downsizing}),
the halo stellar fractions (Section~\ref{section:fstars}), and
the galaxy gas fractions (Section~\ref{section:fgas}).

\subsection{Downsizing and tau models}
\label{section:downsizing}

\begin{figure}
\centering
\plotone{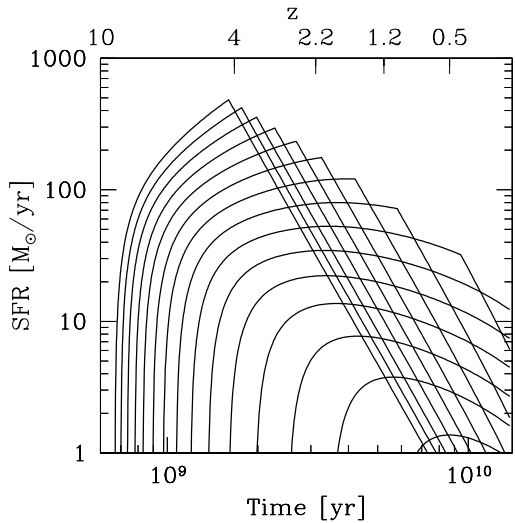}
\caption{Star-formation history SFR$(t)$ for our simulated halos of 
decreasing mass (from left to right).  
The individual tracks simply reflect that more massive halos reach the mass floor $\Mmi$ earlier as shown in Figure~\ref{fig:massgrowth}
and form stars on a shorter time scale (before reaching the mass    ceiling $\Mma$).
}
\label{fig:SFtracks}
\end{figure}

\begin{figure*}
\centering
\plottwo{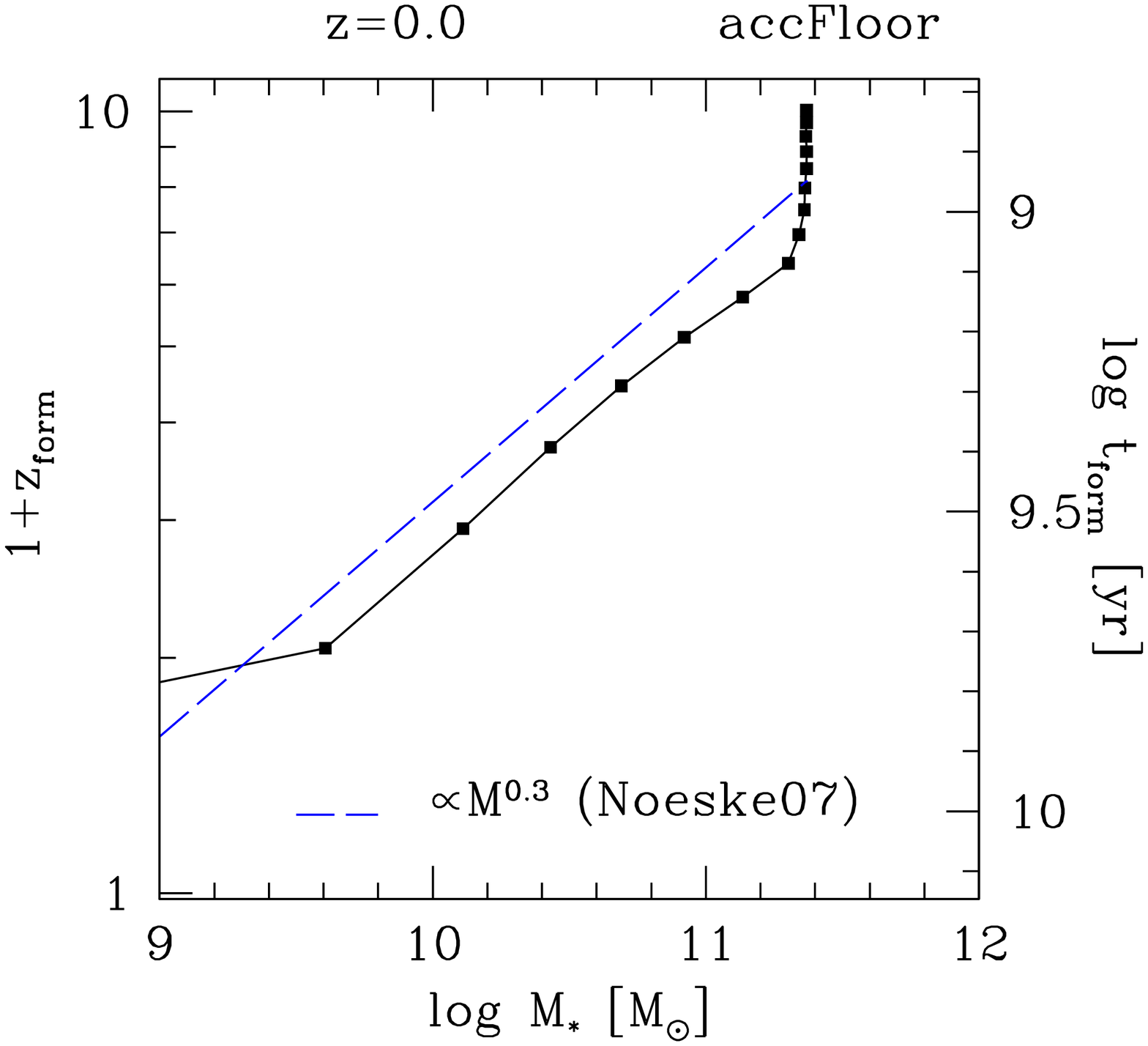}{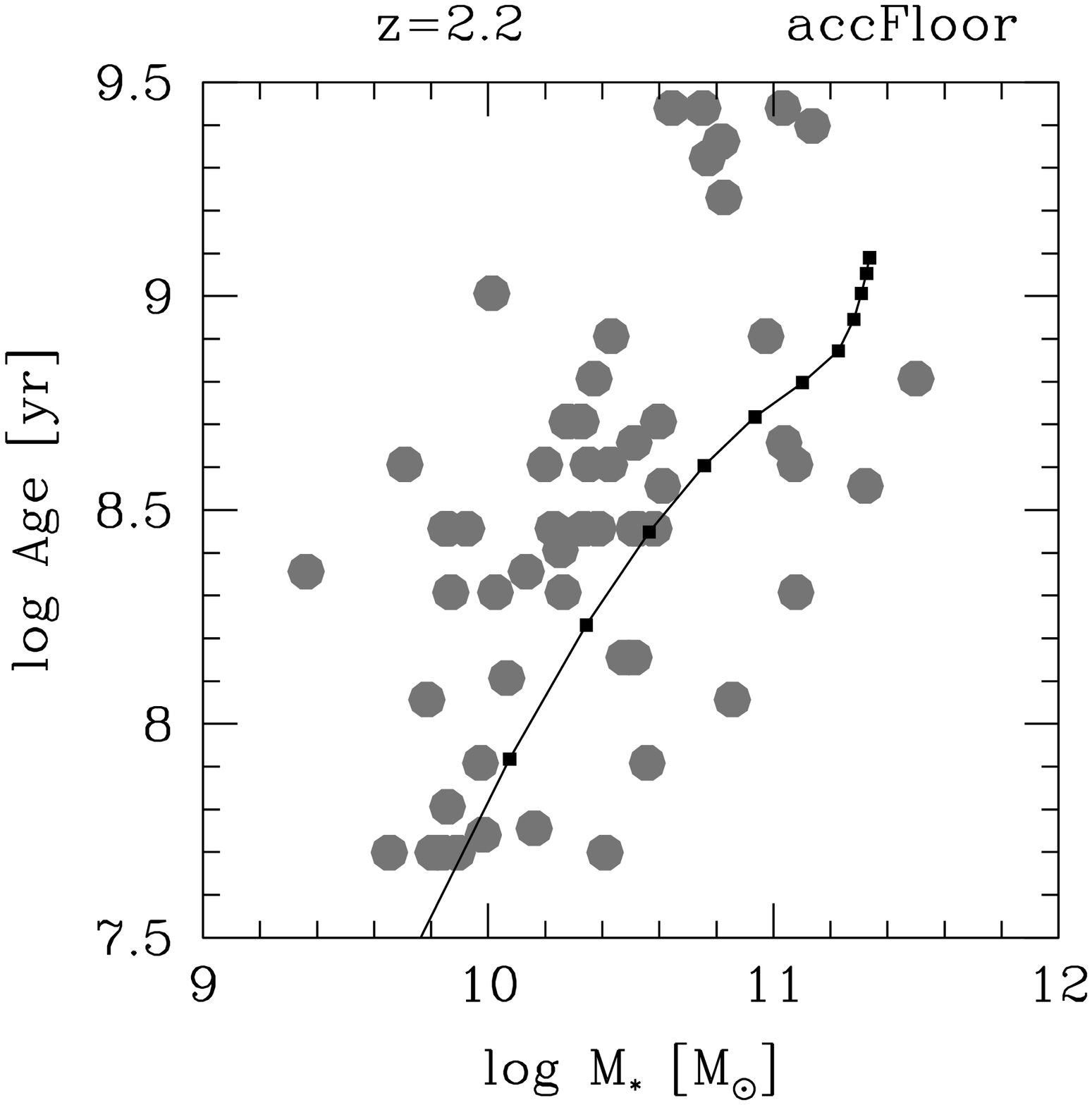}
\caption{Archeological downsizing.
Left: Formation redshift $z_f$ (defined when $\Mv(z)$ reaches $\Mmi$) as a function of stellar mass
for our modeled galaxies (solid squares).
The dashed line shows the $\Ms-z_f$ relation inferred by  \citet{NoeskeK_07b}
from the $z\sim1.0$ SFR sequence, namely, $(1+z_f)\propto \Ms^{0.3}$ (normalization is arbitrary).  
Right: The modeled $z\sim2.2$ age-$\Ms$ relation is shown with the solid squares.
Modelled ages are the luminosity weighted ages (see text).
The gray points show the ages (inferred from SED fitting) for the large sample of $z=2$ SFGs 
available in the SINS survey \citep{ForsterSchreiberN_09a}.
At all epochs, the age-mass dependence is a direct consequence of the mass floor $\Mmi$.
}
\label{fig:agemass}
\end{figure*}

Stellar-population analyses of red galaxies in the local universe
show that stars  tend to have formed earlier and over a shorter time-span
in the most massive ones
\citep[e.g.][]{HeavensA_04a,ThomasD_05a,JimenezR_07a,ThomasD_10a}.
This is sometimes termed as `archeological downsizing'.
This age-mass  correlation is also present  in large samples 
(up to $3\times10^5$) of {\it active} galaxies in the Sloan Digital Sky Survey (SDSS)
 \citep[e.g.][]{GallazziA_05a,JimenezR_07a}.

The SF history (SFH) tracks shown in Figure~\ref{fig:SFtracks} 
demonstrate the origin of the age-mass anti-correlation.
More massive halos reach the mass floor earlier, and spend less
time between the mass floor $\Mmi$ and the mass ceiling $\Mma$.
This is a natural consequence of the halo tracks in Figure~\ref{fig:massgrowth}
combined with our key assumption of $\Mmi$.

Figure~\ref{fig:agemass} shows the resulting age-mass relation
 at $z=0$ (left) and $z=2$ (right).
The left panel shows the formation redshift $z_f$ as a function of stellar mass.
Similarly to \citet{NoeskeK_07b}, 
the formation redshift $z_f$ is defined here when star formation began, 
i.e., when $\Mv(z)$ reaches $\Mmi$.
The mass index of the resulting age-mass relation
is surprisingly close to the $(1+z_f)\propto M^{0.3}$ inferred by \citet{NoeskeK_07b}
shown as the dashed line (normalization arbitrary).

That the SFR sequence leads to the same mass-dependent SFH as in the
archeological downsizing was already argued by \citet{NoeskeK_07b}.
Indeed, they fitted $\tau$-models to the $z\sim1$ SFR-$M_\star$ relation and concluded
the sSFR relation required both $\tau$ and the formation redshift $z_f$ to be mass-dependent 
($\tau \propto M^{-1}$, $(1+z_f)\propto M^{0.3}$). 
They call this mass-dependent 
SFH `Staged Galaxy Formation'
where more massive galaxies are formed first on a shorter time scale.
However, the mass dependency in our model did not have to agree with the results
of \citet{NoeskeK_07a}.

This age-mass relation may already be in place at $z\sim2.2$ (Figure~\ref{fig:agemass}(right)).
Indeed, the large sample of the SINS
survey \citep{ForsterSchreiberN_09a} shows a similar trend  between spectral energy distribution
(SED) derived ages and stellar masses.
In Figure~\ref{fig:agemass}, we compare these observations (gray points) to
the luminosity weighted ages in our model (solid squares), which
are derived according to $\int t_{\rm lbt} \hbox{SFR}(t_{\rm lbt}) \rm d t_{\rm lbt}$
where $t_{\rm lbt}$ is the lookback time since $z=2.2$.

Thus, our model reveals that the key player for downsizing is the accretion mass floor $\Mmi$.
As it was first realized in \citet{NeisteinE_06a},
 the archeological downsizing effect is a direct consequence of a mass floor $\Mmi$.

It has been noted that, for early type galaxies, the more massive ones
reach the red sequence earlier, a phenomenon dubbed `downsizing in mass'
\citep{CimattiA_06a}. In our model, when galaxies
reach the ceiling mass $\Mma$, their gas supply dries out and SF is quenched, i.e.,
the galaxy becomes red and passive.
The same SFH tracks shown in Figure~\ref{fig:SFtracks}
 lead to this other type of downsizing, in this case 
 due to $\Mma$ \citep[see also][]{CattaneoA_08a}.

\subsection{Stellar fractions}
\label{section:fstars}

\begin{figure*}
\centering
\plottwo{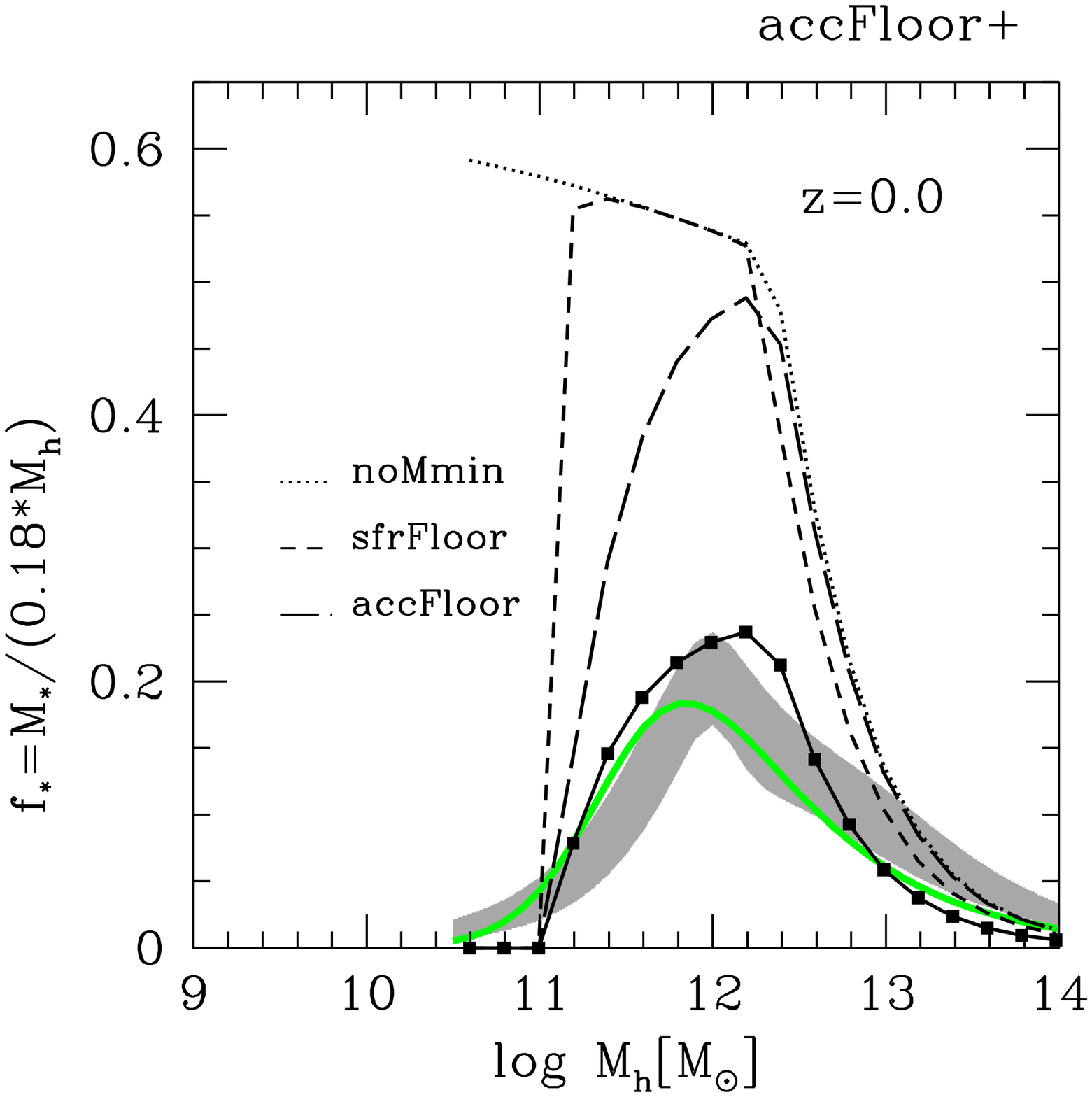}{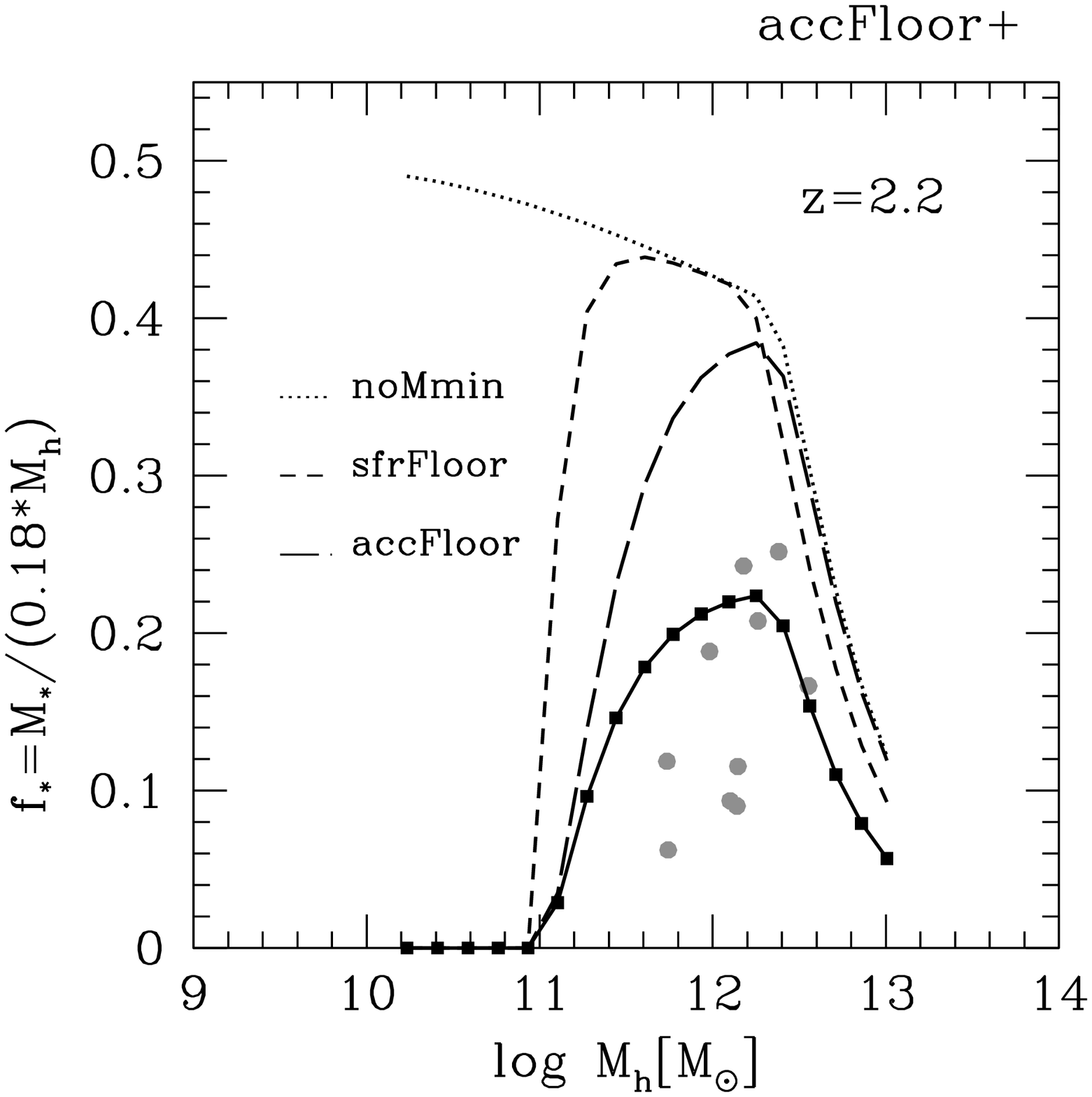}
\caption{Stellar fractions ($f_\star\equiv \Ms/(0.18*\Mv)$)
 at $z=0$ (left) and $z=2$ (right)
predicted from our fiducial model ('accFloor+').
In each panel, 
the alternative models, namely, with no mass floor $M_{\rm min}$  (dotted line), 
with an accretion floor  (long dashed line), and
with an SF floor (short dashed line) are indicated.
The $z=0$ stellar fractions from \citet{MosterB_10a} and \citet{GuoQ_10a}
 are shown as the shaded area and thick line, respectively.
The stellar fractions in $z=2$ SFGs are shown as gray circles \citep{CresciG_09a}.
\label{fig:fstars}}
\end{figure*}

Many have shown that the stellar fraction in galaxies
is a strong function of the total halo mass
\citep[e.g.][]{vandenBoschF_03a,EkeV_05a,ShankarF_06a,vandenBoschF_07a,
KravtsovA_09a,MosterB_10a,GuoQ_10a} owing to the very different shapes of the halo 
mass function and the galaxy stellar mass function.
We  confront our toy model with this additional 
observational constraint.

Figure~\ref{fig:fstars} compares the halo stellar fractions, defined as
\be
f_\star\equiv \frac{M_\star}{0.18*\Mv} \, ,
\ee
predicted from our model to relevant data at $z=0$ (left) and $z=2.2$ (right).
The shaded area shows the results from the SDSS analysis of \citet{GuoQ_10a}, and
the solid line shows the similar result from \citet{MosterB_10a}.
At $z=2.2$,
the observational estimates are from the disk-dominated galaxies of
the SINS survey \citep{ForsterSchreiberN_09a,CresciG_09a}, where
the virial mass is estimated assuming that $\Vv=\Vrot$.

We see that the models  `noMmin' and `sfrFloor'  overpredict the stellar 
fraction for halos below $10^{12}\,\msun$, while the `accFloor' model
recovers the shape $f_\star(\Mv)$, including the drop toward lower masses
occurring at the proper range of $\Mv$.  
Once again, we learn that an effective halo mass floor for accretion 
at $\Mmi \sim 10^{11}\,\msun$ is essential for a fit to the data.

However, the amplitude of the $f_\star(\Mv)$ function predicted in the simple accFloor model
needs to be reduced roughly by a factor two for a match with the data.  
A better fit is obtained if we include in our model (Equation~\ref{eq:bathtub}) an outflow term 
$$\dot M_{\rm out}=\rout\, \dot M_\star.$$ 
This constraint is very consistent with the observations of \citet{HeckmanT_00a} and \citet{MartinC_05a} who found that $\dot M_{\rm out}\simeq$~SFR.
Indeed, the corrections  involved in deriving the outflow rates in galaxies
(for ionization, metallicity, and depletion)
 are usually of several orders of magnitude.
Moreover,  the instantaneous outflow rate needs not be equal to our net outflow
$\dot M_{\rm out}$ term, 
 which only describes the baryons that leave the halo forever.
We note that  this constraint on the outflow rate is degenerate with
the accretion efficiency $\epsin$, i.e., 
$\dot M_{\rm out}=(0.4\hbox{---}0.8)\; \dot M_\star$ for $\epsin$ ranging from 0.5 to 1.0.

It is important to realize that the results presented thus far are independent
of this outflow term. The SFR sequence is not sensitive to feedback 
as lowering the SFR by a factor of two leads to a stellar mass $\Ms$ lower by the same factor.
\citet{DuttonA_10a} demonstrates this in a similar work. 

\subsection{Gas fractions}
\label{section:fgas}

\begin{figure*}
\centering
\plottwo{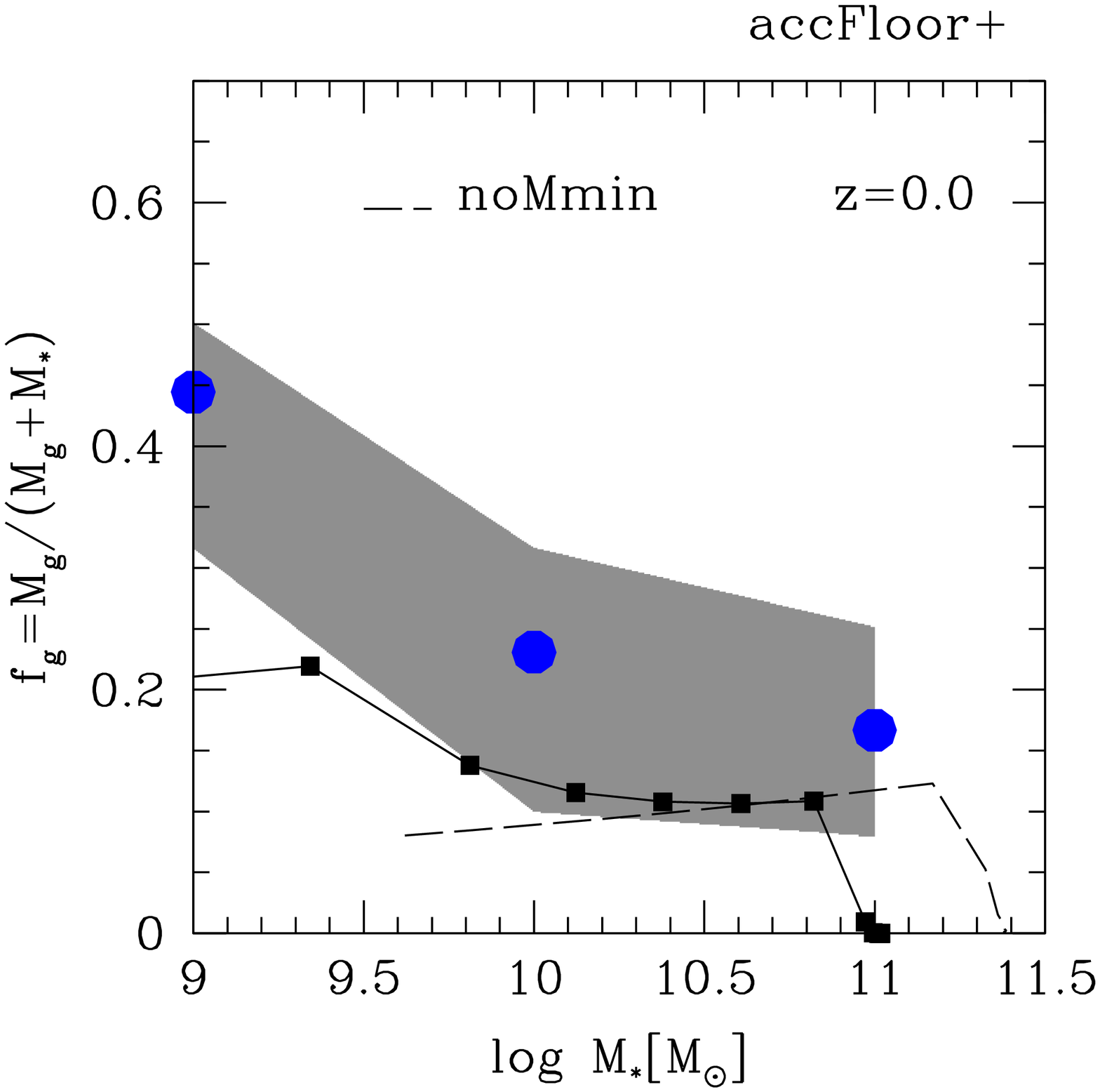}{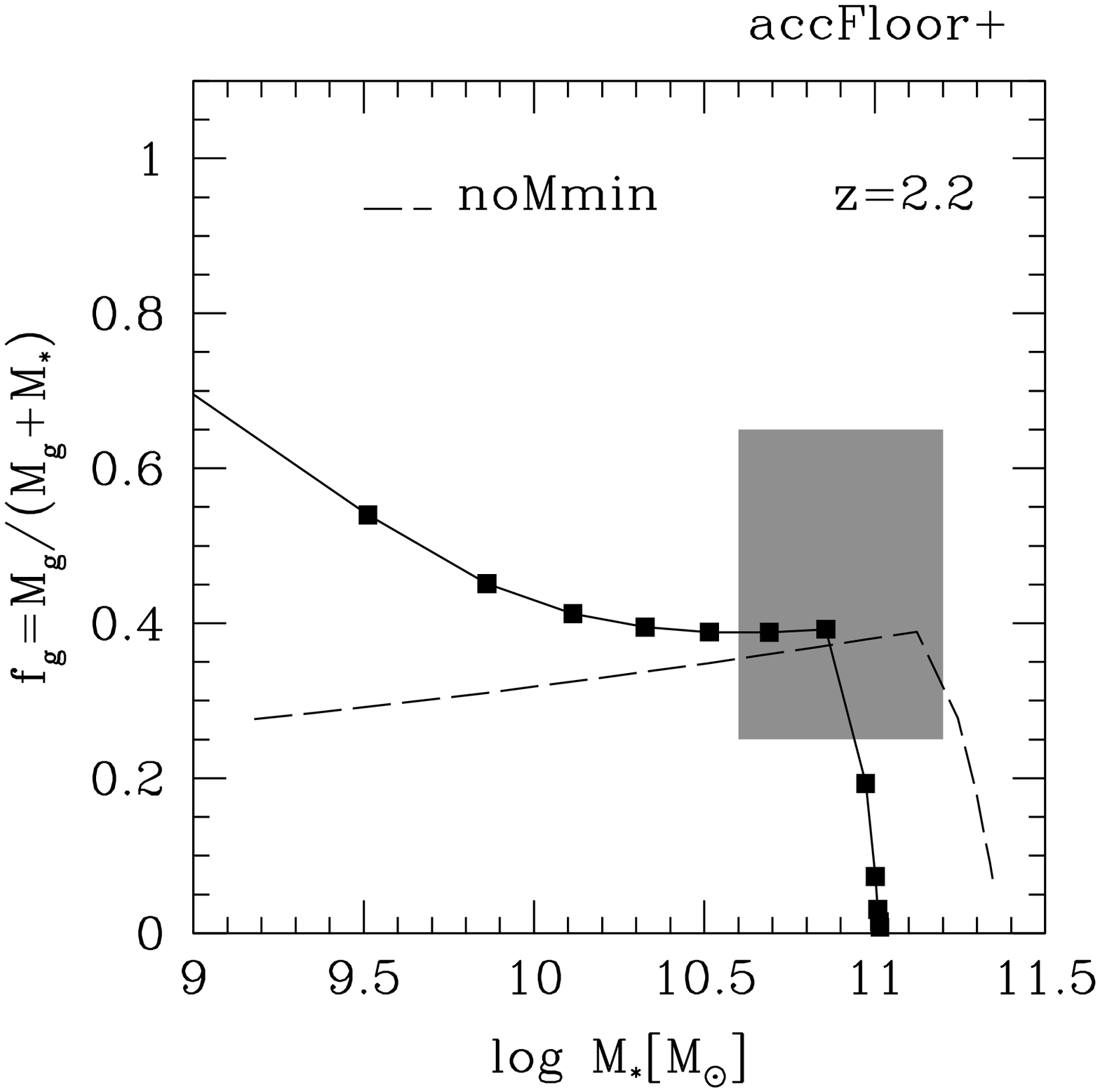}
\caption{The solid squares show the gas fractions  as a function of $M_\star$
predicted from   our fiducial model `accFloor+'
 at $z=0$ (left) and $z=2$ (right).
The observed local gas fractions  from \citet{BaldryI_08a} and
\citet{ZhangW_09a} are shown as the solid circles and shaded area, respectively.
The new $z\sim 2$ gas fraction is represented by the shaded box for dozen  SF galaxies
with $\log M_*>10.6$  \citep[see][]{TacconiL_10a}. 
The upturn of $f_g$ at the low mass end is again a result of the mass floor
$\Mmi$.
\label{fig:fgas}
}
\end{figure*}

Another interesting constraint is provided by the  gas fraction in galaxies,
defined as 
\be
\fg \equiv \frac{\Mg}{\Mg+M_\star} \, .
\ee
At low redshift, we learn from all-sky \HI\ surveys 
\citep[e.g.][]{ZwaanM_03a,RosenbergJ_03a,MeyerM_04a,GiovanelliR_05a,WongO_06a,CatinellaB_10a}
that the gas fractions in SFGs vary systematically
with stellar mass \citep[e.g.][and references therein]{BaldryI_08a}.
These data are presented in the left panel of Figure~\ref{fig:fgas}.
At $z \sim 2$, thanks to the rapid progress in the sensitivity of millimeter 
interferometers (such as the IRAM Plateau de Bure interferometer, PdBI), 
it is now possible to study CO rotational emission lines of normal high 
redshift SFGs at $z>1$ and therefore constrain their gas masses.
The observed gas fraction for $z \sim 1$--2.5 SFGs with $M_{\star}>10^{10.6}\,\msun$ 
ranges from 0.3 to 0.6 \citep{DaddiE_10a,TacconiL_10a}
 with an average of about 0.44 for the large sample of $19$ SFGs of
\citet{TacconiL_10a}.

Figure~\ref{fig:fgas} compares the predicted $\fg(M_\star)$ 
from our model `accFloor+' to those observations at
$z=0$ (left) and $z=2$ (right).
We see that the model is consistent with the data both at 
$z=0$ and at $z \sim 2$. 
Locally, the upturn of $f_g$ at the low mass end is again a result of the mass floor
$\Mmi$.

Contrary to the other results presented so far, the gas fractions are
sensitive to SF efficiency $\esfr$, given that the SFR is set by the steady-state solution
(or by the accretion rate).  A lower SF efficiency leads to higher gas fractions
and vice-versa.
We used a universal $\esfr$ set by the KS relation and
 the redshift evolution of  $f_g$ in our model appears to be very consistent 
with observations. 
Indeed,  the model predicts  $\fg = 0.4-0.55$ at $z=2$ and 
$\fg=0.3-0.45$ at $z=1$, which is in good agreement with the average gas 
fractions estimated from the CO observations, 0.44 and 0.34, respectively
\citep[][see also Daddi et al. 2010]{TacconiL_10a}.

Our simplistic toy model, which was only 
tuned (with $\Mmi$) to reproduce the scaling relations, 
helps us also to understand
the trends of gas fraction with mass and redshift  for SFGs.

\section{Star-formation history: constraining the mass floor}
\label{section:madau}

\begin{figure}
\centering
\ifthenelse{\boolean{preprint}}{  \plotone{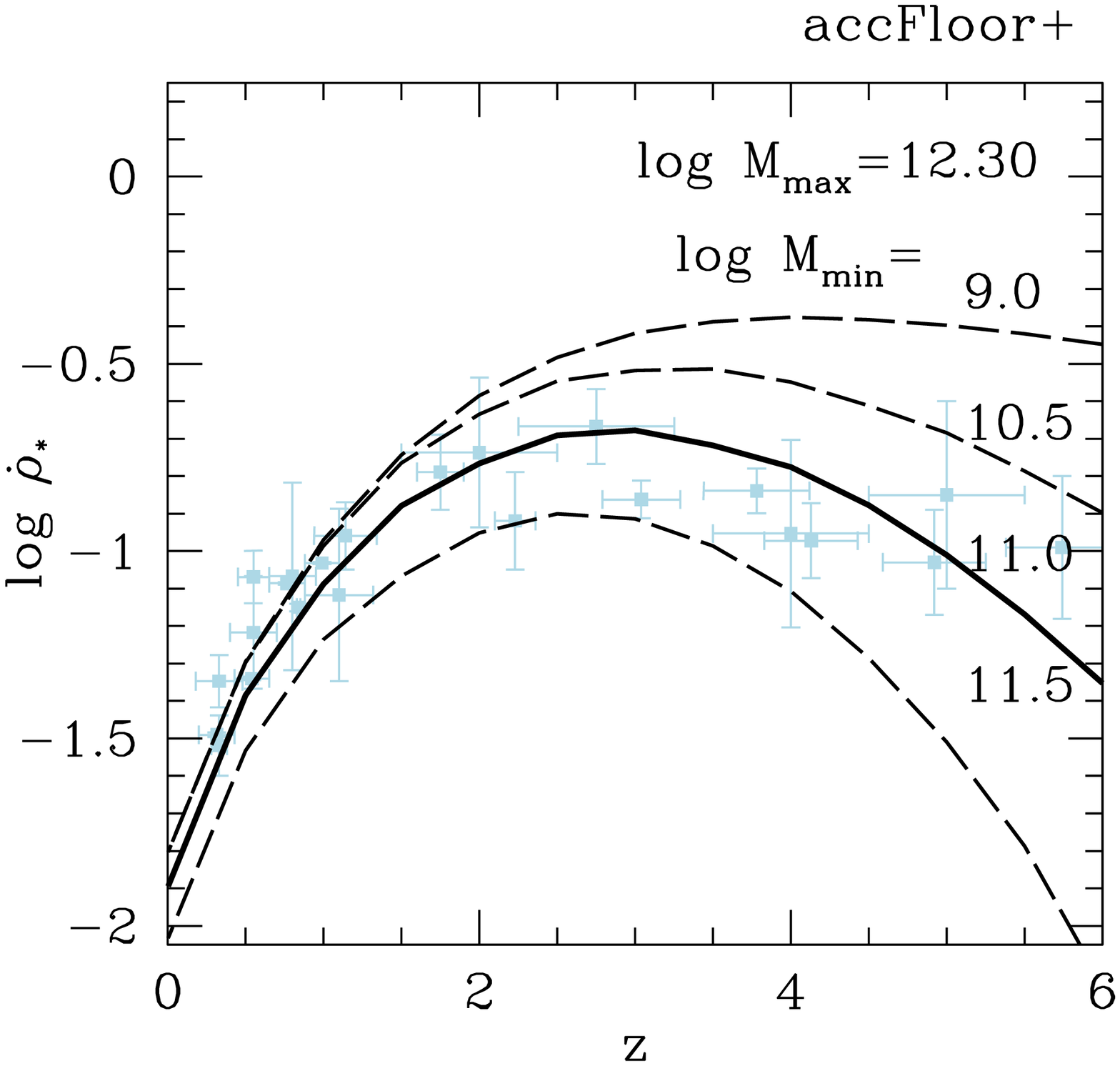} }{\includegraphics[width=8cm]{f24.eps}}
\caption{The star formation history $\dot \rho_\star$.
The dashed curves show that $\dot \rho_\star(z)$ is very sensitive to the minimum mass $M_{\rm min}$. 
The light symbols show the most recent $z<2$ observations
\citep{SchiminovichD_05a,DahlenT_07a,MobasherB_09a} with the $z>2$ data \citep{HopkinsA_06a,VermaA_07a,WilkinsS_08a}
 all converted to a Chabrier IMF following \citet{HopkinsA_08a}. \label{fig:madau}}
\end{figure}

Thus,  based on the two scaling relations, the stellar and gas fractions,
our  model shows that a mass floor $\Mmi$ must play an important role
in galaxy formation and we argued that $\Mmi$ should be $\sim$\mmin~\msun.
We are about to show that the evolution of the
average density of SF, $\dot \rho_\star(z)$,
will give us a  quantitative constraint on $\Mmi$.

Figure~\ref{fig:madau} shows a compilation of data for the cosmic SFH, 
derived from observations under the assumption of a Chabrier IMF.
The average density of SF, $\dot \rho_\star$,
is observed to be rising from $z \sim 6$
to $z \sim 2$, reaching a plateau at about $z \sim 2$, and then dropping by an
order of magnitude between $z \sim 1$ and $z =0$
\citep[e.g.][and references therein]{LillyS_96a,MadauP_96a,HopkinsA_06a}.

Our  `accFloor+' model convolved with the
Sheth-Tormen model for the halo mass function \citep{ShethR_01a}
is shown for 4 different values of $\Mmi$, ranging from
$10^9$ to $3\times 10^{11}\,\msun$. 
We see that the value of $\Mmi$ indeed determines 
the growth of $\dot \rho_\star$ with time at high redshift.
Best agreement to the growth rate observed at $z>2$ is obtained with  
$M_{\rm min}=$\mmin\,$\msun$.
Thus, we conclude that the rise in the cosmic SFH from $z=6$ to $z\sim2$
 confirms the necessity of a mass floor and provides the strongest constraint on the 
value of $\Mmi$.

Our model can also help us to understand the origin of the decline at late times.
Possible reasons for this sharp decline could in principle be
the acceleration of the universe at recent cosmological epochs, 
gas exhaustion into stars and outflows, 
the quenching of SFR above the mass ceiling, or simply the generic
decline in the accretion rate into dark-matter halos due to the expansion 
of the universe.  
The sharp decline at $z<2$ is predominantly driven 
by the decreasing accretion rate as a direct result of the expansion of 
the universe as it can be seen from Figure~\ref{fig:massgrowth} (right)
and discussed in Section~\ref{section:evolution}.  
All the other potential sources for the decline, 
including the acceleration of the universe and the introduction of $\Mma$ 
or gas consumption, have only minor effects on this decline.

Many other groups \citep[e.g.][]{SpringelV_03a,NagamineK_06a,SchayeJ_10a,ChoiJ_09a} have used
cosmological hydrosimulations to investigate the cosmic SFH 
and to identify the physical processes that drive its evolution. 
Similarly to our model, \citet{HernquistL_03a} used an analytical approach 
and reach similar conclusions. They find that, at high-$z$, 
the rise in $\dot \rho_\star(t)$ originates from the gravitationally driven 
growth of halos, while at low-$z$, the decline is
due to the expansion of the universe, through the inhibition of cooling. 
While in our model the SFR at a given halo mass is driven
 by the halo growth rate, their model assumes an ad-hoc form for the 
sSFR that is constant as a function of mass for all halos of virial 
temperature above $10^4$~K. Similarly to our $\Mmi$ assumption,
in their model, stars do not form in halos below $T_{\rm vir}=10^4$~K.
Their assumption also leads to a peak in the SFH, albeit at a higher redshift
$z\sim5$ or 6.  In spite of the differences between their results and our results,
this again illustrates the impact of a mass floor.

\section{Discussion and Conclusions}
\label{section:conclusions}
 
The purpose of this paper is to seek  a basic understanding of the relations
between global properties of SFGs and their time evolution 
between $z \sim 2$ and today. This includes the origin of the SFR sequence 
and the Tully-Fisher  relation.
We achieved this via an idealized and remarkably simple `toy' model, based on the cosmological
evolution of accretion rate into dark-matter halos,
the efficient deep penetration of cold gas streams into the central galaxies
and the standard efficiency of SFR in galactic disks.
The model should be interpreted as a learning tool for gaining insight into the role
played by several key physical processes.

The model naturally leads to the observed  scaling relations,
$\dot M_\star \propto M_\star^{0.8}$ and $M_\star \propto \Vrot^4$
and to the archeological `downsizing' phenomenon,
{\it provided} that there is a mass floor $M_{\rm min} \sim $\mmin~$\msun$ below which 
the accretion of cold gas is substantially suppressed~\footnote{We note that the threshold could have been defined in terms of a constant virial velocity $\Vv$, and this would have lead to similar results.}.
An analysis of the SFH allowed us to put an
 additional strong constraint on the mass floor: $M_{\rm min}\simeq $\mmin~$\msun$,
corresponding to $\Vv\sim 100\,\,\Mv{}_{,11}^{1/3}\,(1+z)_{3.2}^{1/2}$\kms,
best accounts for the observed scaling relations.

Recently, \citet{DuttonA_10a} have used full semi-analytical approach to study the origin of the SFR sequence.
While we agree with the conclusions of \citet{DuttonA_10a} with respect to the steady state, the impact of feedback, and to the evolution of the SFR sequence,
our model differs from theirs on one significant aspect: the lack of a mass floor.
\citet{CattaneoA_10a} used a full semi-analytical model with an effective efficiency \epsin\ similar to ours (compare their Fig.4 to our Fig.1).
Their effective mass floor is due to the combined effect of SN feedback and photo-heating due to reionization.
These authors, however, used a re-ionization threshold at $V_{\rm re}=40$ \kms\ in order to fit 
the luminosity function, which is higher than expected ($V_{\rm re}\sim10$ \kms) from pure reionization \citep[e.g.][]{BarkanaR_99a,IlievI_05a}  
 by an order of magnitude in halo mass.

\subsection{Caveats and limitations}
\label{section:caveats}

Our model, which relies mainly on the cosmological accretion rate,
is based on very few parameters ($\epsin$, $\Mma$,
and $\Mmi$) compared to semi-analytic models  \citep[e.g.][]{SomervilleR_08b}.
Our main free parameter is $\Mmi$ (affecting the cooling/accretion efficiency $\epsin(M)$), as we 
use normal a SF efficiency $\esfr$ and set $\Mma$
 following  \citet[e.g.][]{CattaneoA_06a,CattaneoA_08a}.

The remarkably few parameters in our model are clearly an advantage \citep[see also][]{NeisteinE_10a}, 
but this inevitably leads to  over-simplifications.  For instance, 
modeling the lower and upper thresholds for cold gas accretion  
 as sharp cutoffs is clearly unphysical.
In reality, these transitions  
are likely to be smooth and to depend on other galaxy properties.
For example, $\Mma$ is rather sensitive to the metallicity in the halo 
\citep{DekelA_06a}. 
Indeed, the smoothness of this transition  near $\Mma$ can be derived from hydrodynamical simulations 
\citep{KeresD_05a,BirnboimY_07a,OcvrikP_08a,KeresD_09a}.
In fact, a smooth transition at $\Mma$ would relieve the tension between the model 
and the observed scaling relations at the massive end, 
$M_* \sim 10^{11}\,\msun$, in Figures~\ref{fig:bluesequence:z2} and 
\ref{fig:TFR:z2}.

The toy model presented in this paper does not allow us to address
other observational constraints such as the mass-metallicity relation 
because this is rather sensitive to the exact treatment
 of the metal content of the stellar, SN outflows
combined with the unknown fraction of ISM that is entrained in the outlow.

We also cannot address the scatter in the scaling relations as the model
deals only with the mean global properties.  
Variations in the mass assembly history are to be expected \citep[e.g.][]{GenelS_08a}
and will likely contribute to the scatter in the scaling relations
\citep[e.g.][]{DuttonA_10a}.

Another limitation of our simplified model is that we do not include
bulge growth, which can be an additional drain for SF disks in our model.

\subsection{Conclusions}

In spite of its simplistic assumptions,  our model captures many 
of the observed properties of SF galaxies and their evolution.
This approach helps us gain insight into the following questions:
\vspace{0.4cm}

$\bullet${\it Why are the $z=2$ SFRs so high?} {\it What sets the SFR sequence?}
Once the timescale for SF becomes shorter than the accretion 
timescale, galaxies settle in a quasi-steady state where the SFR is
driven by the accretion rate (Figure~\ref{fig:steadystate}).
As a consequence, the high SFRs for $z=2$ SFGs are driven by the high accretion rate of cold
baryons. 
The steady state also implies that the instantaneous SFR is comparable to the
average SFR ($\langle$SFR$\rangle$) for a few Gyr from $z \sim 4$ to $z\sim1.5$
(see Figure~\ref{fig:massgrowth}(b)). 
This is consistent with
observations from the SINS survey
\citep{ForsterSchreiberN_09a} where the ratio $\langle$SFR$\rangle$/SFR 
is found to be $\simeq1.0$ in $z=2$ SFGs.

$\bullet${\it What is the origin of the scaling relations of
SFGs?}
We find that a single parameter, namely,
the mass floor $\Mmi\sim$\mmin\msun\ for accretion, is key in generating the 
correct slopes of both the SFR sequence and the TF relation 
(Figs.~\ref{fig:bluesequence:z2} and \ref{fig:TFR:z2}). 
Thus, the two scaling relations are not independent relations,
as we show in Section~\ref{section:connections}.

We emphasize that the SFR sequence and its evolution are insensitive 
to the values of the efficiency parameters $\epsin$ and $\esfr$, because
changes in either of these efficiencies affect both $\dot M_\star$ and $M_\star$,
and shift galaxies along the SFR sequence. The same argument holds for the impact of feedback
on the SFR sequence \citep[see][]{DuttonA_10a}.
This is the reason for the failure of any model that attempts to reproduce
the tilts of the scaling relation by varying either the SF efficiency 
or the feedback efficiency
\citep[see][and references therein]{DaveR_08a,DamenM_09b}.

$\bullet${\it What sets the evolution of the SFR-$\Ms$ relation? }
The redshift evolution of the SFR sequence (or sSFR)
is driven predominantly
 by the redshift evolution of the cosmological DM accretion rate.
Both the SFR sequence and the sSFR at a fixed mass follow the cosmological decline of
average accretion rate onto halos as a result of the expansion of the universe
(Figure~\ref{fig:ssfr}). A slight improved match with the observations is obtained when
we force the accretion efficiency to decrease with time at  $z<2$.

$\bullet${\it What is the origin of archeological downsizing?}
Why did more massive galaxies form their stars earlier and over a shorter 
duration \citep[e.g.][]{HeavensA_04a,ThomasD_05a,GallazziA_05a,JimenezR_07a}?
An inspection of the halo trajectories in (Figure~\ref{fig:massgrowth})
reveals that more massive halos reach the mass floor earlier, and spend less
time between the mass floor and the mass ceiling (see also Figure~\ref{fig:SFtracks}).
It is the mass floor $\Mmi$ that naturally produces the age-mass relation
(Figure~\ref{fig:agemass}) as \citet{NeisteinE_06a} already pointed out.

$\bullet${\it Is the observed low stellar fraction compatible with the
high accretion efficiency of cold gas?}
Having a high accretion efficiencies ($\epsin>$50\%) is entirely compatible
with the observed stellar mass fractions $f_\star$
that are low ($<50$\%), i.e., $f_\star\sim 0.20$ (Figure~\ref{fig:fstars})
provided a net outflow rate $M_{\rm out}\propto \rout\, \dot M_\star$ is included.
This is consistent with observations $M_{\rm out}\simeq 1.0\, \dot M_\star$
given that a significant fraction of the out-flowing gas 
may be returned into the interstellar medium  by $z=0$ \citep[e.g.][]{OppenheimerB_08a};

$\bullet${\it Why are the gas fractions large at $z>1$?}
The observed high gas fractions at $>1$
arise naturally from the high, mass-dependent accretion efficiency 
and the low star-formation efficiency. 
The SFR is set by the accretion rate in the steady-state solution, 
while the gas fraction is determined by the SFR efficiency \esfr.
The model predicts mean gas fractions of 30\%--45\%\ at $z=1.2$
and 40\%--55\%\ at $z=2.2$ (Figure~\ref{fig:fgas})
in good agreement with the recent observations of \citet{TacconiL_10a} and \citet{DaddiE_10a}.

$\bullet${\it What is the origin of the shape of the SFH?} 
The decline in the SFH from $z=2$ to the present is primordially
due to the decline in accretion rate associated with the
cosmological expansion. The rise of $\dot \rho_\star$ from $z=6$ to $z=2$ 
necessitate a mass floor near $\Mmi \sim 10^{11}\,\msun$ 
(Figure~\ref{fig:madau}).

\vspace{0.4cm}

The successes of our model are evidence for the central role played by the key ingredient of
our reservoir model, namely, the halo growth rate in the standard
cosmology.
While our model reveals the importance of a mass floor $\Mmi$ for accretion,
it does not address its physical origin.

\section{On the origin of the mass floor}
\label{section:accretionfloor}

The exact mechanism that suppresses or holds accretion below $\Mmi$ may be due 
either to photo-heating associated with the fresh UV background after 
reionization
\citep[e.g.][]{ThoulA_96a,QuinnT_96a,BarkanaR_99a,GnedinN_00a,DijkstraM_04a},
 or to SN feedback  \citep[e.g.][]{DekelA_86a,DuttonA_09a,OppenheimerB_08a}~\footnote{Note, in the context of momentum driven winds \citep{OppenheimerB_08a},
 our mass floor corresponds to a threshold in the mass loading factor $\eta$
where $\eta\rightarrow\infty$ below $\Mmi$. }.
Photo-heating associated with re-ionization increases the mean
temperature of the IGM to $\sim 10^4$~K.
This reduces the cooling rate of the hot gas and prevents the
assembly of new low-mass galaxies with $\Vv<10$\kms (or $\Mv \sim 10^{8}$\msun), and
those that have already collapsed are relatively quickly photo-evaporated
\citep[e.g.][]{BarkanaR_99a,IlievI_05a}.
SN explosions,  like photo-heating from reionization, provide a negative 
feedback on SF.  If present simultaneously, photo-heating and SN feedback effects
can mutually amplify each other’s ability to suppress the SFR
\citep{PieriM_07a,PawlikA_09a}.
{Lastly, the formation of cold ($T\sim 10$--30~K) molecular gas (necessary for SF)
requires specific physical conditions in addition to high gas densities, such as 
high gas pressure \citep{BlitzL_06a} and pre-enrichment.
Whether these conditions are met in low mass halos is an open question.
Furthermore, radiative feedback effects  from massive stars 
during intense SF episodes could play a crucial role in  disrupting the GMCs
 as proposed by \citet{MurrayN_10a}.}

It is possible that no single mechanism can produce such a mass floor.
For instance, since simple arguments on SN feedback give $f_\star \propto  \Ms/\Mv \propto \Mv^{\beta}$ with
$\beta=2/3$ \citep{DekelA_86a},  an additional physical process must be at play.
The combination of SN feedback and
 a (very high)  reionization threshold can produce a steeper $f_\star$ as shown in \citet{CattaneoA_10a}.
Similarly, \citet{vandenBoschF_03b} and \citet{MoH_05a} showed that artificially 
induced re-heating of the IGM to $T\sim$~$10^5$~K directly leads to a 
mass-dependent efficiency $f_{\star}$.
However,  this effects alone also cannot account for the steep mass dependence
of $f_\star$ \citep[e.g.][]{LuY_07a}.

Lastly, \citet{CantalupoS_10a} argued  
that the missing ingredient could be the effect of photoionization by local sources on
the  incoming cooling gas. He showed that soft X-ray generated by SF efficiently alters
the ionization state of atoms, such as O. This effectively removes the main coolants
and increases the cooling times by orders of magnitude, essentially stopping accretion, preferentially in low-mass halos.  
While this is an attractive mechanism, further work is needed in order to understand quantitatively
 whether photo-ionization from local
sources are able to regulate the cosmological gas accretion rates.

One way to gain insight into which of these possible mechanisms are at
play is to constrain the transition   in accretion efficiency
near $\Mmi$ ($\epsin \propto \Mv^{\eta}$), which is related to the slope $\beta$ of the relation
$f_\star \propto  \Ms/\Mv \propto \Mv^{\beta}$ under the quasi-steady state solution with $\eta\sim\beta$.
Our modelling indicates that $\eta$ should be steeper than unity.
Observationally, various groups have determined that  $\beta$ is $\geq2$
\citep{ShankarF_06a,BaldryI_08a, KravtsovA_09a,MosterB_10a,GuoQ_10a}.

\acknowledgments
We thank the anonymous referee for his or her constructive report.
We specially thank A. Renzini for  stimulating comments on a earlier draft
that have led to significant improvements. We also thank D. Schaerer and K. Finlator for useful discussions.
This research has been supported by the German-Israeli Foundation (GIF)
grant I-895-207.7/2005,
by the German Research Foundation (DFG) via German-Israeli 
Project Cooperation grant STE1869/1-1.GE625/15-1,
and by a grant from the Israel Science Foundation.
N.B. acknowledges the hospitality at the Racah Institute of Physics of the
Hebrew University of Jerusalem where much of this work has been performed.
N.M.F.S. acknowledges support from the Minerva Program of
 the Max-Planck-Gesellschaft.

\bibliography{references}

\begin{thebibliography}{155}
\expandafter\ifx\csname natexlab\endcsname\relax\def\natexlab#1{#1}\fi

\bibitem[{{Baldry} {et~al.}(2008){Baldry}, {Glazebrook}, \&
  {Driver}}]{BaldryI_08a}
{Baldry}, I.~K., {Glazebrook}, K., \& {Driver}, S.~P. 2008, \mnras, 388, 945

\bibitem[{{Barkana} \& {Loeb}(1999)}]{BarkanaR_99a}
{Barkana}, R., \& {Loeb}, A. 1999, \apj, 523, 54

\bibitem[{{Bauermeister} {et~al.}(2010){Bauermeister}, {Blitz}, \&
  {Ma}}]{BauermeisterA_10a}
{Bauermeister}, A., {Blitz}, L., \& {Ma}, C. 2010, \apj, 717, 323

\bibitem[{{Bell} {et~al.}(2005){Bell}, {Papovich}, {Wolf}, {Le Floc'h},
  {Caldwell}, {Barden}, {Egami}, {McIntosh}, {Meisenheimer},
  {P{\'e}rez-Gonz{\'a}lez}, {Rieke}, {Rieke}, {Rigby}, \& {Rix}}]{BellE_05a}
{Bell}, E.~F. {et~al.} 2005, \apj, 625, 23

\bibitem[{{Birnboim} \& {Dekel}(2003)}]{BirnboimY_03a}
{Birnboim}, Y., \& {Dekel}, A. 2003, \mnras, 345, 349

\bibitem[{{Birnboim} {et~al.}(2007){Birnboim}, {Dekel}, \&
  {Neistein}}]{BirnboimY_07a}
{Birnboim}, Y., {Dekel}, A., \& {Neistein}, E. 2007, \mnras, 380, 339

\bibitem[{{Blitz} \& {Rosolowsky}(2006)}]{BlitzL_06a}
{Blitz}, L., \& {Rosolowsky}, E. 2006, \apj, 650, 933

\bibitem[{{Blumenthal} {et~al.}(1986){Blumenthal}, {Faber}, {Flores}, \&
  {Primack}}]{BlumenthalG_86a}
{Blumenthal}, G.~R., {Faber}, S.~M., {Flores}, R., \& {Primack}, J.~R. 1986,
  \apj, 301, 27

\bibitem[{{Bouch{\'e}} {et~al.}(2007{\natexlab{a}}){Bouch{\'e}}, {Cresci},
  {Davies}, {Eisenhauer}, {F{\"o}rster Schreiber}, {Genzel}, {Gillessen},
  {Lehnert}, {Lutz}, {Nesvadba}, {Shapiro}, {Sternberg}, {Tacconi}, {Verma},
  {Cimatti}, {Daddi}, {Renzini}, {Erb}, {Shapley}, \& {Steidel}}]{BoucheN_07b}
{Bouch{\'e}}, N. {et~al.} 2007{\natexlab{a}}, \apj, 671, 303

\bibitem[{{Bouch{\'e}} {et~al.}(2007{\natexlab{b}}){Bouch{\'e}}, {Murphy},
  {P{\'e}roux}, {Davies}, {Eisenhauer}, {F{\"o}rster Schreiber}, \&
  {Tacconi}}]{BoucheN_07a}
{Bouch{\'e}}, N., {Murphy}, M.~T., {P{\'e}roux}, C., {Davies}, R.,
  {Eisenhauer}, F., {F{\"o}rster Schreiber}, N.~M., \& {Tacconi}, L.
  2007{\natexlab{b}}, \apjl, 669, L5

\bibitem[{{Bournaud} {et~al.}(2007){Bournaud}, {Elmegreen}, \&
  {Elmegreen}}]{BournaudF_07a}
{Bournaud}, F., {Elmegreen}, B.~G., \& {Elmegreen}, D.~M. 2007, \apj, 670, 237

\bibitem[{{Brinchmann} {et~al.}(2004){Brinchmann}, {Charlot}, {White},
  {Tremonti}, {Kauffmann}, {Heckman}, \& {Brinkmann}}]{BrinchmannJ_04a}
{Brinchmann}, J., {Charlot}, S., {White}, S.~D.~M., {Tremonti}, C.,
  {Kauffmann}, G., {Heckman}, T., \& {Brinkmann}, J. 2004, \mnras, 351, 1151

\bibitem[{{Bruzual} \& {Charlot}(2003)}]{BruzualG_03a}
{Bruzual}, G., \& {Charlot}, S. 2003, \mnras, 344, 1000

\bibitem[{{Burkert} {et~al.}(2010){Burkert}, {Genzel}, {Bouche}, {Cresci},
  {Khochfar}, {Sommer-Larsen}, {Sternberg}, {Naab}, {Foerster-Schreiber},
  {Tacconi}, {Shapiro}, {Hicks}, {Lutz}, {Davies}, {Buschkamp}, \&
  {Genel}}]{BurkertA_09a}
{Burkert}, A. {et~al.} 2010, \apj, submitted (arXiv:0907.4777)

\bibitem[{{Cantalupo}(2010)}]{CantalupoS_10a}
{Cantalupo}, S. 2010, \mnras, 403, L16

\bibitem[{{Catinella} {et~al.}(2010){Catinella}, {Schiminovich}, {Kauffmann},
  {Fabello}, {Wang}, {Hummels}, {Lemonias}, {Moran}, {Wu}, {Giovanelli},
  {Haynes}, {Heckman}, {Basu-Zych}, {Blanton}, {Brinchmann}, {Budav{\'a}ri},
  {Gon{\c c}alves}, {Johnson}, {Kennicutt}, {Madore}, {Martin}, {Rich},
  {Tacconi}, {Thilker}, {Wild}, \& {Wyder}}]{CatinellaB_10a}
{Catinella}, B. {et~al.} 2010, \mnras, 403, 683

\bibitem[{{Cattaneo} {et~al.}(2006){Cattaneo}, {Dekel}, {Devriendt},
  {Guiderdoni}, \& {Blaizot}}]{CattaneoA_06a}
{Cattaneo}, A., {Dekel}, A., {Devriendt}, J., {Guiderdoni}, B., \& {Blaizot},
  J. 2006, \mnras, 370, 1651

\bibitem[{{Cattaneo} {et~al.}(2008){Cattaneo}, {Dekel}, {Faber}, \&
  {Guiderdoni}}]{CattaneoA_08a}
{Cattaneo}, A., {Dekel}, A., {Faber}, S.~M., \& {Guiderdoni}, B. 2008, \mnras,
  389, 567

\bibitem[{{Cattaneo} {et~al.}(2010){Cattaneo}, {Mamon}, {Warnick}, \&
  {Knebe}}]{CattaneoA_10a}
{Cattaneo}, A., {Mamon}, G.~A., {Warnick}, K., \& {Knebe}, A. 2010, MNRAS,
  submitted (arXiv:1002.3257)

\bibitem[{{Chen} {et~al.}(2009){Chen}, {Wild}, {Kauffmann}, {Blaizot}, {Davis},
  {Noeske}, {Wang}, \& {Willmer}}]{ChenY_09a}
{Chen}, Y., {Wild}, V., {Kauffmann}, G., {Blaizot}, J., {Davis}, M., {Noeske},
  K., {Wang}, J., \& {Willmer}, C. 2009, \mnras, 393, 406

\bibitem[{{Choi} \& {Nagamine}(2009)}]{ChoiJ_09a}
{Choi}, J., \& {Nagamine}, K. 2009, MNRAS, submitted (arXiv:0909.5425)

\bibitem[{{Cimatti} {et~al.}(2006){Cimatti}, {Daddi}, \&
  {Renzini}}]{CimattiA_06a}
{Cimatti}, A., {Daddi}, E., \& {Renzini}, A. 2006, \aap, 453, L29

\bibitem[{{Cowie} {et~al.}(1995){Cowie}, {Songaila}, {Kim}, \&
  {Hu}}]{CowieL_95b}
{Cowie}, L.~L., {Songaila}, A., {Kim}, T., \& {Hu}, E.~M. 1995, \aj, 109, 1522

\bibitem[{{Cresci} {et~al.}(2009){Cresci}, {Hicks}, {Genzel}, {Schreiber},
  {Davies}, {Bouch{\'e}}, {Buschkamp}, {Genel}, {Shapiro}, {Tacconi},
  {Sommer-Larsen}, {Burkert}, {Eisenhauer}, {Gerhard}, {Lutz}, {Naab},
  {Sternberg}, {Cimatti}, {Daddi}, {Erb}, {Kurk}, {Lilly}, {Renzini},
  {Shapley}, {Steidel}, \& {Caputi}}]{CresciG_09a}
{Cresci}, G. {et~al.} 2009, \apj, 697, 115

\bibitem[{{Daddi} {et~al.}(2010{\natexlab{a}}){Daddi}, {Bournaud}, {Walter},
  {Dannerbauer}, {Carilli}, {Dickinson}, {Elbaz}, {Morrison}, {Riechers},
  {Onodera}, {Salmi}, {Krips}, \& {Stern}}]{DaddiE_10a}
{Daddi}, E. {et~al.} 2010{\natexlab{a}}, \apj, 713, 686

\bibitem[{{Daddi} {et~al.}(2008){Daddi}, {Dannerbauer}, {Elbaz}, {Dickinson},
  {Morrison}, {Stern}, \& {Ravindranath}}]{DaddiE_08a}
{Daddi}, E., {Dannerbauer}, H., {Elbaz}, D., {Dickinson}, M., {Morrison}, G.,
  {Stern}, D., \& {Ravindranath}, S. 2008, \apjl, 673, L21

\bibitem[{{Daddi} {et~al.}(2009){Daddi}, {Dannerbauer}, {Stern}, {Dickinson},
  {Morrison}, {Elbaz}, {Giavalisco}, {Mancini}, {Pope}, \&
  {Spinrad}}]{DaddiE_09a}
{Daddi}, E. {et~al.} 2009, \apj, 694, 1517

\bibitem[{{Daddi} {et~al.}(2007){Daddi}, {Dickinson}, {Morrison}, {Chary},
  {Cimatti}, {Elbaz}, {Frayer}, {Renzini}, {Pope}, {Alexander}, {Bauer},
  {Giavalisco}, {Huynh}, {Kurk}, \& {Mignoli}}]{DaddiE_07a}
---. 2007, \apj, 670, 156

\bibitem[{{Daddi} {et~al.}(2010{\natexlab{b}}){Daddi}, {Elbaz}, {Walter},
  {Bournaud}, {Salmi}, {Carilli}, {Dannerbauer}, {Dickinson}, {Monaco}, \&
  {Riechers}}]{DaddiE_10b}
---. 2010{\natexlab{b}}, \apjl, 714, L118

\bibitem[{{Dahlen} {et~al.}(2007){Dahlen}, {Mobasher}, {Dickinson}, {Ferguson},
  {Giavalisco}, {Kretchmer}, \& {Ravindranath}}]{DahlenT_07a}
{Dahlen}, T., {Mobasher}, B., {Dickinson}, M., {Ferguson}, H.~C., {Giavalisco},
  M., {Kretchmer}, C., \& {Ravindranath}, S. 2007, \apj, 654, 172

\bibitem[{{Damen} {et~al.}(2009{\natexlab{a}}){Damen}, {F{\"o}rster Schreiber},
  {Franx}, {Labb{\'e}}, {Toft}, {van Dokkum}, \& {Wuyts}}]{DamenM_09b}
{Damen}, M., {F{\"o}rster Schreiber}, N.~M., {Franx}, M., {Labb{\'e}}, I.,
  {Toft}, S., {van Dokkum}, P.~G., \& {Wuyts}, S. 2009{\natexlab{a}}, \apj,
  705, 617

\bibitem[{{Damen} {et~al.}(2009{\natexlab{b}}){Damen}, {Labb{\'e}}, {Franx},
  {van Dokkum}, {Taylor}, \& {Gawiser}}]{DamenM_09a}
{Damen}, M., {Labb{\'e}}, I., {Franx}, M., {van Dokkum}, P.~G., {Taylor},
  E.~N., \& {Gawiser}, E.~J. 2009{\natexlab{b}}, \apj, 690, 937

\bibitem[{{Dav{\'e}}(2008)}]{DaveR_08a}
{Dav{\'e}}, R. 2008, \mnras, 385, 147

\bibitem[{{Dekel} \& {Birnboim}(2006)}]{DekelA_06a}
{Dekel}, A., \& {Birnboim}, Y. 2006, \mnras, 368, 2

\bibitem[{{Dekel} {et~al.}(2009{\natexlab{a}}){Dekel}, {Birnboim}, {Engel},
  {Freundlich}, {Goerdt}, {Mumcuoglu}, {Neistein}, {Pichon}, {Teyssier}, \&
  {Zinger}}]{DekelA_09a}
{Dekel}, A. {et~al.} 2009{\natexlab{a}}, \nat, 457, 451

\bibitem[{{Dekel} {et~al.}(2009{\natexlab{b}}){Dekel}, {Sari}, \&
  {Ceverino}}]{DekelA_09b}
{Dekel}, A., {Sari}, R., \& {Ceverino}, D. 2009{\natexlab{b}}, \apj, 703, 785

\bibitem[{{Dekel} \& {Silk}(1986)}]{DekelA_86a}
{Dekel}, A., \& {Silk}, J. 1986, \apj, 303, 39

\bibitem[{{Dijkstra} {et~al.}(2004){Dijkstra}, {Haiman}, {Rees}, \&
  {Weinberg}}]{DijkstraM_04a}
{Dijkstra}, M., {Haiman}, Z., {Rees}, M.~J., \& {Weinberg}, D.~H. 2004, \apj,
  601, 666

\bibitem[{{Drory} \& {Alvarez}(2008)}]{DroryN_08a}
{Drory}, N., \& {Alvarez}, M. 2008, \apj, 680, 41

\bibitem[{{Dutton} \& {van den Bosch}(2009)}]{DuttonA_09a}
{Dutton}, A.~A., \& {van den Bosch}, F.~C. 2009, \mnras, 396, 141

\bibitem[{{Dutton} {et~al.}(2010){Dutton}, {van den Bosch}, \&
  {Dekel}}]{DuttonA_10a}
{Dutton}, A.~A., {van den Bosch}, F.~C., \& {Dekel}, A. 2010, \mnras, in press
  (arXiv/0912.2169)

\bibitem[{{Dutton} {et~al.}(2007){Dutton}, {van den Bosch}, {Dekel}, \&
  {Courteau}}]{DuttonA_07a}
{Dutton}, A.~A., {van den Bosch}, F.~C., {Dekel}, A., \& {Courteau}, S. 2007,
  \apj, 654, 27

\bibitem[{{Efstathiou} {et~al.}(1985){Efstathiou}, {Davis}, {White}, \&
  {Frenk}}]{EfstathiouG_85a}
{Efstathiou}, G., {Davis}, M., {White}, S.~D.~M., \& {Frenk}, C.~S. 1985,
  \apjs, 57, 241

\bibitem[{{Eke} {et~al.}(2005){Eke}, {Baugh}, {Cole}, {Frenk}, {King}, \&
  {Peacock}}]{EkeV_05a}
{Eke}, V.~R., {Baugh}, C.~M., {Cole}, S., {Frenk}, C.~S., {King}, H.~M., \&
  {Peacock}, J.~A. 2005, \mnras, 362, 1233

\bibitem[{{Elbaz} {et~al.}(2007){Elbaz}, {Daddi}, {Le Borgne}, {Dickinson},
  {Alexander}, {Chary}, {Starck}, {Brandt}, {Kitzbichler}, {MacDonald},
  {Nonino}, {Popesso}, {Stern}, \& {Vanzella}}]{ElbazD_07a}
{Elbaz}, D. {et~al.} 2007, \aap, 468, 33

\bibitem[{{Elmegreen} {et~al.}(2008){Elmegreen}, {Bournaud}, \&
  {Elmegreen}}]{ElmegreenB_08a}
{Elmegreen}, B.~G., {Bournaud}, F., \& {Elmegreen}, D.~M. 2008, \apj, 688, 67

\bibitem[{{Elmegreen} {et~al.}(2007){Elmegreen}, {Elmegreen}, {Ravindranath},
  \& {Coe}}]{ElmegreenD_07a}
{Elmegreen}, D.~M., {Elmegreen}, B.~G., {Ravindranath}, S., \& {Coe}, D.~A.
  2007, \apj, 658, 763

\bibitem[{{Epinat} {et~al.}(2009){Epinat}, {Contini}, {Le Fevre}, {Vergani},
  {Garilli}, {Amram}, {Queyrel}, {Tasca}, \& {Tresse}}]{EpinatB_09a}
{Epinat}, B. {et~al.} 2009, \aap, 504, 789

\bibitem[{{Erb}(2008)}]{ErbD_08a}
{Erb}, D.~K. 2008, \apj, 674, 151

\bibitem[{{Erb} {et~al.}(2006{\natexlab{a}}){Erb}, {Steidel}, {Shapley},
  {Pettini}, {Reddy}, \& {Adelberger}}]{ErbD_06c}
{Erb}, D.~K., {Steidel}, C.~C., {Shapley}, A.~E., {Pettini}, M., {Reddy},
  N.~A., \& {Adelberger}, K.~L. 2006{\natexlab{a}}, \apj, 647, 128

\bibitem[{{Erb} {et~al.}(2006{\natexlab{b}}){Erb}, {Steidel}, {Shapley},
  {Pettini}, {Reddy}, \& {Adelberger}}]{ErbD_06b}
---. 2006{\natexlab{b}}, \apj, 646, 107

\bibitem[{{F{\"o}rster Schreiber} {et~al.}(2009){F{\"o}rster Schreiber},
  {Genzel}, {Bouche}, {Cresci}, {Davies}, {Buschkamp}, {Shapiro}, {Tacconi},
  {Hicks}, {Genel}, {Shapley}, {Erb}, {Steidel}, {Lutz}, {Eisenhauer},
  {Gillessen}, {Sternberg}, {Renzini}, {Cimatti}, {Daddi}, {Kurk}, {Lilly},
  {Kong}, {Lehnert}, {Nesvadba}, {Verma}, {McCracken}, {Arimoto}, {Mignoli}, \&
  {Onodera}}]{ForsterSchreiberN_09a}
{F{\"o}rster Schreiber}, N.~M. {et~al.} 2009, \apj, 706, 1364

\bibitem[{{F{\"o}rster Schreiber} {et~al.}(2006){F{\"o}rster Schreiber},
  {Genzel}, {Lehnert}, {Bouch{\'e}}, {Verma}, {Erb}, {Shapley}, {Steidel},
  {Davies}, {Lutz}, {Nesvadba}, {Tacconi}, {Eisenhauer}, {Abuter}, {Gilbert},
  {Gillessen}, \& {Sternberg}}]{ForsterSchreiberN_06a}
---. 2006, \apj, 645, 1062

\bibitem[{{Gallazzi} {et~al.}(2005){Gallazzi}, {Charlot}, {Brinchmann},
  {White}, \& {Tremonti}}]{GallazziA_05a}
{Gallazzi}, A., {Charlot}, S., {Brinchmann}, J., {White}, S.~D.~M., \&
  {Tremonti}, C.~A. 2005, \mnras, 362, 41

\bibitem[{{Genel} {et~al.}(2010){Genel}, {Bouch{\'e}}, {Naab}, {Sternberg}, \&
  {Genzel}}]{GenelS_10a}
{Genel}, S., {Bouch{\'e}}, N., {Naab}, T., {Sternberg}, A., \& {Genzel}, R.~.
  2010, \apj, in press (arXiv/1005.4058)

\bibitem[{{Genel} {et~al.}(2008){Genel}, {Genzel}, {Bouch{\'e}}, {Sternberg},
  {Naab}, {Schreiber}, {Shapiro}, {Tacconi}, {Lutz}, {Cresci}, {Buschkamp},
  {Davies}, \& {Hicks}}]{GenelS_08a}
{Genel}, S. {et~al.} 2008, \apj, 688, 789

\bibitem[{{Genzel} {et~al.}(2008){Genzel}, {Burkert}, {Bouch{\'e}}, {Cresci},
  {F{\"o}rster Schreiber}, {Shapley}, {Shapiro}, {Tacconi}, {Buschkamp},
  {Cimatti}, {Daddi}, {Davies}, {Eisenhauer}, {Erb}, {Genel}, {Gerhard},
  {Hicks}, {Lutz}, {Naab}, {Ott}, {Rabien}, {Renzini}, {Steidel}, {Sternberg},
  \& {Lilly}}]{GenzelR_08a}
{Genzel}, R. {et~al.} 2008, \apj, 687, 59

\bibitem[{{Genzel} {et~al.}(2006){Genzel}, {Tacconi}, {Eisenhauer},
  {F{\"o}rster Schreiber}, {Cimatti}, {Daddi}, {Bouch{\'e}}, {Davies},
  {Lehnert}, {Lutz}, {Nesvadba}, {Verma}, {Abuter}, {Shapiro}, {Sternberg},
  {Renzini}, {Kong}, {Arimoto}, \& {Mignoli}}]{GenzelR_06a}
---. 2006, \nat, 442, 786

\bibitem[{{Genzel} {et~al.}(2010){Genzel}, {Tacconi}, {Gracia-Carpio},
  {Sternberg}, {Cooper}, {Shapiro}, {Bolatto}, {Bouche}, {Bournaud}, {Burkert},
  {Combes}, {Comerford}, {Cox}, {Davis}, {Foerster Schreiber},
  {Garcia-Burillo}, {Lutz}, {Naab}, {Neri}, {Omont}, {Shapley}, \&
  {Weiner}}]{GenzelR_10a}
---. 2010, \mnras, in press (arXiv:1003.5180)

\bibitem[{{Giovanelli} {et~al.}(2005){Giovanelli}, {Haynes}, {Kent},
  {Perillat}, {Catinella}, {Hoffman}, {Momjian}, {Rosenberg}, {Saintonge},
  {Spekkens}, {Stierwalt}, {Brosch}, {Masters}, {Springob}, {Karachentsev},
  {Karachentseva}, {Koopmann}, {Muller}, {van Driel}, \& {van
  Zee}}]{GiovanelliR_05a}
{Giovanelli}, R. {et~al.} 2005, \aj, 130, 2613

\bibitem[{{Gnedin}(2000)}]{GnedinN_00a}
{Gnedin}, N.~Y. 2000, \apj, 542, 535

\bibitem[{{Gonz{\'a}lez} {et~al.}(2010){Gonz{\'a}lez}, {Labb{\'e}}, {Bouwens},
  {Illingworth}, {Franx}, {Kriek}, \& {Brammer}}]{GonzalezV_10a}
{Gonz{\'a}lez}, V., {Labb{\'e}}, I., {Bouwens}, R.~J., {Illingworth}, G.,
  {Franx}, M., {Kriek}, M., \& {Brammer}, G.~B. 2010, \apj, 713, 115

\bibitem[{{Grazian} {et~al.}(2007){Grazian}, {Salimbeni}, {Pentericci},
  {Fontana}, {Nonino}, {Vanzella}, {Cristiani}, {de Santis}, {Gallozzi},
  {Giallongo}, \& {Santini}}]{GrazianA_07a}
{Grazian}, A. {et~al.} 2007, \aap, 465, 393

\bibitem[{{Guo} {et~al.}(2010){Guo}, {White}, {Li}, \&
  {Boylan-Kolchin}}]{GuoQ_10a}
{Guo}, Q., {White}, S., {Li}, C., \& {Boylan-Kolchin}, M. 2010, \mnras, 404,
  1111

\bibitem[{{Heavens} {et~al.}(2004){Heavens}, {Panter}, {Jimenez}, \&
  {Dunlop}}]{HeavensA_04a}
{Heavens}, A., {Panter}, B., {Jimenez}, R., \& {Dunlop}, J. 2004, \nat, 428,
  625

\bibitem[{{Heckman} {et~al.}(2000){Heckman}, {Lehnert}, {Strickland}, \&
  {Armus}}]{HeckmanT_00a}
{Heckman}, T.~M., {Lehnert}, M.~D., {Strickland}, D.~K., \& {Armus}, L. 2000,
  \apjs, 129, 493

\bibitem[{{Hernquist} \& {Springel}(2003)}]{HernquistL_03a}
{Hernquist}, L., \& {Springel}, V. 2003, \mnras, 341, 1253

\bibitem[{{Hopkins} \& {Beacom}(2006)}]{HopkinsA_06a}
{Hopkins}, A.~M., \& {Beacom}, J.~F. 2006, \apj, 651, 142

\bibitem[{{Hopkins} \& {Beacom}(2008)}]{HopkinsA_08a}
---. 2008, \apj, 682, 1486

\bibitem[{{Iliev} {et~al.}(2005){Iliev}, {Shapiro}, \& {Raga}}]{IlievI_05a}
{Iliev}, I.~T., {Shapiro}, P.~R., \& {Raga}, A.~C. 2005, \mnras, 361, 405

\bibitem[{{Immeli} {et~al.}(2004){Immeli}, {Samland}, {Gerhard}, \&
  {Westera}}]{ImmeliA_04a}
{Immeli}, A., {Samland}, M., {Gerhard}, O., \& {Westera}, P. 2004, \aap, 413,
  547

\bibitem[{{James} {et~al.}(2008){James}, {Prescott}, \& {Baldry}}]{JamesP_08a}
{James}, P.~A., {Prescott}, M., \& {Baldry}, I.~K. 2008, \aap, 484, 703

\bibitem[{{Jimenez} {et~al.}(2007){Jimenez}, {Bernardi}, {Haiman}, {Panter}, \&
  {Heavens}}]{JimenezR_07a}
{Jimenez}, R., {Bernardi}, M., {Haiman}, Z., {Panter}, B., \& {Heavens}, A.~F.
  2007, \apj, 669, 947

\bibitem[{{Kassin} {et~al.}(2007){Kassin}, {Weiner}, {Faber}, {Koo}, {Lotz},
  {Diemand}, {Harker}, {Bundy}, {Metevier}, {Phillips}, {Cooper}, {Croton},
  {Konidaris}, {Noeske}, \& {Willmer}}]{KassinS_07a}
{Kassin}, S.~A. {et~al.} 2007, \apjl, 660, L35

\bibitem[{{Kennicutt}(1998)}]{KennicuttR_98a}
{Kennicutt}, R.~C. 1998, \araa, 36, 189

\bibitem[{{Kere\v{s}} {et~al.}(2009){Kere\v{s}}, {Katz}, {Fardal}, {Dave}, \&
  {Weinberg}}]{KeresD_09a}
{Kere\v{s}}, D., {Katz}, N., {Fardal}, M., {Dave}, R., \& {Weinberg}, D.~H.
  2009, \mnras, 395, 160

\bibitem[{{Kere\v{s}} {et~al.}(2005){Kere\v{s}}, {Katz}, {Weinberg}, \&
  {Dav{\'e}}}]{KeresD_05a}
{Kere\v{s}}, D., {Katz}, N., {Weinberg}, D.~H., \& {Dav{\'e}}, R. 2005, \mnras,
  363, 2

\bibitem[{{Kravtsov}(2010)}]{KravtsovA_09a}
{Kravtsov}, A.~V. 2010, Advances in Astronomy, 2010

\bibitem[{{Krumholz} \& {Thompson}(2007)}]{KrumholzM_07a}
{Krumholz}, M.~R., \& {Thompson}, T.~A. 2007, \apj, 669, 289

\bibitem[{{Larson}(1974)}]{LarsonR_74a}
{Larson}, R.~B. 1974, \mnras, 166, 585

\bibitem[{{Law} {et~al.}(2007){Law}, {Steidel}, {Erb}, {Larkin}, {Pettini},
  {Shapley}, \& {Wright}}]{LawD_07a}
{Law}, D.~R., {Steidel}, C.~C., {Erb}, D.~K., {Larkin}, J.~E., {Pettini}, M.,
  {Shapley}, A.~E., \& {Wright}, S.~A. 2007, \apj, 669, 929

\bibitem[{{Law} {et~al.}(2009){Law}, {Steidel}, {Erb}, {Larkin}, {Pettini},
  {Shapley}, \& {Wright}}]{LawD_09a}
---. 2009, \apj, 697, 2057

\bibitem[{{Lilly} {et~al.}(1996){Lilly}, {Le Fevre}, {Hammer}, \&
  {Crampton}}]{LillyS_96a}
{Lilly}, S.~J., {Le Fevre}, O., {Hammer}, F., \& {Crampton}, D. 1996, \apjl,
  460, L1

\bibitem[{{Lu} \& {Mo}(2007)}]{LuY_07a}
{Lu}, Y., \& {Mo}, H.~J. 2007, \mnras, 377, 617

\bibitem[{{Madau} {et~al.}(1996){Madau}, {Ferguson}, {Dickinson}, {Giavalisco},
  {Steidel}, \& {Fruchter}}]{MadauP_96a}
{Madau}, P., {Ferguson}, H.~C., {Dickinson}, M.~E., {Giavalisco}, M.,
  {Steidel}, C.~C., \& {Fruchter}, A. 1996, \mnras, 283, 1388

\bibitem[{{Martin}(2005)}]{MartinC_05a}
{Martin}, C.~L. 2005, \apj, 621, 227

\bibitem[{{Martin} \& {Kennicutt}(2001)}]{MartinC_01a}
{Martin}, C.~L., \& {Kennicutt}, Jr., R.~C. 2001, \apj, 555, 301

\bibitem[{{McBride} {et~al.}(2009){McBride}, {Fakhouri}, \&
  {Ma}}]{McBrideM_09a}
{McBride}, J., {Fakhouri}, O., \& {Ma}, C. 2009, \mnras, 398, 1858

\bibitem[{{McGaugh}(2005)}]{McGaughS_05a}
{McGaugh}, S.~S. 2005, \apj, 632, 859

\bibitem[{{Meyer} {et~al.}(2008){Meyer}, {Zwaan}, {Webster}, {Schneider}, \&
  {Staveley-Smith}}]{MeyerM_08a}
{Meyer}, M.~J., {Zwaan}, M.~A., {Webster}, R.~L., {Schneider}, S., \&
  {Staveley-Smith}, L. 2008, \mnras, 391, 1712

\bibitem[{{Meyer} {et~al.}(2004){Meyer}, {Zwaan}, {Webster}, {Staveley-Smith},
  {Ryan-Weber}, {Drinkwater}, {Barnes}, {Howlett}, {Kilborn}, {Stevens},
  {Waugh}, {Pierce}, {Bhathal}, {de Blok}, {Disney}, {Ekers}, {Freeman},
  {Garcia}, {Gibson}, {Harnett}, {Henning}, {Jerjen}, {Kesteven}, {Knezek},
  {Koribalski}, {Mader}, {Marquarding}, {Minchin}, {O'Brien}, {Oosterloo},
  {Price}, {Putman}, {Ryder}, {Sadler}, {Stewart}, {Stootman}, \&
  {Wright}}]{MeyerM_04a}
{Meyer}, M.~J. {et~al.} 2004, \mnras, 350, 1195

\bibitem[{{Mo} {et~al.}(1998){Mo}, {Mao}, \& {White}}]{MoH_98a}
{Mo}, H.~J., {Mao}, S., \& {White}, S.~D.~M. 1998, \mnras, 295, 319

\bibitem[{{Mo} {et~al.}(2005){Mo}, {Yang}, {van den Bosch}, \&
  {Katz}}]{MoH_05a}
{Mo}, H.~J., {Yang}, X., {van den Bosch}, F.~C., \& {Katz}, N. 2005, \mnras,
  363, 1155

\bibitem[{{Mobasher} {et~al.}(2009){Mobasher}, {Dahlen}, {Hopkins}, {Scoville},
  {Capak}, {Rich}, {Sanders}, {Schinnerer}, {Ilbert}, {Salvato}, \&
  {Sheth}}]{MobasherB_09a}
{Mobasher}, B. {et~al.} 2009, \apj, 690, 1074

\bibitem[{{Moster} {et~al.}(2010){Moster}, {Somerville}, {Maulbetsch}, {van den
  Bosch}, {Macci{\`o}}, {Naab}, \& {Oser}}]{MosterB_10a}
{Moster}, B.~P., {Somerville}, R.~S., {Maulbetsch}, C., {van den Bosch}, F.~C.,
  {Macci{\`o}}, A.~V., {Naab}, T., \& {Oser}, L. 2010, \apj, 710, 903

\bibitem[{{Murray} {et~al.}(2010){Murray}, {Quataert}, \&
  {Thompson}}]{MurrayN_10a}
{Murray}, N., {Quataert}, E., \& {Thompson}, T.~A. 2010, \apj, 709, 191

\bibitem[{{Nagamine} {et~al.}(2006){Nagamine}, {Ostriker}, {Fukugita}, \&
  {Cen}}]{NagamineK_06a}
{Nagamine}, K., {Ostriker}, J.~P., {Fukugita}, M., \& {Cen}, R. 2006, \apj,
  653, 881

\bibitem[{{Navarro} \& {Steinmetz}(2000)}]{NavarroJ_00a}
{Navarro}, J.~F., \& {Steinmetz}, M. 2000, \apj, 538, 477

\bibitem[{{Neistein} \& {Dekel}(2008)}]{NeisteinE_08a}
{Neistein}, E., \& {Dekel}, A. 2008, \mnras, 388, 1792

\bibitem[{{Neistein} {et~al.}(2006){Neistein}, {van den Bosch}, \&
  {Dekel}}]{NeisteinE_06a}
{Neistein}, E., {van den Bosch}, F.~C., \& {Dekel}, A. 2006, \mnras, 372, 933

\bibitem[{{Neistein} \& {Weinmann}(2009)}]{NeisteinE_10a}
{Neistein}, E., \& {Weinmann}, S.~M. 2009, \mnras, in press (arXiv:0911.3147)

\bibitem[{{Noeske} {et~al.}(2007{\natexlab{a}}){Noeske}, {Faber}, {Weiner},
  {Koo}, {Primack}, {Dekel}, {Papovich}, {Conselice}, {Le Floc'h}, {Rieke},
  {Coil}, {Lotz}, {Somerville}, \& {Bundy}}]{NoeskeK_07a}
{Noeske}, K.~G. {et~al.} 2007{\natexlab{a}}, \apjl, 660, L47

\bibitem[{{Noeske} {et~al.}(2007{\natexlab{b}}){Noeske}, {Weiner}, {Faber},
  {Papovich}, {Koo}, {Somerville}, {Bundy}, {Conselice}, {Newman},
  {Schiminovich}, {Le Floc'h}, {Coil}, {Rieke}, {Lotz}, {Primack}, {Barmby},
  {Cooper}, {Davis}, {Ellis}, {Fazio}, {Guhathakurta}, {Huang}, {Kassin},
  {Martin}, {Phillips}, {Rich}, {Small}, {Willmer}, \& {Wilson}}]{NoeskeK_07b}
---. 2007{\natexlab{b}}, \apjl, 660, L43

\bibitem[{{Noguchi}(1998)}]{NoguchiM_98a}
{Noguchi}, M. 1998, \nat, 392, 253

\bibitem[{{Ocvirk} {et~al.}(2008){Ocvirk}, {Pichon}, \&
  {Teyssier}}]{OcvrikP_08a}
{Ocvirk}, P., {Pichon}, C., \& {Teyssier}, R. 2008, \mnras, 390, 1326

\bibitem[{{Oesch} {et~al.}(2010){Oesch}, {Bouwens}, {Illingworth}, {Carollo},
  {Franx}, {Labb{\'e}}, {Magee}, {Stiavelli}, {Trenti}, \& {van
  Dokkum}}]{OeschP_10a}
{Oesch}, P.~A. {et~al.} 2010, \apjl, 709, L16

\bibitem[{{Oliver} {et~al.}(2010){Oliver}, {Frost}, {Farrah},
  {Gonzalez-Solares}, {Shupe}, {Henriques}, {Roseboom}, {Afonso Luis},
  {Babbedge}, {Frayer}, {Lencz}, {Lonsdale}, {Masci}, {Padgett}, {Polletta},
  {Rowan-Robinson}, {Siana}, {Smith}, {Surace}, \& {Vaccari}}]{OliverS_10a}
{Oliver}, S. {et~al.} 2010, \mnras, in press (arXiv/1003.24460)

\bibitem[{{Oppenheimer} \& {Dav{\'e}}(2008)}]{OppenheimerB_08a}
{Oppenheimer}, B.~D., \& {Dav{\'e}}, R. 2008, \mnras, 387, 577

\bibitem[{{Pannella} {et~al.}(2009){Pannella}, {Carilli}, {Daddi}, {Mc
  Cracken}, {Owen}, {Renzini}, {Strazzullo}, {Civano}, {Koekemoer},
  {Schinnerer}, {Scoville}, {Smolcic}, {Taniguchi}, {Aussel}, {Kneib},
  {Ilbert}, {Mellier}, {Salvato}, {Thompson}, \& {Willott}}]{PannellaM_09a}
{Pannella}, M. {et~al.} 2009, \apjl, 698, L116

\bibitem[{{Pawlik} \& {Schaye}(2009)}]{PawlikA_09a}
{Pawlik}, A.~H., \& {Schaye}, J. 2009, \mnras, 396, L46

\bibitem[{{Pieri} \& {Martel}(2007)}]{PieriM_07a}
{Pieri}, M.~M., \& {Martel}, H. 2007, \apjl, 662, L7

\bibitem[{{Puech} {et~al.}(2008){Puech}, {Flores}, {Hammer}, {Yang}, {Neichel},
  {Lehnert}, {Chemin}, {Nesvadba}, {Epinat}, {Amram}, {Balkowski}, {Cesarsky},
  {Dannerbauer}, {di Serego Alighieri}, {Fuentes-Carrera}, {Guiderdoni},
  {Kembhavi}, {Liang}, {{\"O}stlin}, {Pozzetti}, {Ravikumar}, {Rawat},
  {Vergani}, {Vernet}, \& {Wozniak}}]{PuechM_08a}
{Puech}, M. {et~al.} 2008, \aap, 484, 173

\bibitem[{{Quinn} {et~al.}(1996){Quinn}, {Katz}, \& {Efstathiou}}]{QuinnT_96a}
{Quinn}, T., {Katz}, N., \& {Efstathiou}, G. 1996, \mnras, 278, L49

\bibitem[{{Rosenberg} \& {Schneider}(2003)}]{RosenbergJ_03a}
{Rosenberg}, J.~L., \& {Schneider}, S.~E. 2003, \apj, 585, 256

\bibitem[{{Rupke} {et~al.}(2005){Rupke}, {Veilleux}, \& {Sanders}}]{RupkeD_05a}
{Rupke}, D.~S., {Veilleux}, S., \& {Sanders}, D.~B. 2005, \apjs, 160, 115

\bibitem[{{Santini} {et~al.}(2009){Santini}, {Fontana}, {Grazian}, {Salimbeni},
  {Fiore}, {Fontanot}, {Boutsia}, {Castellano}, {Cristiani}, {de Santis},
  {Gallozzi}, {Giallongo}, {Menci}, {Nonino}, {Paris}, {Pentericci}, \&
  {Vanzella}}]{SantiniP_09a}
{Santini}, P. {et~al.} 2009, \aap, 504, 751

\bibitem[{{Sawicki} {et~al.}(2007){Sawicki}, {Iwata}, {Ohta}, {Thompson},
  {Tamura}, {Akiyama}, {Aoki}, {Ando}, \& {Kiuchi}}]{SawickiM_07a}
{Sawicki}, M. {et~al.} 2007, in Astronomical Society of the Pacific Conference
  Series, Vol. 380, Deepest Astronomical Surveys, ed. {J.~Afonso,
  H.~C.~Ferguson, B.~Mobasher, \& R.~Norris}, 433--+

\bibitem[{{Schaerer} \& {de Barros}(2009)}]{SchaererD_09a}
{Schaerer}, D., \& {de Barros}, S. 2009, \aap, 502, 423

\bibitem[{{Schaerer} \& {de Barros}(2010)}]{SchaererD_10a}
---. 2010, \aap, 515, A73+

\bibitem[{{Schaye} {et~al.}(2010){Schaye}, {Dalla Vecchia}, {Booth}, {Wiersma},
  {Theuns}, {Haas}, {Bertone}, {Duffy}, {McCarthy}, \& {van de
  Voort}}]{SchayeJ_10a}
{Schaye}, J. {et~al.} 2010, \mnras, 402, 1536

\bibitem[{{Schiminovich} {et~al.}(2005){Schiminovich}, {Ilbert}, {Arnouts},
  {Milliard}, {Tresse}, {Le F{\`e}vre}, {Treyer}, {Wyder}, {Budav{\'a}ri},
  {Zucca}, {Zamorani}, {Martin}, {Adami}, {Arnaboldi}, {Bardelli}, {Barlow},
  {Bianchi}, {Bolzonella}, {Bottini}, {Byun}, {Cappi}, {Contini}, {Charlot},
  {Donas}, {Forster}, {Foucaud}, {Franzetti}, {Friedman}, {Garilli},
  {Gavignaud}, {Guzzo}, {Heckman}, {Hoopes}, {Iovino}, {Jelinsky}, {Le Brun},
  {Lee}, {Maccagni}, {Madore}, {Malina}, {Marano}, {Marinoni}, {McCracken},
  {Mazure}, {Meneux}, {Morrissey}, {Neff}, {Paltani}, {Pell{\`o}}, {Picat},
  {Pollo}, {Pozzetti}, {Radovich}, {Rich}, {Scaramella}, {Scodeggio},
  {Seibert}, {Siegmund}, {Small}, {Szalay}, {Vettolani}, {Welsh}, {Xu}, \&
  {Zanichelli}}]{SchiminovichD_05a}
{Schiminovich}, D. {et~al.} 2005, \apjl, 619, L47

\bibitem[{{Schmidt}(1959)}]{SchmidtM_59a}
{Schmidt}, M. 1959, \apj, 129, 243

\bibitem[{{Schmidt}(1963)}]{SchmidtM_63a}
---. 1963, \apj, 137, 758

\bibitem[{{Shankar} {et~al.}(2006){Shankar}, {Lapi}, {Salucci}, {De Zotti}, \&
  {Danese}}]{ShankarF_06a}
{Shankar}, F., {Lapi}, A., {Salucci}, P., {De Zotti}, G., \& {Danese}, L. 2006,
  \apj, 643, 14

\bibitem[{{Shapiro} {et~al.}(2008){Shapiro}, {Genzel}, {F{\"o}rster Schreiber},
  {Tacconi}, {Bouch{\'e}}, {Cresci}, {Davies}, {Eisenhauer}, {Johansson},
  {Krajnovi{\'c}}, {Lutz}, {Naab}, {Arimoto}, {Arribas}, {Cimatti}, {Colina},
  {Daddi}, {Daigle}, {Erb}, {Hernandez}, {Kong}, {Mignoli}, {Onodera},
  {Renzini}, {Shapley}, \& {Steidel}}]{ShapiroK_08a}
{Shapiro}, K.~L. {et~al.} 2008, \apj, 682, 231

\bibitem[{{Shapiro} {et~al.}(2009){Shapiro}, {Genzel}, {Quataert}, {F{\"o}rster
  Schreiber}, {Davies}, {Tacconi}, {Armus}, {Bouch{\'e}}, {Buschkamp},
  {Cimatti}, {Cresci}, {Daddi}, {Eisenhauer}, {Erb}, {Genel}, {Hicks}, {Lilly},
  {Lutz}, {Renzini}, {Shapley}, {Steidel}, \& {Sternberg}}]{ShapiroK_09a}
---. 2009, \apj, 701, 955

\bibitem[{{Shapley} {et~al.}(2003){Shapley}, {Steidel}, {Pettini}, \&
  {Adelberger}}]{ShapleyA_03a}
{Shapley}, A.~E., {Steidel}, C.~C., {Pettini}, M., \& {Adelberger}, K.~L. 2003,
  \apj, 588, 65

\bibitem[{{Sheth} {et~al.}(2001){Sheth}, {Mo}, \& {Tormen}}]{ShethR_01a}
{Sheth}, R.~K., {Mo}, H.~J., \& {Tormen}, G. 2001, \mnras, 323, 1

\bibitem[{{Somerville} {et~al.}(2008){Somerville}, {Hopkins}, {Cox},
  {Robertson}, \& {Hernquist}}]{SomervilleR_08b}
{Somerville}, R.~S., {Hopkins}, P.~F., {Cox}, T.~J., {Robertson}, B.~E., \&
  {Hernquist}, L. 2008, \mnras, 391, 481

\bibitem[{{Springel} {et~al.}(2006){Springel}, {Frenk}, \&
  {White}}]{SpringelV_06a}
{Springel}, V., {Frenk}, C.~S., \& {White}, S.~D.~M. 2006, \nat, 440, 1137

\bibitem[{{Springel} \& {Hernquist}(2003)}]{SpringelV_03a}
{Springel}, V., \& {Hernquist}, L. 2003, \mnras, 339, 312

\bibitem[{{Stark} {et~al.}(2009){Stark}, {Ellis}, {Bunker}, {Bundy}, {Targett},
  {Benson}, \& {Lacy}}]{StarkD_09a}
{Stark}, D.~P., {Ellis}, R.~S., {Bunker}, A., {Bundy}, K., {Targett}, T.,
  {Benson}, A., \& {Lacy}, M. 2009, \apj, 697, 1493

\bibitem[{{Stewart} {et~al.}(2008){Stewart}, {Bullock}, {Wechsler}, {Maller},
  \& {Zentner}}]{StewartK_08a}
{Stewart}, K.~R., {Bullock}, J.~S., {Wechsler}, R.~H., {Maller}, A.~H., \&
  {Zentner}, A.~R. 2008, \apj, 683, 597

\bibitem[{{Tacconi} {et~al.}(2010){Tacconi}, {Genzel}, {Neri}, {Cox}, {Cooper},
  {Shapiro}, {Bolatto}, {Bouch{\'e}}, {Bournaud}, {Burkert}, {Combes},
  {Comerford}, {Davis}, {Schreiber}, {Garcia-Burillo}, {Gracia-Carpio}, {Lutz},
  {Naab}, {Omont}, {Shapley}, {Sternberg}, \& {Weiner}}]{TacconiL_10a}
{Tacconi}, L.~J. {et~al.} 2010, \nat, 463, 781

\bibitem[{{Thomas} {et~al.}(2005){Thomas}, {Maraston}, {Bender}, \& {Mendes de
  Oliveira}}]{ThomasD_05a}
{Thomas}, D., {Maraston}, C., {Bender}, R., \& {Mendes de Oliveira}, C. 2005,
  \apj, 621, 673

\bibitem[{{Thomas} {et~al.}(2010){Thomas}, {Maraston}, {Schawinski}, {Sarzi},
  \& {Silk}}]{ThomasD_10a}
{Thomas}, D., {Maraston}, C., {Schawinski}, K., {Sarzi}, M., \& {Silk}, J.
  2010, \mnras, 404, 1775

\bibitem[{{Thoul} \& {Weinberg}(1996)}]{ThoulA_96a}
{Thoul}, A.~A., \& {Weinberg}, D.~H. 1996, \apj, 465, 608

\bibitem[{{Tully} \& {Fisher}(1977)}]{TullyB_77a}
{Tully}, R.~B., \& {Fisher}, J.~R. 1977, \aap, 54, 661

\bibitem[{{Tully} \& {Pierce}(2000)}]{TullyB_00a}
{Tully}, R.~B., \& {Pierce}, M.~J. 2000, \apj, 533, 744

\bibitem[{{van den Bergh}(1962)}]{VandenberghS_62a}
{van den Bergh}, S. 1962, \aj, 67, 486

\bibitem[{{van den Bosch}(2002)}]{vandenBoschF_02a}
{van den Bosch}, F.~C. 2002, \mnras, 331, 98

\bibitem[{{van den Bosch} {et~al.}(2003{\natexlab{a}}){van den Bosch}, {Abel},
  \& {Hernquist}}]{vandenBoschF_03b}
{van den Bosch}, F.~C., {Abel}, T., \& {Hernquist}, L. 2003{\natexlab{a}},
  \mnras, 346, 177

\bibitem[{{van den Bosch} {et~al.}(2003{\natexlab{b}}){van den Bosch}, {Mo}, \&
  {Yang}}]{vandenBoschF_03a}
{van den Bosch}, F.~C., {Mo}, H.~J., \& {Yang}, X. 2003{\natexlab{b}}, \mnras,
  345, 923

\bibitem[{{van den Bosch} {et~al.}(2007){van den Bosch}, {Yang}, {Mo},
  {Weinmann}, {Macci{\`o}}, {More}, {Cacciato}, {Skibba}, \&
  {Kang}}]{vandenBoschF_07a}
{van den Bosch}, F.~C. {et~al.} 2007, \mnras, 376, 841

\bibitem[{{van Starkenburg} {et~al.}(2008){van Starkenburg}, {van der Werf},
  {Franx}, {Labb{\'e}}, {Rudnick}, \& {Wuyts}}]{VanstarkenburgL_08a}
{van Starkenburg}, L., {van der Werf}, P.~P., {Franx}, M., {Labb{\'e}}, I.,
  {Rudnick}, G., \& {Wuyts}, S. 2008, \aap, 488, 99

\bibitem[{{Verma} {et~al.}(2007){Verma}, {Lehnert}, {F{\"o}rster Schreiber},
  {Bremer}, \& {Douglas}}]{VermaA_07a}
{Verma}, A., {Lehnert}, M.~D., {F{\"o}rster Schreiber}, N.~M., {Bremer}, M.~N.,
  \& {Douglas}, L. 2007, \mnras, 377, 1024

\bibitem[{{Wechsler} {et~al.}(2002){Wechsler}, {Bullock}, {Primack},
  {Kravtsov}, \& {Dekel}}]{WechslerR_02a}
{Wechsler}, R.~H., {Bullock}, J.~S., {Primack}, J.~R., {Kravtsov}, A.~V., \&
  {Dekel}, A. 2002, \apj, 568, 52

\bibitem[{{White} \& {Frenk}(1991)}]{WhiteS_91a}
{White}, S.~D.~M., \& {Frenk}, C.~S. 1991, \apj, 379, 52

\bibitem[{{Wilkins} {et~al.}(2008){Wilkins}, {Trentham}, \&
  {Hopkins}}]{WilkinsS_08a}
{Wilkins}, S.~M., {Trentham}, N., \& {Hopkins}, A.~M. 2008, \mnras, 385, 687

\bibitem[{{Wong} {et~al.}(2006){Wong}, {Ryan-Weber}, {Garcia-Appadoo},
  {Webster}, {Staveley-Smith}, {Zwaan}, {Meyer}, {Barnes}, {Kilborn},
  {Bhathal}, {de Blok}, {Disney}, {Doyle}, {Drinkwater}, {Ekers}, {Freeman},
  {Gibson}, {Gurovich}, {Harnett}, {Henning}, {Jerjen}, {Kesteven}, {Knezek},
  {Koribalski}, {Mader}, {Marquarding}, {Minchin}, {O'Brien}, {Putman},
  {Ryder}, {Sadler}, {Stevens}, {Stewart}, {Stootman}, \& {Waugh}}]{WongO_06a}
{Wong}, O.~I. {et~al.} 2006, \mnras, 371, 1855

\bibitem[{{Wong} \& {Blitz}(2002)}]{WongT_02a}
{Wong}, T., \& {Blitz}, L. 2002, \apj, 569, 157

\bibitem[{{Wright} {et~al.}(2007){Wright}, {Larkin}, {Barczys}, {Erb},
  {Iserlohe}, {Krabbe}, {Law}, {McElwain}, {Quirrenbach}, {Steidel}, \&
  {Weiss}}]{WrightS_07a}
{Wright}, S.~A. {et~al.} 2007, \apj, 658, 78

\bibitem[{{Yabe} {et~al.}(2009){Yabe}, {Ohta}, {Iwata}, {Sawicki}, {Tamura},
  {Akiyama}, \& {Aoki}}]{YabeK_09a}
{Yabe}, K., {Ohta}, K., {Iwata}, I., {Sawicki}, M., {Tamura}, N., {Akiyama},
  M., \& {Aoki}, K. 2009, \apj, 693, 507

\bibitem[{{Zhang} {et~al.}(2009){Zhang}, {Li}, {Kauffmann}, {Zou}, {Catinella},
  {Shen}, {Guo}, \& {Chang}}]{ZhangW_09a}
{Zhang}, W., {Li}, C., {Kauffmann}, G., {Zou}, H., {Catinella}, B., {Shen}, S.,
  {Guo}, Q., \& {Chang}, R. 2009, \mnras, 397, 1243

\bibitem[{{Zwaan} \& {et~al.}(2003)}]{ZwaanM_03a}
{Zwaan}, M.~A., \& {et~al.} 2003, \aj, 125, 2842

\end{thebibliography}
\bibliographystyle{apj}

\pagebreak

\appendix

\section{Specific SFR in the early universe}

Upon completion of this work, 
several groups have used deep {\it Hubble Space Telescope} data
to their limit and published constraints on the SFR sequence beyond $z=4$
\citep{StarkD_09a}, at $z=5$ \citep{YabeK_09a},  and $z=7$ \citep{GonzalezV_10a}.
As shown in Fig.~\ref{fig:SFH:highz}, some groups \citep{StarkD_09a,GonzalezV_10a}
find that the SFR sequence does no longer evolve, i.e., the
sSFR remains constant from $z\sim4$ to $z\sim7$, while \citet{SchaererD_10a} find moderate
and  \citet{YabeK_09a}  strong evolution \citep[see also][]{SawickiM_07a}.
The analysis of \citet{SchaererD_10a}   includes the sample of \citet{GonzalezV_10a} with the added WFC3 data
from \citet{OeschP_10a}.
Given that it is a challenge to derive accurate (dust-corrected) SFR 
and stellar masses at those redshifts in a self-consistent way, 
 it may be premature to view the (non-)evolution of \citet{GonzalezV_10a} beyond $z=2$ as definitely understood. 


\subsection{Assumptions in the model}

\begin{figure}
\plotone{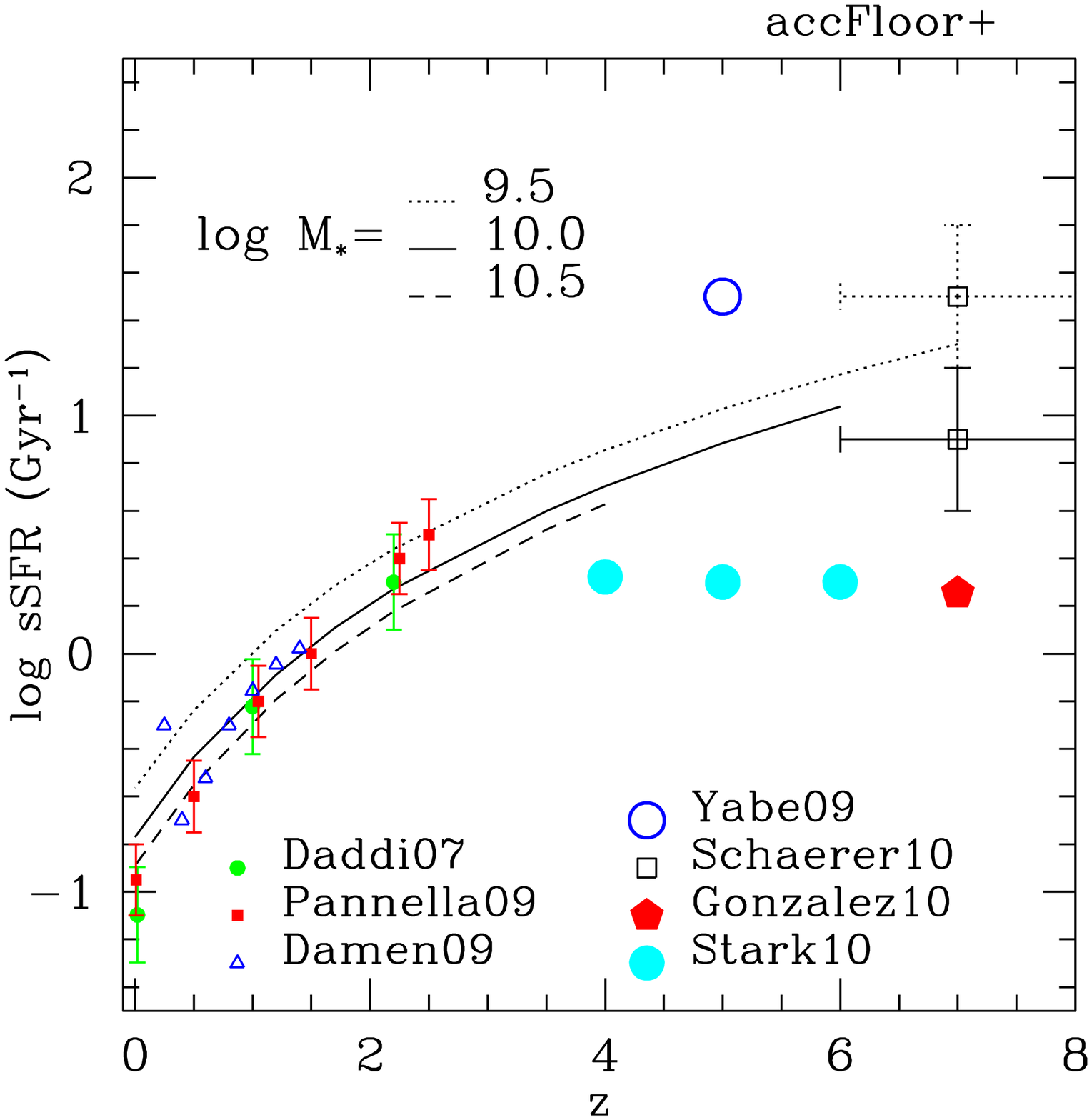}
\caption{The evolution of sSFR  at high-redshifts.
The lines show the predicted predicted evolution by our model (`accFloor+')  at a fixed $\Ms$ as in Fig.~\ref{fig:ssfr}.
At $z<4$, the data points are as in Fig.~\ref{fig:ssfr}.
At $z>4$, the \citet{GonzalezV_10a} and \citet{StarkD_09a} results indicate no-evolution of the sSFR, while \citet{YabeK_09a} indicates strong evolution.
This   clear lack of agreement among the various groups reflects the challenges in deriving accurate (dust-corrected) SFR 
and stellar masses at those redshifts in a self-consistent way (see text).
The re-analysis of the \citet{GonzalezV_10a} sample by \citet{SchaererD_10a} using the recent WFC3 data
 is shown with the open square with solid error bars
(with $\log \Ms\simeq 9.5$).
The `faint' sample of \citet{SchaererD_10a}  with $\log \Ms\simeq 8.0$ is shown with the square with dotted error bars.
\label{fig:SFH:highz}}
\end{figure}

Would the non-evolution of sSFR be real claimed by \citet{GonzalezV_10a}, could the discrepant results of our model be
due to some of our assumptions?
Our model fails to indicate  the turn over of \citet{GonzalezV_10a} in the evolution of
 sSFR($z$)  beyond $z=2$  as shown in Figure~\ref{fig:SFH:highz}.
We turn to some of our assumptions in order to examine the robustness of this failed prediction.
Could our model be modified in such a way to account for this apparent non-evolution
of sSFR?  Our main assumption, the accretion mass floor $\Mmi$,
only alters the mass index of the SFR sequence, but does not alter its redshift evolution.
An assumption certain to break down at $z>4$ is our assumed constant recycling fraction $R$.
Indeed, $R$ is a strong function of stellar ages \citep{BruzualG_03a} for ages $<1$~Gyr.
In order to understand how this affects our calculations in the first Gyr, we note that $R$ affects the SF efficiency. 
The recycled fraction $R$ is low at early times, meaning that the gas consumption is more efficient
via the $(1-R)$SFR factor in Equation~\ref{eq:bathtub}.
Since the SF efficiency is increased, the steady state solution is reached faster, i.e.,
sSFR$(z)$ is going to be set by the redshift dependence of the accretion rate sooner.
Therefore, a more realistic $R(t)$ will  not work in the right direction.

At those early times, the main assumption to break down in our model
is related to the steady-state solution.
As detailed in Section~\ref{section:steadystate}, galaxies in our model
are out of the steady-state  solution beyond $z>5$.
Therefore, the assumption made for the SF timescale
(or efficiency), namely that $t_{\rm sfr}$ scales with the halo dynamical time $t_{\rm dyn}$
 may break down: gas (or fuel for SF) is pouring at a rate
faster than it can be consumed implying that the SF efficiency may be far from the local value.
A more efficient gas consumption (shorter SF time scale) at early times means that
 the steady-state solution is reached more rapidly.
This will only exacerbate the discrepancy between the $z=$4--8 data and the model.
A less efficient gas consumption (longer SF time scale) at early times means that
the SFR is increasing to some power of time $t$ or redshift $(1+z)$.
Thus, the sSFR is always going to be
proportional to sSFR$(t)\propto 1/t$.   

Therefore, upon examining our assumptions, we find that the non-evolution of sSFR beyond $z=2$  cannot be achieved easily. 
In other words, the evolution of sSFR beyond $z=4$ is giving strong constraints on early galaxy formation.
 Taken at face-value, the \citet{GonzalezV_10a}
 results are going to be extremely challenging for any theoretical model.  

\subsection{Assumptions in the observations}

In the observations, in order to fit the observed SEDs, one has to  assume 
(i) a SFH, and (ii) treat the nebular emission lines consistently \citep{SchaererD_09a}.
These two can affect the sSFR significantly.
For instance, at high SFRs and young ages, the nebular emission lines (e.g., [\ion{O}{3}], $H\beta$)
 can affect the color significantly \citep[][Finlator, in prep.]{SchaererD_09a,YabeK_09a}.
When this is omitted, the resulting Balmer break is overestimated, and hence the stellar mass, leading
to an underestimated sSFR.

Similarly, the SFH can have a strong impact on the derived sSFR.
At $z=2$, the often assumed `constant SFH' works generally well \citep[e.g.][]{ForsterSchreiberN_09a} and this assumption
is in good agreement with the growth predicted   (see Fig.~\ref{fig:ssfr} and Section~\ref{section:steadystate}).
At $z>5$, different groups make different assumptions.
For example, \citet{GonzalezV_10a} prefer a constant SFH, while \citet{StarkD_09a}   and \citet{SchaererD_10a}  
 use an exponentially   declining SFH with a range of ages. 
Unfortunately, the SFH is difficult to constrain from the data themselves, but has a significant impact on the sSFR.
Indeed, if   SFR$(t)$ is  constant, then sSFR goes as $1/t$.
If SFR$(t)$ is declining exponentially, then $\Ms(t)$ is dominated by the initial SFR at $t=t_0$ ($\Ms\simeq \exp(-t_0/\tau)$), 
and sSFR is dominated  by the age $\Delta t=t-t_0$, with sSFR$\propto \exp(-\Delta t/\tau)$.
If SFR$(t)$ is increasing exponentially, then $\Ms(t)$ is proportional to $\exp(+t/\tau)$, and the current SFR is also $\propto\exp(+t/\tau)$. Hence,
the sSFR is constant in this case.

Given the short times between $z=5$, $z=6$, and $z=7$, the various populations observed at these epochs are likely descendants
and progenitors of one another and the inferred involution of sSFR$(z)$ ought to be consistent with the assumed SFH.
Currently, it appears that this self-consistency in the observational results is not present.

\subsection{Star formation histories}

\begin{figure}
\plotone{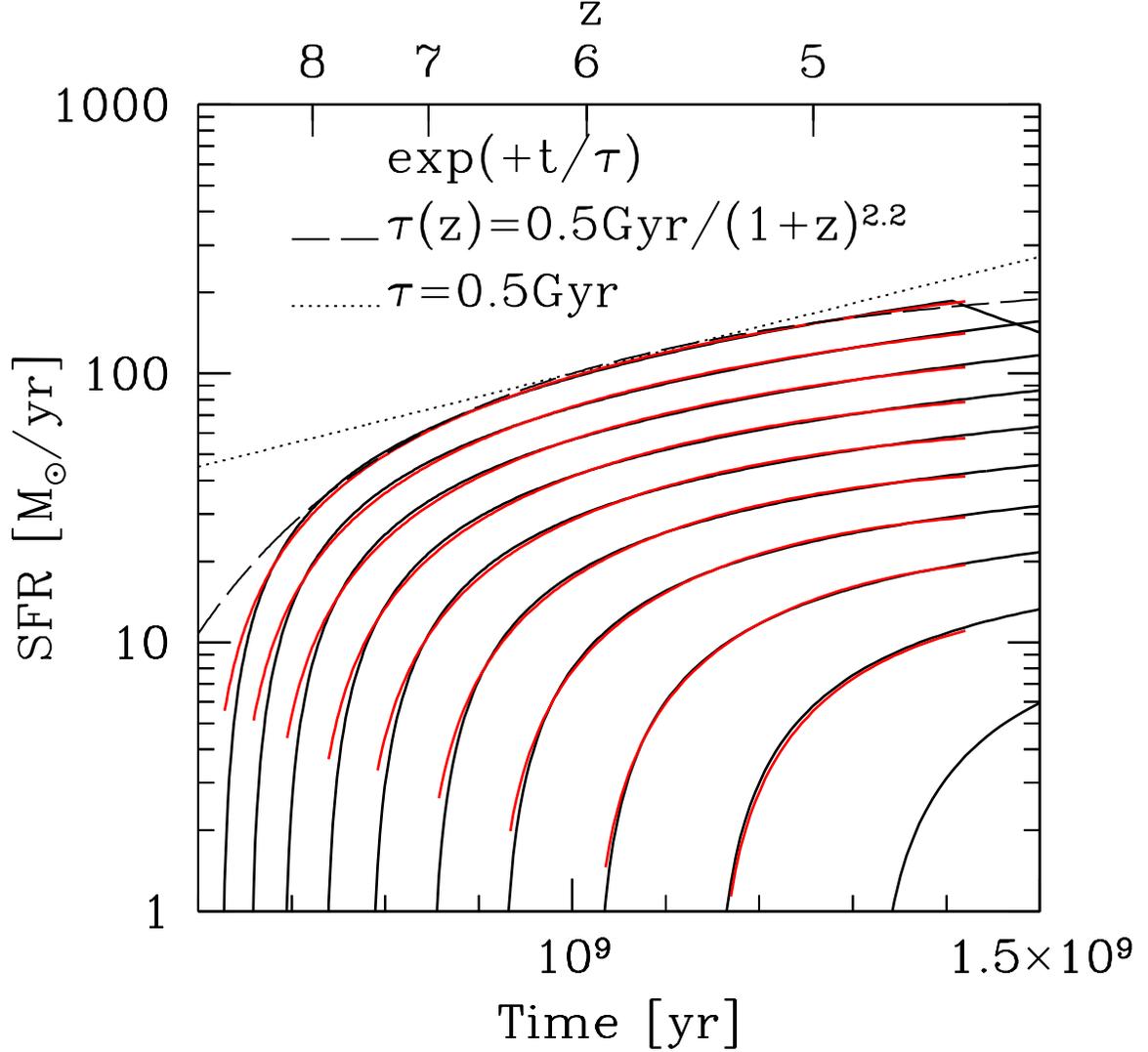}
\caption{At $z>4.5$, the star-formation histories are close to being {\it increasing} exponential, SFR$(t)\propto e^{+t/\tau}$ with $\tau\sim0.5$~Gyr (dotted line). 
Motivated by the halo growth, the dashed line shows an evolving tau $\tau(z)\simeq0.5$Gyr$\left(\frac{1+z}{1+5.5}\right)^{-2.2}$.
The red curves shows  second order polynomial fits.
\label{fig:SFH:fits}}
\end{figure}

At $z>5$, galaxy formation is surely more uncertain. Proto-galaxies are more metal poor, the universe was just re-ionized, two facts that
will impact  molecular gas formation, hence the KS relation itself.  Hence, several of our assumptions are likely to fail at those early epochs.
However, a robust and generic feature of any model at those epochs is the rapid growth rate $\dot M$, which implies that the SFHs are very far from
 exponentially declining.  
In order to guide interested readers with SFHs appropriate in this regime of steady growth (at $z>4$), we note
that in the redshift range $z=5$ to $z=7$, our SFHs are approximate {\it increasing} exponential, SFR(t)$\propto\exp(+t/\tau)$, with
a time scale $\tau$ close to $\sim0.5$~Gyr as shown in Fig.~\ref{fig:SFH:fits} with the dotted line. 
Using $\tau=0.5$~Gyr, one can approximate the SFR$(t)$ with  

\begin{eqnarray}
\hbox{SFR}(t)&\propto& e^{\left (+t/\tau \right)} \hbox{for times $t>t_f$},
\end{eqnarray}
where the formation time $t_f$ is given by $(1+z_f)\simeq0.4\times\Ms^{0.13}$.
As  Fig.~\ref{fig:SFH:fits} shows that $\tau$ varies from $0.3$~Gyr to $0.6$~Gyr
between $z=5$ and $z=7$, a physically motivated form (dashed line) is

\begin{eqnarray}
\hbox{SFR}(t)&=& 100 e^{\left(+(t-t_c)/\tau(z) \right)} \hbox{with $\tau(z)=0.5$Gyr$\left(\frac{1+z}{1+z_b}\right)^{-2.2}$},\label{eq:SFH:tauz}
\end{eqnarray}
where $t_c$ and $z_b$ are the fitted parameters. Table~\ref{table:SFH:taufits} lists the fitted parameters.
A more accurate fit is obtained using polynomials  of the type $a_0+a_1U^1+a_2U^2$,
where $U=t/(10^9$yr), and shown as the red curves in Fig.~\ref{fig:SFH:fits}.

\begin{table*}
\centering
\begin{tabular}{cccc}
Halo ID	&   	$t_c$	(Gyr)	& 	$z_b$ &	$\Ms(z=4)$\\
\hline\\
       8   &   1.73$\pm$0.03 	& 	1.63$\pm$0.10 & 6.9E+8\\
       9   &   2.38$\pm$0.10 	& 	4.32$\pm$0.23 & 2.1E+8\\
      10   &   3.01$\pm$0.19 	& 	6.69$\pm$0.39 & 4.3E+9 \\
      11   &   3.19$\pm$0.20	 & 	8.31$\pm$0.43 & 7.4E+9 \\
      12   &   2.67$\pm$0.10 & 		8.37$\pm$0.30 & 1.1E+10 \\ 
      13   &   2.08$\pm$0.04 	& 	7.82$\pm$0.17 & 1.7E+10 \\
      14   &   1.63$\pm$0.01 	& 	7.04$\pm$0.09 & 2.4E+10 \\
      15   &   1.336$\pm$0.004 &	6.39$\pm$0.04 & 3.3E+10\\
      16   &   1.141$\pm$0.001 &	5.84$\pm$0.03 & 4.3E+10\\
      17   &   1.008$\pm$0.001 &	5.38$\pm$0.02 & 5.1E+10 \\
\hline
\end{tabular}
\caption{Parameters for the SFHs using Equation~\ref{eq:SFH:tauz} \label{table:SFH:taufits}.}
\end{table*}


\end{document}